\documentclass[floatfix,aps,prx,reprint]{revtex4-2}
\usepackage{silence}
\usepackage{palatino}
\usepackage{graphicx}
\usepackage{amssymb}
\usepackage{amsmath}

\usepackage[hidelinks]{hyperref} 
\usepackage{cleveref} 

\usepackage{bm}
\usepackage{float}
\usepackage{mathrsfs}
\usepackage{mathtools}
\usepackage{url}
\usepackage{bbold}
\usepackage{upgreek}
\usepackage{physics}

\usepackage{amsthm}
\newtheorem{lemma}{Lemma}
\newtheorem{corollary}{Corollary}
\newtheorem{remark}{Remark}
\newtheorem{theorem}{Theorem}
\usepackage[normalem]{ulem}

\usepackage{tikz}
\usepackage{circuitikz}
\usetikzlibrary{calc}
\usetikzlibrary{matrix}
\usetikzlibrary{shapes}

\usepackage{commands} 

\Crefname{figure}{Figure}{Figures}
\crefname{figure}{Fig.}{Figs.}
\Crefname{equation}{Equation}{Equations}
\crefname{equation}{Eq.}{Eqs.}
\Crefname{section}{Section}{Sections}
\crefname{section}{Sec.}{Secs.}
\crefname{appendix}{Appendix}{Appendices}
\Crefname{appendix}{Appendix}{Appendices}

\begin{document}


\title{Measurement-induced phase transitions in quantum inference problems and quantum hidden Markov models}
\author{Sun Woo P. Kim}
\email{swk34@cantab.ac.uk}
\affiliation{Department of Physics, King's College London, Strand, London WC2R 2LS, United Kingdom}
\author{Austen Lamacraft}
\affiliation{TCM Group, Cavendish Laboratory, University of Cambridge, Cambridge CB3 0HE, United Kingdom}
\author{Curt von Keyserlingk}
\affiliation{Department of Physics, King's College London, Strand, London WC2R 2LS, United Kingdom}

\begin{abstract}
Recently, there is interest in coincident `sharpening' and `learnability' transitions in monitored quantum systems. In the latter, an outside observer's ability to infer properties of a quantum system from measurements undergoes a phase transition. Such transitions appear to be related to the decodability transition in quantum error correction, but the precise connection is not clear. Here, we study these problems under one framework we call the general quantum inference problem. In cases as above where the system has a Markov structure, we say that the inference is on a quantum hidden Markov model. We show a formal connection to classical hidden Markov models and that they coincide for certain setups. For example, we prove this for those involving Haar-random unitaries and measurements. We introduce the notion of Bayes non-optimality, where parameters used for inference differs from true ones. This allows us to expand the phase diagrams of above models. At Bayes optimality, we obtain an explicit relation between `sharpening' and `learnability' order parameters, explicitly showing that the two transitions coincide. Next, we study concrete examples. We review quantum error correction on the toric and repetition code and their mapping to 2D random-bond Ising model (RBIM) through our framework. We study the Haar-random $\mathrm{U}(1)$-symmetric monitored quantum circuit and tree, mapping each to inference models that we call the planted SSEP and planted XOR, respectively, and expanding the phase diagram to Bayes non-optimality. For the circuit, we deduce the phase boundary numerically and analytically argue that it is of a single universality class. For the tree, we present an exact solution of the entire phase boundary, which displays re-entrance as does the 2D RBIM. We discuss these phase diagrams, with their interpretations for quantum inference problems and rigorous arguments on their shapes.
\end{abstract}

\maketitle

\section{Introduction}
Recently, there has been considerable attention on measurement-induced phase transitions on certain monitored quantum circuits. Here, as the rate or strength of measurements is increased, there is a coincident `observable sharpening' and `learnability' transition \cite{agrawal2022entanglement,agrawal2023observing,barratt2022field,barratt2022transitions}. In the former case, the variance of an observable on the final quantum state undergoes a phase transition. While for the latter, the phase transition is for the ability for a `decoder' to correctly predict the observable given the measurement outcomes. The appearance of inference in the latter case suggests a connection to other active areas in physics: quantum error correction, and statistical physics of inference: the latter is less discussed in the quantum community.

\paragraph*{Sharpening/learnability transition in monitored circuits ---} More concretely, in the `sharpening' transition, one prepares an initial state $\ket{\psi_0}$ and performs a monitored evolution characterised by measurement strength or rate $\epsilon$, obtaining a particular trajectory of measurements and final state $\ket{\psi_t}$. Then, for typical trajectories and a particular observable $\hat O$, the variance $\delta O^2 = \bra{\psi_t} \hat O^2 \ket{\psi_t} - \bra{\psi_t} \hat O \ket{\psi_t}^2$ undergoes a phase transition from a finite value to zero as $\epsilon$ increases \cite{agrawal2022entanglement}. In the `learnability' transition, we consider `Alice' who chooses an initial state from an ensemble of states with definite values of the observable $\{\ket{\psi_O}\}$ with probability $p(\ket{\psi_O})$, performs the monitored evolution, then obtains the measurement outcomes $Y$. An eavesdropper `Eve' intercepts \emph{only} the measurement outcomes and tries to guess $O$. The classification error of Eve's guess undergoes a phase transition from a finite value to zero with $\epsilon$ \cite{agrawal2023observing}.

\paragraph*{Quantum error correction ---} A simplified model for quantum error correction (QEC) can be described as follows \cite{dennis2002topological}. First, we prepare a quantum state $\ket{\psi}$ that encode some quantum information. Then, we suppose that, due to the environment, errors $E$ occur with some probability $p(E)$, sending the state to $\hat{E} \ket{\psi}$. Then, we measure some syndromes $S$ that depends on the errors $E$. We then use a `decoder' to infer the operation to correct the errors and revert the state back to the original $\ket{\psi}$. A natural question is whether there is a transition in the ability to successfully error-correct as the rate of errors varies. The point at which this occurs is the \emph{threshold}. Within this general setting, we can generalise to a case where our readout of the syndromes is unreliable, and/or the case where we accumulate errors $e_t$ and measure syndromes $s_t$ at each timestep $t$ before we apply an operation at some final time.

\paragraph*{Bayesian inference ---}
In classical inference problems, we are interested in a state of the system $\bm{x}$. As the inferrer, we assume a prior distribution $p(\bm{x})$ for the state. We do not have direct access to the state $\bm{x}$, but instead have access to observations $\bm{y}$ that we assume are correlated with $\bm{x}$ according to a likelihood/`observation model' $p(\bm{y} \vert \bm{x})$. Our goal is to infer $\bm{x}$ given $\bm{y}$. To do this we compute the posterior $p(\bm{x} \vert \bm{y})$ according to Bayes' rule,
\begin{align}
    p(\bm{x} \vert \bm{y}) = \frac{p(\bm{y} \vert \bm{x}) p(\bm{x})}{p(\bm{y})},
\end{align}
where the marginal likelihood is given by $p(\bm{y}) = \sum_{\bm{x}} p(\bm{y} \vert \bm{x}) p(\bm{x})$.

We may also assume additional structure, such as that of a hidden Markov model (HMM) \cite{rabiner1986introduction, pkim2025planted}. Here, the state at time $t$, $\bm{x}_t$ undergoes a Markov process, and therefore the prior for the entire trajectory is given by
\begin{align}
    p(\bm{x}_{1:t}) = \left(\prod_{\tau=2}^t p(\bm{x}_\tau \vert {\bm{x}}_{\tau-1})\right) p(\bm{x}_1),
\end{align}
where we have introduced the notation $\bm{x}_{1:t}$ to denote a sequence of variables ${\bm{x}_1, \ldots \bm{x}_t}$. At every timestep, information about the current state $\bm{x}_t$ is transmitted through observations $\bm{y}_t$ according to $p(\bm{y}_t \vert \bm{x}_t)$, leading to total likelihood
\begin{align}
    p(\bm{y}_{1:t} \vert \bm{x}_{1:t}) = \prod_{\tau=1}^t p(\bm{y}_t \vert \bm{x}_t).
\end{align}
With the additional structure, one can consider various inference tasks. One is a `filtering' task, where the posterior for the current state is inferred from a history of measurements, $p(\bm{x}_t \vert \bm{y}_{1:t})$. It can be shown that the unnormalised posterior distribution $q(\bm{x}_t \vert \bm{y}_{1:t})$ of the \emph{current} state $\bm{x}_t$ given the history of measurements $\bm{y}_{1:t}$, evolves linearly and as a Markov process according to the \emph{forward algorithm} \cite{stratonovich1965non}
\begin{align} \label{eq:forward-equation}
    q(\bm{x}_t \vert \bm{y}_{1:t}) = \sum_{\bm{x}_{t-1}} p(\bm{y}_t \vert \bm{x}_t) p(\bm{x}_t \vert \bm{x}_{t-1}) q(\bm{x}_{t-1} \vert \bm{y}_{1:t-1}),
\end{align}
with the initial condition $q(\bm{x}_1 \vert \bm{y}_{1:1}) = p(\bm{y}_1 \vert \bm{x}_1) p(\bm{x}_1)$. In fact, the unnormalised posterior is just the joint distribution $p(\bm{x}_t, \bm{y}_{1:t}) = q(\bm{x}_t \vert \bm{y}_{1:t})$. Importantly, the evolution of the posterior for the filtering task is Markovian, i.e., the posterior for the future timestep only depends on that of the current timestep $q(\cdot \vert \bm{y}_{1:t-1})$ and new observation  $\bm{y}_t$. This is a consequence of the structure of HMMs \cite{rabiner1986introduction}.

\paragraph*{General quantum inference problem --- } In this work, we consider an inference problem that we call the general quantum inference problem. Here, we consider a quantum system where certain channels records their outcomes in classical registers. Some of these registers are revealed to the observer, while others are hidden. Then, the task for the observer is to infer the hidden register given the revealed ones. Within this general framework, one can also define quantum hidden Markov models (qHMMs), which have a similar structure as classical HMMs on the level of density matrices. All the aforementioned inference problems are special cases of the general quantum inference problem.

\paragraph*{Outline --- }
In \cref{sec:teacher-student}, we introduce the teacher-student framework for classical inference problems where a `student' infers from data provided by the `teacher'. This terminology is taken from the field of \emph{statistical physics of inference} \cite{zdeborova2016statistical}. Then, in \cref{sec:bayes-optimality}, we discuss the notion Bayes optimality in inference problems, equivalent to the `Nishimori condition', which is where the model assumed by the student agrees with the teacher. We show that we have an exact relationship between order parameters of inference (or `learnability') and uncertainty (or `sharpening') in this case. We then introduce the general quantum inference problem in \cref{sec:general-quant-inference-problem}. In \cref{sec:qhmms}, we consider an additional Markov structure similarly to classical inference problems and introduce quantum Hidden Markov models (qHMMs), which can be viewed as a generalisation of HMMs, and discuss under what conditions they reduce to HMMs. As concrete examples, in \cref{sec:haar-case}, we prove that inference for qHMMs with Haar random unitaries under certain conditions reduce to HMMs.

Next, we study three examples through the lens of inference. First, in \cref{sec:2d-rbim}, we review two QEC problems under bit-flip errors: the toric code, and the quantum repetition code with read-out errors on the syndrome measurements. We show that both setups can be cast as inference problems and map to the 2D random-bond Ising model (RBIM), and that the repetition code case has a structure of a HMM. Second, in \cref{sec:planted-ssep} we study the Haar-random $\rm{U}(1)$-symmetric monitored quantum circuit, which maps to a classical inference problem that we call the \emph{planted simple symmetric exclusion process (planted SSEP)}, where the state of a SSEP is inferred from its noisy `images'. Under the teacher-student scenario, we go beyond existing work and expand the phase diagram to Bayes non-optimality in student and teacher parameter space. Using a combination of numerical and analytic arguments, we deduce its phase boundary and universality class. Third, in \cref{sec:planted-xor}, we go the other way and generate a quantum inference problem from a classical one. We start from the \emph{planted XOR} model, a classical inference problem on a HMM where a Boolean variable, resulting from XOR operations with a tree structure, is inferred from a series of noisy measurements. This maps to a Haar-random $\rm U(1)$-symmetric monitored quantum tree. We exploit the structure of the tree to analytically obtain the full expanded phase diagram in the space of student and teacher parameters. In particular, we uncover a re-entrance in the phase diagram, which is also present in the 2D RBIM. Finally, in \cref{sec:discussion}, we discuss the phase diagrams of these inference problems, with a rigorous argument on their shape and their interpretation in the quantum case.


\textit{We note that while this manuscript was being finalised, a preprint Ref. \cite{nahum2025bayesian} appeared on arXiv. Its observations have some overlap with ours under the condition of Bayes optimality, while our use of the teacher-student scenario allows non-optimal inference to be explored.}

\tableofcontents

\begin{figure*}[t!]
  \centering
  \includegraphics[width=\linewidth]{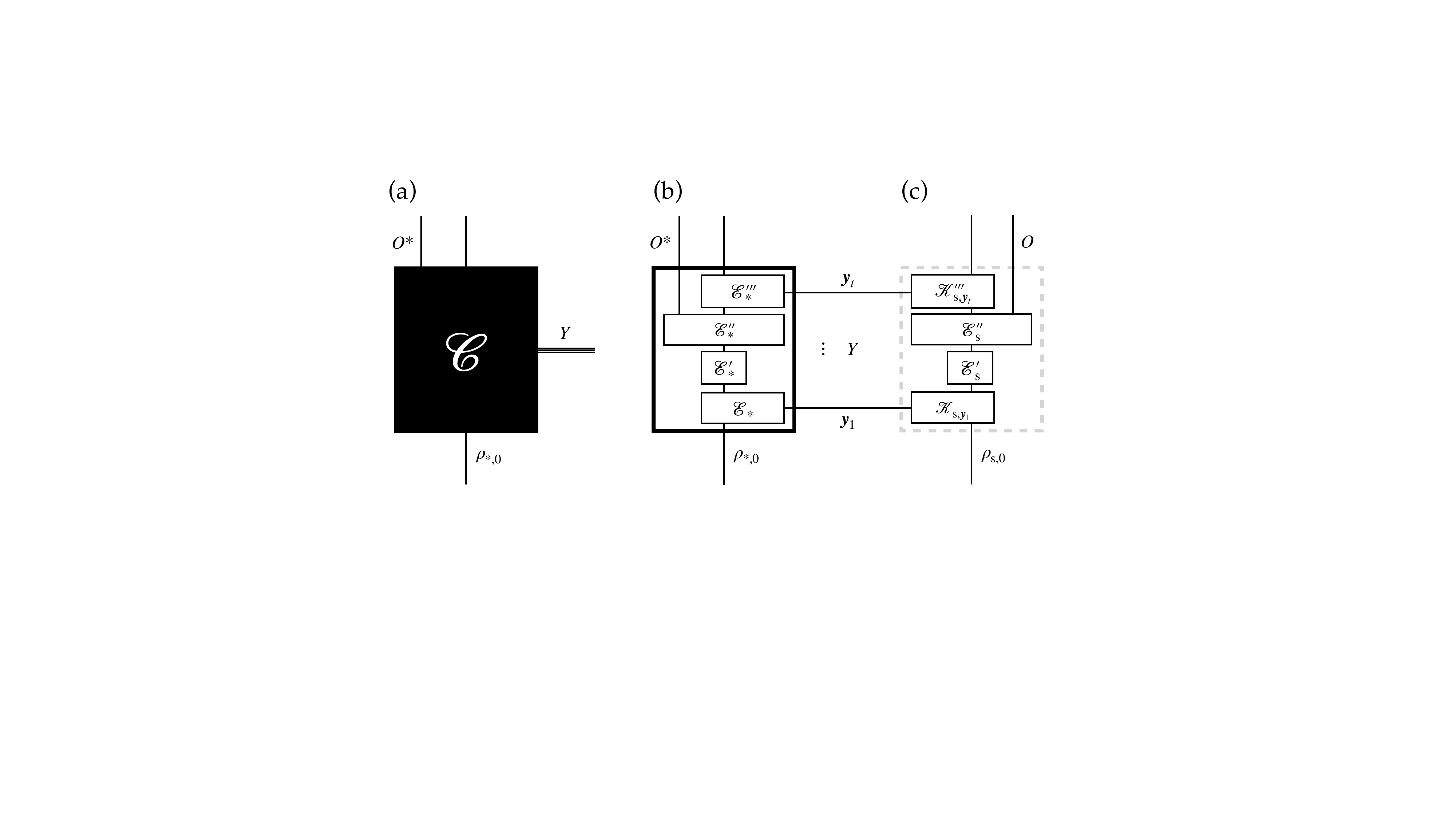}
  \caption{\textbf{General quantum inference problems and on quantum hidden Markov models (qHMMs).} (a), (b): In the general quantum inference problem, the `teacher' (a) applies a quantum channel $\mathcal{C}_*$ to their quantum state, which generates a hidden register $O^*$ and revealed registers $Y$. Given $Y$, the `student' (b) conditions their channel by projecting on to $Y$ via $\mathcal{T}_Y[\cdot] := \tr_{\mathcal{Y}}[\mathcal{P}_Y(\cdot)]$. The student assumes a channel $\mathcal{C}_\mathrm{s}$, which may differ from the teacher's. Given their conditioned density matrix $\rho_\mathrm{s}(Y) = \mathcal{T}_Y \mathcal{C}_\mathrm{s} [\rho_{\mathrm{s}, 0}]$, the student can compute a posterior on $O$ given $Y$, $p_\mathrm{s}(O \vert Y) = \tr[O \rho_\rm{s}(Y)]$. (c), (d): Additional causal structure can be put on quantum inference problems. If the channel at each timestep does not take previous revealed registers as input, then the evolution of the student's density matrix is Markovian and inference is on a qHMM. 
  } \label{fig:qip-qhmm}
\end{figure*}

\section{Teacher-student framework and the planted ensemble} \label{sec:teacher-student}

In the above discussion, we considered a highly idealised scenario we have perfect information about the distribution that generates the measurements/syndromes/observations. A more accurate situation would be when they differ. In statistical physics of inference, there is a standard language to describe the inference problem, namely the teacher-student scenario \cite{zdeborova2016statistical}. Here, we discuss the framework in the classical Bayesian inference setting.

We consider the ``teacher'', an underlying phenomenon that produces some random variable $\hat{\bm{x}}^*$ which we call the teacher variable. 
The teacher variable is picked from $p_*(\cdot)$, which we will call the teacher's prior for reasons that will follow. Our phenomenon is then observed, giving rise to a new random variable $\hat{y}$ which is a noisy function of $\hat{\bm{x}}^*$; it is encoded by a conditional distribution $p_*(\hat{\bm{y}} = \bm{y} \vert \hat{\bm{x}}^*=\bm{x}^*)$; henceforth we will denote such expressions as $p_*(\bm{y} \vert \bm{x}^*)$ for brevity. This is a convenient abuse of notation, as we are distinguishing different distributions by their arguments. We will refer to this conditional distribution as the likelihood or the obsevation model.

We the observer, or "student", are given $\hat{\bm{y}}$. We have a probabilistic model, which is not necessarily correct, for how $\hat{\bm{y}}$ relates to the underling phenomenon $\hat{\bm{x}}^*$. We call the corresponding random variable $\hat{\bm{x}}$, and whose distribution is encoded by the conditional probability $p_\mathrm{s}(\bm{x} \vert \bm{y})$. If the student had a perfect model of the experiment, then we would have $p_s(\cdot\vert\bm{y})=p_*(\cdot\vert\bm{y})$.

As the student, our probabilistic model of the experiment would be the \emph{posterior} assuming a particular likelihood $p_\mathrm{s}(\bm{x} \vert \bm{y})$ and prior $p_\mathrm{s}(\bm{x})$. For the remainder of the article, we will consider the case where the student has access to the functional forms of the distributions but does not necesarily the correct parameters. Therefore we will denote the general functional forms without subscripts, for example likelihood as $p(\bm{x} \vert \bm{y})$, and use $\mathrm{s}$ or $*$ to specify if it is the student's or the teacher's.

The joint distribution
\begin{equation} \label{eq:planted-ensemble}
    \begin{aligned}
        p(\bm{x}, \bm{y}, \bm{x}^*) & = p_\mathrm{s}(\bm{x} \vert \bm{y}) p_*(\bm{y} \vert \bm{x}^*) p_*(\bm{x}^*) \\
        & = \frac{p_\mathrm{s}(\bm{y} \vert \bm{x}) p_\mathrm{s}(\bm{x}) p_*(\bm{y} \vert \bm{x}^*) p_*(\bm{x}^*)}{p_\mathrm{s}(\bm{y})}
    \end{aligned}
\end{equation}
is known as the \emph{planted ensemble} \cite{zdeborova2016statistical}. In some cases, it may be beneficial to regard the student's posterior $p_\rm{s}(\bm x \vert \bm y)$ as a Gibbs distribution. This arises, for example, in the study of spin glasses. In this language, we can map $p_\mathrm{s}(\bm{y} \vert \bm{x}) p_\mathrm{s}(\bm{x} )\leftrightarrow e^{-\beta H(\bm{x} \vert \bm{y})}$ and the marginal likelihood/evidence $p_\mathrm{s}(\bm{y}) \leftrightarrow Z(\bm{y})$ as a partition function, dependent on some `disorder' $\bm{y}$. In disordered statistical mechanical systems, $\bm{y}$ is often chosen iid. In the planted ensemble, by contrast, it is distributed as

\begin{align}
    p(\bm{y}) = \sum_{\bm{x}^*} p_*(\bm{y} \vert \bm{x}^*) p_*(\bm{x}^*),
\end{align}
so the distribution is affected by the teacher configurations $\bm{x}^*$ that give rise to it. $\bm{x}^*$ is sometimes said to be "planted" in $\bm{y}$, which gives rise to the name of the ensemble.

\section{Bayes optimality and moment relations} \label{sec:bayes-optimality}
In statistical physics of Bayesian inference, when the student and teacher distributions coincide $(*=\mathrm{s})$, the student's model is said to be `Bayes optimal'. The naming is apt for the following reasons. Consider a classifier $h(\bm{y})$ that predicts the hidden state given the observations $\bm{y}$. Then the classification performance is given by the average overlap between $h(\bm{y})$ and $\bm{x}^*$, $\E_{\bm{y}, \bm{x}^*}[\delta_{h(\bm{y}), \bm{x}^*}]$. Then, the optimal classifier that maximises the overlap is $\mathrm{argmax}$ of the true posterior \cite{bishop2006pattern},
\begin{align} \label{eq:optimal-classifier}
    h_\mathrm{optimal}(\bm{y}) = \mathop{\mathrm{argmax}}_{\bm{x}} p_*(\bm{x} \vert \bm{y}).
\end{align}
For real-valued state observables $O(\bm x)$, consider an estimator $\tilde{O}(\bm y)$ given $\bm{y}$.
One possible figure of merit 
would be the mean-squared-error $\E_{\bm y, \bm x^*}[(\tilde{O}(\bm y) - O(\bm x^*))^2]$. 
Then, the optimal estimator that minimises mean-squared-error of the estimator is the expectation of the true posterior \cite{pishro2014introduction},
\begin{align} \label{eq:optimal-estimator}
    \tilde{O}_\mathrm{optimal}(\bm{y}) = \E_{\bm{x} \sim p_*(\cdot \vert \bm{y})}[O(\bm{x})].
\end{align}
At Bayes optimality, the joint distribution \cref{eq:planted-ensemble} is symmetric with respect to $\bm{x} \leftrightarrow \bm{x}^*$. This means that the marginal distribution for $\bm{x}$ must have the same distribution as that of $\bm{x}^*$. In turn, sampling from the student's posterior distribution is identical to sampling from the true distribution. 

In the spin glass and statistical physics of inference literature, Bayes optimality is also known as the Nishimori condition \cite{zdeborova2016statistical}. 

Given an inferred posterior $p_\mathrm{s}(\bm{x} \vert \bm{y})$ and an observable $O(\bm{x})$, there are a few natural `order parameters' that quantify the performance of the inference. 

The first natural quantity is the mean-squared-error of the mean estimator from the inferred posterior (MSEM),
\begin{align} \label{eq:msem}
    \mathrm{MSEM} := \E_{\bm{y}, \bm{x}^*} \left[ \left( \left\langle O(\bm{x}) \right\rangle_{\mathrm{s}} - O(\bm{x}^*) \right)^2\right],
\end{align}
where $\left\langle O(\bm{x}) \right\rangle_{\mathrm{s}} = \E_{\bm{x} \sim p_\mathrm{s}(\cdot \vert \bm{y})}[O(\bm{x})]$ is the expectation on the inferred posterior. It can be viewed as the student's estimator $\tilde O(\bm y)$. Note that at Bayes optimality, the mean on the inferred posterior is the optimal estimator.

The second natural quantity in the context of the planted ensemble, is the mean-squared error (MSE),
\begin{align} \label{eq:mse}
    \mathrm{MSE} := \E_{\bm{x}, \bm{y}, \bm{x}^*}\left[\left(O(\bm{x}) - O(\bm{x}^*)\right)^2\right],
\end{align}
where the expectation is over the planted ensemble \cref{eq:planted-ensemble}. This case can be thought of as the student obtaining their estimator $\tilde O(\bm{y})$ by sampling from their posterior.

The third quantity is the variance of $O$ on the inferred posterior distribution,
\begin{align} \label{eq:dO2}
    \delta O^2_\mathrm{s} := \mathbb{E}_{\bm{y}, \bm{x}^*} \left[ \langle O(\bm{x})^2 \rangle_\mathrm{s} - \langle O(\bm{x}) \rangle^2_\mathrm{s} \right],
\end{align}
which measures the uncertainty of the student on their estimate of the observable. 

At Bayes optimality, exploiting the symmetry between the student and teacher replicas, we can derive an exact relation between \cref{eq:mse,eq:dO2,eq:msem},
\begin{align}
    \delta O^2_\mathrm{s}(\mathrm{Bayes \; optimal}) & = \mathrm{MSEM}(\mathrm{Bayes \; optimal}) \nonumber \\ 
    & = \frac{1}{2} \mathrm{MSE}(\mathrm{Bayes \; optimal}). \label{eq:accuracy-precision}
\end{align}
Therefore, on average, uncertainty in the student's posterior is equal to its error. The derivation is detailed in \cref{apdx:moment-relatons}.

\section{The general quantum inference problem} \label{sec:general-quant-inference-problem}
In this section, we use the language of the teacher-student scenario discussed in \cref{sec:teacher-student} to unify inference problems in quantum systems. We call this framework the general quantum inference problem.

Consider the teacher who prepares a quantum state $\rho_{*, 0}$. This may be a mixed state, if they were to choose from an ensemble of pure state with some probability, or a pure state. The teacher applies a quantum channel $\mathcal{C}_*$ to the state. The internal structure of the channel may be complicated, but let us assume that the teacher records some of the outcomes in classical registers. For example, they may be outcomes of measurements, or records of application of error channels. Out of all the registers, the teacher will reveal some of these $\ketbra{Y}{Y}$ to the student, while hiding others $\ketbra{O^*}{O^*}$. This is shown schematically in \cref{fig:qip-qhmm}(a). After the application of the channels the teacher's density matrix can be written as
\begin{align}
    \mathcal{C}_* \rho_{*, 0} & = \sum_{O^*, Y} \rho_*(O^*, Y) \otimes \ketbra{Y}{Y} \otimes \ketbra{O^*}{O^*} \\
    & = \sum_Y \rho_*(Y) \otimes \ketbra{Y}{Y}.
\end{align}
In the second line, we have grouped $\rho_*(Y) = \sum_{O^*} \rho_*(O^*, Y) \otimes \ketbra{O^*}{O^*}$. Its trace gives the probability of the teacher recording $Y$, $p_*(Y)$. Additionally, the joint probability of recording both $O^*$ and $Y$ is given by $p_*(O^*, Y) = \tr[\rho_*(O^*, Y)]$.

The goal of the student is to infer the hidden register $O^*$ given the revealed registers $Y$. To do this, the student simulates the density matrix conditioned --- in a manner we describe below --- on the revealed registers.
In their simulation, the student applies a channel $\mathcal{C}_\mathrm{s}$. This channel may differ from the teacher's in general. As discussed before, we assume that it has the same functional form as $\cl C_*$ and only differs in its parameters, unless stated otherwise. After $\mathcal{C}_\mathrm{s}$, the student conditions on the revealed register $Y$ by applying $\mathcal{T}_Y[\cdot] = \tr_{\mathcal{Y}} [\mathcal{P}_Y(\cdot)]$. Here, $\mathcal{P}_Y$ is a projector onto the revealed registers and the trace is over the register degrees of freedom $\mathcal{Y}$. This in general will reduce the norm of the student's density matrix. The student's procedures are shown in \cref{fig:qip-qhmm} (b). Then the student's unnormalised density matrix conditioned on $Y$ is
\begin{equation}
  \begin{aligned} \label{eq:students-dm}
      \rho_\mathrm{s}(Y) & = \mathcal{T}_Y \mathcal{C}_\mathrm{s} \rho_{\mathrm{s},0} \\
      & = \sum_{O} \rho_\mathrm{s}(O, Y) \otimes \ketbra{O}{O},
  \end{aligned}
\end{equation}
where $O$ denotes the student's hidden register. The trace of \cref{eq:students-dm} is the probability that the student's model produces register $Y$.

The student can calculate their posterior distribution for the observable as
\begin{align}
p_\mathrm{s}(O \vert Y) = \frac{\tr\big[\mathcal{P}_O \rho_\mathrm{s}(Y)\big]}{\tr[\rho_\mathrm{s}(Y)]},
\end{align}
where $\mathcal{P}_O$ projects the student's hidden registers onto a particular outcome $O$.

For the purposes of analysis, it will be convenient to consider the student's density matrix simulation as an actual density matrix. Then the joint density matrix between the student, the revealed registers, and the teacher, is given by
\begin{align} \label{eq:joint-density-matrix}
  \rho = \sum_{Y} \frac{\rho_\mathrm{s}(Y)}{\tr[\rho_\mathrm{s}(Y)]} \otimes \ketbra{Y}{Y} \otimes \rho_{*}(Y),
\end{align}
where similarly to the student, $\rho_*(Y) = \mathcal{T}_Y \mathcal{C}_* \rho_{*, 0}$. Here we have normalised the student's density matrix so that the joint density matrix is properly normalised. \cref{eq:joint-density-matrix} is the quantum analogue of the planted ensemble \cref{eq:planted-ensemble}.

At the end, the teacher can `grade' the student's inference. One possibility is that the student calculates an estimator $\tilde O(Y)$ by taking the expectation over their posterior, $\tilde O(Y) = \sum_O O p_\mathrm{s}(O \vert Y)$, then hands it to the teacher. The teacher then calculates the MSEM. $\tilde O(Y)$ can be expressed in terms of \cref{eq:joint-density-matrix} as $\tilde O_Y = \tr[O_\mathrm{s} \mathcal{P}_Y \rho]$, where the operator $O_\mathrm{s}$ acts on the student's register. Then the MSEM averaged over the experimental runs is found by
\begin{align}
\mathrm{MSEM} = \sum_Y \tr[\left(O_*-\tilde O(Y)\right)^2 \mathcal{P}_Y \rho],
\end{align}
where $O_*$ is an operator acting on the teacher's hidden register. 
Another way the student could produce an estimator is by sampling from their posterior, $\tilde O(Y) \sim p_\rm{s}(\cdot \vert Y)$, which is then averaged over the experimental runs. This gives the MSE, which can be written in terms of the joint density matrix as
\begin{align}
\mathrm{MSE} = \sum_Y \tr[(O_* - O_\mathrm{s})^2 \mathcal{P}_Y \rho].
\end{align}

Within this formalism we can also study the observable sharpening transition. In this case we can simply ignore the student by tracing them out. Then the observable variance of the state can be calculated by
\begin{align} \label{eq:quantum-dO2}
\delta O^2_* = \sum_Y \left(\frac{\tr[O^2_* \mathcal{P}_Y \rho]}{\tr[\mathcal{P}_Y \rho ]} - \frac{\tr[O_* \mathcal{P}_Y \rho]^2}{\tr[\mathcal{P}_Y \rho ]^2} \right).
\end{align}
One can also consider the observable variance of the student's posterior $\delta O^2_\mathrm{s}$. This is given by replacing $*\rightarrow\mathrm{s}$ in \cref{eq:quantum-dO2}. Note that at Bayes optimality the joint density matrix \cref{eq:joint-density-matrix} is symmetric with respect to exchange of student and teacher. Therefore,
\begin{align} \label{eq:dO2s-dO2t}
  \delta O^2_* = \delta O^2_\mathrm{s}(\mathrm{Bayes \; optimal}),
\end{align}
i.e. the observable variance of the student's posterior is equal to the observable variance of the true quantum state.

Moreover, using the moment relations of \cref{eq:accuracy-precision}, this exactly implies that the `observable-sharpening' transition corresponds to the `learnability' transition at Bayes optimality.

\section{Inference on quantum hidden Markov models} \label{sec:qhmms}
Now consider the case where the whole channel $\mathcal{C}$ can be broken down into a sequence of channels in time, as in \cref{fig:qip-qhmm}(c). There are two kinds of channels in this sequence. For the first kind, denoted as $\cl R_*$, the outcomes are recorded, while for the second kind, $\cl E_*$, they are not.

For the recorded channels, the teacher records the outcome in a register as
\begin{align}
    \mathcal{R}_* : \rho_* \rightarrow \sum_{\bm{a}} K_{*, \bm{a}} \rho_* K_{*, \bm{a}}^\dagger \otimes \ketbra{\bm{a}}{\bm{a}}.
\end{align}
Here, $\{K_{*,\bm{a}}\}_{\bm a}$ are Kraus operators which obey the canonical Kraus map condition. Out of all the registers, some $Y = \bm{y}_{1:t}$ will be revealed to the student, while others $O^*$ will not.

Now we assume that the channel applied by the teacher at each step does not depend on the outcomes of the previous \emph{revealed} registers. Then the student can condition each channel with the revealed register $\bm{y}_t$, separately, as
\begin{align}
  \mathcal{T}_{\bm{y}_t} \mathcal{R}_\rm{s} = \mathcal{K}_{\rm{s}, \bm{y}_t} : \rho_\rm{s} \rightarrow K_{\rm{s}, \bm{y}_t} \rho_\rm{s} K_{\rm{s}, \bm{y}_t}^\dagger.
\end{align}
In this case, we say that the inference is on a quantum hidden Markov model (qHMM) \cite{monras2010hidden}.

We can formally connect qHMMs to HMMs as follows. Consider one `timestep' as the interval between applying a operator conditioned on a revealed register $\mathcal{K}_{\bm{y}_t}$. We group all channels between such Kraus maps as $\mathcal{E}_t$. Then, in one timestep, the student's density matrix evolves as
\begin{align} \label{eq:qhmm-evolution}
  \rho_\mathrm{s}(\bm{y}_{1:t}) = \mathcal{K}_{\rm{s}, \bm{y}_t}\mathcal{E}_{\rm{s}, t} \rho_\mathrm{s}(\bm{y}_{1:t-1}).
\end{align}
Now consider the forward equation \cref{eq:forward-equation} for HMMs on a student's model. If we represent the unnormalised posterior distribution as a probability vector, $\lvert q_\rm{s} (\bm{y}_{1:t}))$, such that $q_\rm{s} (\bm{x}_t \vert \bm{y}_{1:t}) = (\bm{x}_t \vert q_\rm{s} (\bm{y}_{1:t}))$, then the forward equation can be written as
\begin{align} \label{eq:hmm-vector-evolution}
  \keT{q_\rm{s} (\bm{y}_{1:t})} = \mathsf{K}_{\rm{s}, \bm{y}_t} \mathsf{E}_{\rm{s}, t} \keT{q_\rm{s} (\bm{y}_{1:t-1})}.
\end{align}
Here, $\sf{K}_{\rm{s}, \bm y_t}$ is  is the diagonal measurement model/likelihood matrix defined as $(\bm{x}_t \vert \mathsf{K}_{\rm{s}, \bm{y}_t} \vert \bm{x}_t) = p_\rm{s} (\bm{y}_t \vert \bm{x}_t)$, and $\sf{E}_{\rm{s}, t}$ is the prior Markov kernel matrix defined as $(\bm{x}_{t+1} \vert \mathsf{E}_{\rm{s}, t} \vert \bm{x}_t) = p_\rm{s} (\bm{x}_{t+1} \vert \bm{x}_t)$.

Therefore the unconditioned channel $\mathcal{E}_t$ and conditioned channel $\mathcal{K}_{\bm{y}_t}$ of qHMM corresponds to the prior Markov kernels and measurement model/likelihood in HMM, respectively. As mentioned earlier, the only Bayes non-optimal situations we consider are where the student's model has the same functional form as an optimal model of the system but perhaps the wrong parameters. Note, however, that the student is free to use a qHMM model described by \cref{eq:qhmm-evolution} irrespective of the true structure of the teacher's channel, which could even be non-Markovian.  In this case the teacher would have the general structure of \cref{fig:qip-qhmm}(a) while the student uses \cref{fig:qip-qhmm}(d).

In general, the student may need to simulate the full density matrix in order to obtain an accurate estimate for the hidden register $O^*$. However, depending on the particular hidden register that the student is trying to infer and the structure of the channels, it may be sufficient for the student to evolve a reduced density matrix on a subset of degrees of freedom whose evolution is still Markovian. Furthermore, it may be that the reduced density matrix is completely diagonal in a certain basis, $\rho_s^{\mathrm{reduced}} = \sum_{\bm{x}_t} q_\mathrm{s}(\bm{x}_t \vert \bm{y}_{1:t}) \ketbra{\bm{x}_t}{\bm{x}_t}$. In these cases, where the quantum system is accurately captured by a classical HMM, the diagonal matrix elements $q_\mathrm{s}(\bm{x}_t \vert \bm{y}_{1:t})$ are exactly described by \cref{eq:hmm-vector-evolution}. Therefore, HMMs are a subset of qHMMs. In the remainder of this work, we will describe many such cases where qHMMs reduce to HMMs.

\section{The case of Haar random unitaries} \label{sec:haar-case}
In this section, we give some explicit examples of where classical HMMs provide good, even optimal, predictions for interacting quantum systems. Specifically, we will consider the case where the teacher's evolution involves applying Haar-random gates and measurements to a many-qubit system. It may initially seem surprising that classical models can capture the full complexity of a quantum many-body system. However, in one limit, the ignorance of the student, and in the other, (that of large on-site Hilbert space dimension $d$) the self-averaging qualities of the dynamics give rise to effectively classical behavior.






Consider Haar random unitary gates \cite{collins2022weingarten} with some block diagonal structure in the computational basis. This could be due to an imposed symmetry. For example, taking the local Hilbert space to be $\cl{H}_\rm{loc}=\bb{C}^2 \otimes \bb{C}^d$ (a qubit-qudit pair), we can consider $\rm{U}(1)$-symmetric gates that preserve the particle number in the qubit sector when acting on the overall Hilbert space $\cl{H}_\rm{loc}^{\otimes N}$. A two qubit gate would have the block diagonal structure, 
\begin{align} \label{eq:u-u-one}
  u_{\rm{U}(1)} = u^{(2)} \oplus u^{(1)} \oplus u^{(0)}.
\end{align}
Each block acts on the 2, 1, and 0 particle sector and has dimension $d^2 \times d^2$, $2d^2 \times 2d^2$, and $d^2 \times d^2$ respectively.


Then, we combine such Haar random unitary gates in sequence with measurements of the observables that are also diagonal the computational basis. For example, a weak measurement channel on qubits with continuous outcomes could be
\begin{align}
  \rho \rightarrow \int dy Q(y) \rho Q(y) \otimes \ketbra{y}{y},
\end{align}
where the weak measurement operator (or partial projector) $Q(y)$ is \cite{barchielli1982model}
\begin{align} \label{eq:weak-measurement}
Q(y) = \frac{1}{\sqrt[4]{2 \pi \sigma^2}} \exp\left[ -\frac{(y - \epsilon \hat{Z})^2}{4 \sigma^2} \right].
\end{align}
Here, $y$, $\epsilon$, and $\sigma$ are measurement outcome, strength, and width, respectively.

In this setting, we consider two scenarios of inference problems, and discuss when they can be mapped to classical HMMs.

\subsection{Hidden Haar unitaries}
The Haar unitary channel where the teacher records the choice of unitary is given by
\begin{align} \label{eq:haar-channel-record}
  \rho \rightarrow \int d\mu(u) u \rho u^\dagger \otimes \ketbra{u}{u}.
\end{align}

In the first case, we consider when the teacher does not reveal their choice of Haar unitary. In this case, the classical register does not participate in any future operations. Therefore we are free to trace over these degrees of freedom, leaving us with the memoryless Haar unitary channel,
\begin{align} \label{eq:haar-channel-memoryless}
  \rho \rightarrow \int d\mu(u) u \rho u^\dagger.
\end{align}

\cref{eq:haar-channel-memoryless} is completely depolarising on each sector and returns an equally weighted ensemble of diagonal density matrices in the sector. The measurements on the qubits are also on diagonal observables. The result is that charge diagonal observables obey a closed set of equations of motion amongst themselves. Therefore other degrees of freedom can be safely traced out, and the circuit can be mapped to a classical HMM.

For example, our two-site $u_{\rm{U}(1)}$ Haar random-unitary channel is equal to
\begin{align}
    \rho \rightarrow \mathcal{E}_{\mathrm{SSEP}} \otimes \mathcal{D}_\rm{qudits} [\rho],
\end{align}
where $\mathcal{E}_{\mathrm{SSEP}}$ is the incoherent simple symmetric exclusion process (SSEP) \cite{binder2010theoretical} channel on the qubits, and $\mathcal{D}_\rm{qudits}$ is the depolarising channel on the qudits. On the diagonal entries of density matrices in the qubit computational basis, $\mathcal{E}_{\mathrm{SSEP}}$ acts as a SSEP Markov kernel on two Ising variables $s$ and $s'$ corresponding to on-site charges
\begin{align} \label{eq:ssep-kernel}
  p_{\mathrm{SSEP}}(s_t, s'_t \vert s_{t-1}, s'_{t-1})
  = (s_t s_t'\rvert 
  \begin{psmallmatrix}
  1 & 0 & 0 & 0 \\
  0 & \frac{1}{2} & \frac{1}{2} & 0 \\
  0 & \frac{1}{2} & \frac{1}{2} & 0 \\
  0 & 0 & 0 & 1
  \end{psmallmatrix}
  \lvert s_{t-1} s'_{t-1}). 
\end{align}
The measurement operator maps to likelihood/measurement model. For example, the weak measurement \cref{eq:weak-measurement} maps to the measurement model
\begin{align}
  p(y \vert s) = \frac{1}{\sqrt{2 \pi \sigma^2}} \exp\left[ -\frac{(y - \epsilon s)^2}{2 \sigma^2} \right].
\end{align}
Lastly, the prior at initial time is $p(\bm{s}_1) = \tr[\cl{P}_{\bm{s}_1} \rho_0]$. Thus, in the setting where the teacher does not reveal their specific choices of U(1) Haar unitaries, the teacher's experimentally measured outputs are indistinguishable from those generated by an effectively classical HMM. Thus, for the purposes of prediction, the student cannot do any better than to use this classical HMM.

\subsection{Revealed Haar unitaries}
In the next case, the teacher stores the choice of Haar unitary in a classical register and reveals it to the student as in \cref{eq:haar-channel-record}.

In this case, coherence is intact for each trajectory. One might imagine that the student will have to simulate the full quantum evolution for an optimal posterior. Now we consider that the Haar unitary acts on qubit-qudit pairs (with qudit dimension $d$), with the same block diagonal structure on the qubits as before. In the limit of $d \gg \rm{e}^{LT}$, the typical gate $u$ is well approximated by its Haar-average. For example, for our two-site $u_{\rm{U}(1)}$,
\begin{align}\label{eq:self_avg}
  u_{\rm{U}(1)}^\rm{typ}[\; \cdot \;] (u_{\rm{U}(1)}^\rm{typ})^\dagger \mathop{\approx}^{d \gg \rm{e}^{LT}} \mathcal{E}_{\mathrm{SSEP}} \otimes \mathcal{D}_{\mathrm{qudits}}[\; \cdot \;].
\end{align}
Precisely, in \cref{apdx:concentration-of-measure}, we rigorously prove that, for any circuit with Haar-random unitary gates acting on $k$ qubit-qudit pairs with block diagonal structure (with each block acting on all qudits) and  weak measurements on number of qubits such that there are extensive number ($\sim VT$) of such elements, we have
\begin{align}
  \E_{U, Y} \left[\abs{p(\bm{s} \vert Y) - \bar{p}(\bm{s} \vert Y)}\right] \leq C \frac{VT r_1^{V} r_2^{VT}}{d^{k/2}},
\end{align}
where $C$, $r_{1,2}$ are some constants independent of $d$, $V$ is the volume, $T$ is the total number of timesteps, and $\bar p(\bm s \vert Y)$ is the posterior derived from the Haar-averaged channels, i.e. using the channel on the RHS of \cref{eq:self_avg} in place of the unitaries in order to deduce the posterior.

Previously, that $\bar p(\bm s \vert Y)$ well-approximate $p(\bm s \vert Y)$ was checked at the level of the replica trick and in the formal limit $d \rightarrow \infty$ for a quantum circuit with two-qubit $u_{\rm{U}(1)}$ local unitaries with onsite projective measurements, in Ref. \cite{agrawal2022entanglement}.

In the $d \gg \rm{e}^{VT}$ limit, Haar-unitaries are self-averaging. For that reason, the student's inferences are not improved further if the teacher reveals the particular unitaries used; optimal inference is achieved with the ensemble-averaged circuit, which is a classical HMM model. 

Lastly, note that disregarding (tracing out) the student sector leaves behind a monitored quantum circuit. In the large $d$ limit, any observables on qubits diagonal in the computational basis are given by a classical stochastic model (a HMM), while the ``maximally quantum'' $d=1$ limit cannot be reduced to such classical models. Therefore, qudit dimension $d$ can be used to interpolate between the two limits.

\section{Random bond Ising model and quantum error correction} \label{sec:2d-rbim}
In this section, we review the emergence of the 2D random-bond Ising model (RBIM) for two QEC scenarios originally discussed in Ref. \cite{dennis2002topological}: QEC with bit-flip errors on the toric code and one-shot syndrome measurements, and on the repetition code and repeated noisy syndrome measurements with read-out errors. For a detailed introduction, see Ref. \cite{dennis2002topological}.

First, we look at these QEC scenarios as quantum inference problems on qHMMs. The toric code case has a trivial Markov structure, as there is only one cycle of errors followed by syndrome measurements. On the other hand, the repetition code case has an extended temporal structure as measurements and errors channels are repeated. The choice of bit-flip errors will mean these scenarios reduce to inference on classical HMMs.

With the mapping to HMMs established, we adopt the language of Bayesian inference in the teacher/student setting for HMMs and discuss each scenario.

\subsection{Quantum error correction as a quantum inference problem}
We consider scenarios where the teacher prepares an encoded logical qubit $\ket{\psi} = \alpha \ket{\bar 0} + \beta \ket{\bar 1}$ comprised of many physical qubits. For the toric and repetition codes, the physical qubits live on links $l$ of the 2D square lattice and 1D chain respectively. 

The teacher randomly applies bit-flip errors on the physical qubits. Importantly, the teacher keeps a record of links where a bit-flip error is \emph{currently} applied in an `environment' $\ketbra{\bm{f}}{\bm{f}}$ where $\bm{f} = (f_l)_l$. Here we introduced the compact tuple-builder notation $(f_l)_l := (f_l : l \in \rm{all \; links})$. 

When there is a bit-flip is applied on link $l$, the register is set to $f_l = -1$, and to $f_l = 1$ otherwise. Therefore, the environment is initialised as $\ketbra{\bm{1}}{\bm{1}}$.

This is modelled by a bit-flip error channel that acts with rate $\uppi_*$ and also updates the record of the currently applied errors, given by
\begin{align} \label{eq:bit-flip-errors}
  \rho_* & \rightarrow \uppi_* (X_l \otimes X^\rm{env}_l) \rho_* (X_l \otimes X^\rm{env}_l) + (1-\uppi_*) \rho_* \\
  & = \sum_{e_l} K_{*, e_l} \rho_* K^\dag_{*, e_l}.
\end{align}
Here, $X_l$ acts on the physical qubits, whereas $X_l^\rm{env}$ acts on the environment registers. In the following sections, we will relate error rate $\uppi_*$ with fidelity $\beta_*$ as $\uppi_* = e^{-\beta_*} / 2 \cosh \beta_*$. In the second line, we wrote the channel in terms of Kraus operators, $K_{e_l} = \sqrt{p(e_l)} (X_l \otimes X_l^\rm{env})$. There, occurrence of an error and no error is denoted as $e_l=-1$ and $e_l=1$ respectively. Its probability is given by $p(e_l) = e^{-\beta_* e_l}/2 \cosh(\beta_*)$.

Next, the teacher measures the syndromes and records the outcomes. For the toric code, the syndrome is $\hat s_P = \hat Z_P = \prod_{l \in P} \hat Z_l$, where $P$ is a plaquette on the square lattice and $l \in P$ denotes the links neighbouring $P$. For the repetition code, the syndrome lives on the vertex and is $\hat s_v = \hat Z_v = \prod_{l \in v} \hat Z_l$. The syndrome measurement channel only acts on the quantum state and is given by
\begin{align} \label{eq:syndrome-measurements}
  \rho_* \rightarrow \sum_{s} {\bb P}_s \rho_* {\bb P}_{s} \otimes \ketbra{s}{s}.
\end{align}
For the repetition code, we will consider read-out errors. There, there is a further error channel on the syndrome register, i.e. the \emph{true} syndrome register $s^*$ is now a hidden register, and only the (potentially) corrupted syndrome register $s = e \cdot s^*$ is revealed. This corresponds to the channel
\begin{align} \label{eq:noisy-syndrome-measurements}
  \rho_* 
  & \rightarrow \sum_{s} \sum_{s^*} {\bb P}_{s^*} \rho_* {\bb P}_{s^*} \otimes \sqrt{p_*(s \vert s^*)} \ketbra{s}{s} \sqrt{p_*(s \vert s^*)}.
\end{align}
We consider the case where the probability of occurrence of read-out error is the same as that of the bit-flip error, $\uppi_*$ \cite{dennis2002topological}. Therefore we have $p(s=e \cdot s^* \vert s^*) = p(e) = e^{-\beta_* e} / 2 \cosh \beta_*$.

Then, the syndromes $S$ are revealed to the student, while the environment register is hidden from them. The goal of the student is to infer the hidden environment register from the revealed syndromes. 

Although the above discussion was restricted to toric and repetition codes, the above can be easily generalised to any QEC models, replacing the bit-flip errors, codes, and syndrome measurements to arbitrary ones. As long as the revealed syndrome measurements do not participate in subsequent channels, as is the case above, the quantum inference problem is on a qHMM.

In the special case of bit-flip errors on toric and repetition codes discussed, the inference problem can be reduced to that on classical HMMs. Application of bit-flip errors will result in a statistical mixture of states $\ket{\bm{\tau}} := \prod_l X^{(1+\tau_l)/2} \ket{\psi}$. If the environment register faithfully records the errors, the density matrix will be a statistical mixture $\sum_{\alpha} p_\alpha \ketbra*{\bm{\tau}^{(\alpha)}}{\bm{\tau}^{(\alpha)}} \otimes \ketbra*{\bm{f}^{(\alpha)}}{\bm{f}^{(\alpha)}}$, where $\tau_l^{(\alpha)} = f_l^{(\alpha)}$. On such states, tracing out the Hilbert space of the quantum state, we reduce to a HMM on the hidden environment registers. The error channel \cref{eq:bit-flip-errors} reduces to
\begin{align}
  p_*(f^*_{l, t} \vert f^*_{l, t-1}) = \frac{\exp(-\beta_* f^*_{l,t}, f^*_{l, t-1})}{2 \cosh \beta_*},
\end{align}
whilst for the syndrome measurement channels \cref{eq:syndrome-measurements,eq:noisy-syndrome-measurements}, after tracing out the quantum state, simply moves the action of projectors to the environment register, resulting in measurement models $p_*(s^*_P \vert (f^*_l)_{l \in P})$ as in \cref{eq:toric-code-measurement-model} for the toric code and $p_*(s^*_v \vert (f^*_l)_{l \in v})$ as in \cref{eq:repetition-code-measurement-model} for the repetition code.

\subsection{Toric code}
For the toric code scenario, we consider single round of errors, followed by projective measurements of syndromes. Therefore we can just consider the occurrence of errors $E^* = (e^*_l)_l$. In the mapped classical inference problem, teacher first picks the true error chain $E^*$ from their prior
\begin{align}
  p_*(E^*) = \prod_l p_*(e^*_l) \propto \prod_l \exp\left(\beta_* e^*_l \right),
\end{align}
then generates syndromes $S = (s_P)_P$ from their likelihood/observation model. As we consider noiseless projective measurements, this is given by
\begin{align} \label{eq:toric-code-measurement-model}
  \begin{aligned}
      p_*(S \vert E_*) & = \prod_P p_*(s_P \vert (e^*_{l \in P})_l) \\ & = \prod_P \delta\left\{{s_P, \prod_{l \in P} e^*_l}\right\} \\ & = \delta_{S, \partial E^*},
  \end{aligned}
\end{align}
where $\delta \{ \cdot, \cdot \}$, $\delta_{\cdot, \cdot}$ are both Kronecker deltas, and $\partial E^*$ denotes the boundary of the error chain. Here, boundary is defined on the dual lattice and therefore lives on the plaquettes. Note that the process of generating observations (syndromes in this case) is completely deterministic. The failure of perfect inference stems from the fact that the map $E^* \rightarrow S$ is many-to-one.

Given the syndromes $S = (s_P)_P$, the student can infer a posterior distribution assuming student parameters $\uppi = e^{-\beta}/ 2 \cosh(\beta)$,
$$
p_\mathrm{s}(E \vert S) = \frac{p_\mathrm{s}(S \vert E) p_\mathrm{s}(E)}{p_\mathrm{s}(S)}.
$$
Marginalising over the syndromes, the conditional probability of the student inferring $E$ given that the teacher picked $E^*$ is
$$
\begin{aligned}
p(E \vert E^*) & = \sum_S p_\mathrm{s}(E \vert S) p_*(S \vert E^*) \\
& = \sum_S \frac{\delta_{S, \partial E} p_\mathrm{s}(E) \delta_{S, \partial E^*}}{p_\mathrm{s}(S)} \\
& \propto \delta_{\partial E, \partial E^*} p_\mathrm{s}(E).
\end{aligned}
$$
The set of inferred errors $E$ that respect the delta function only contains error chains such that $E = C \odot E^*$, where $\odot$ is element-wise multiplication, such that $e_l = e^*_l c_l$. Here $C = (c_l)_l$, $c_l \in \{-1, +1\}$ and links where $c_l=-1$ form a cycle. We can automatically satisfy such configurations by parameterising $c_l = \sigma_{P_1(l)} \sigma_{P_2(l)}$, where $P_{1,2}(l)$ are the two plaquettes neighboured by link $l$ and $\sigma_P$ is a fictitious Ising variable living on plaquette $P$. Finally we can write down
\begin{align} \label{eq:rbim-boltzmann-prob}
p\left(E = C(\Sigma) \odot E^* \vert E^*\right) = \frac{\exp \left[ - \beta \sum_l e^*_l \sigma_{P_1(l)} \sigma_{P_2(l)} \right]}{\mathcal{Z}(E^*)},
\end{align}
for $\Sigma = (\sigma_v)_v$ and $\cl Z(E^*)$ is the partition function that normalises the distribution. This is the Gibbs distribution for the 2D RBIM. Finally, to calculate expectation values, we do so with respect to the \emph{true/teacher's} prior distribution $p_*(E^*) \propto \exp\left[\beta_* \sum_l e^*_l\right]$.

A natural order parameter for successful inference and therefore error correction is the MSE between the true/teacher's error $E^*_l$ and inferred/student's error $E_l$ at the same link averaged over realisations, $\E[(E_l - E_l^*)^2] = 2 - 2 \mathbb{E}[E_l E^*_l]$, or equivalently, their correlations,
\begin{align}
    \E_{E, S, E^*} \left[ E_l E_l^* \right] 
    & = \E_{E, S, E^*} \left[ E^*_l \sigma_{P_1(l)} \sigma_{P_2(l)} E_l^* \right] \nonumber \\
    & = \E_{E, S, E^*}\left[ \sigma_{P_1(l)} \sigma_{P_2(l)} \right],
\end{align}
which corresponds to ferromagnetic (FM) order for the 2D RBIM. In fact, as long as the inferred error is in the same homological class as that of the true error, the error correction is successful. Therefore the more relevant order parameter would be the probability of inferring the correct homology class. However, it turns out the transitions for both order parameters coincide \cite{dennis2002topological}.

\subsection{Quantum repetition code}
In the repetition code case, the teacher prepares the initial register as $f^*_{l, t=0} = 1$, therefore $p_*(\bm{f}^*_0) = \delta_{\bm{f}^*_0, \bm{1}}$. For subsequent times, $\boldsymbol{f}^*_{t}$ undergoes a Markov chain as
$$
p_*(\boldsymbol{f}^*_t \vert \boldsymbol{f}^*_{t-1}) = \prod_l p_*(f^*_{l,t} \vert f^*_{l,t-1}) \propto \prod_l \exp\left[\beta_* f^*_{l,t} f^*_{l, t-1}\right],
$$
and the prior for entire trajectory $F^* := \boldsymbol{f}^*_{1:t}$, becomes
\begin{equation}
  \begin{aligned}
    p_*(F^*) & = \left( \prod_{\tau=2}^t p_*(\boldsymbol{f}^*_t \vert \boldsymbol{f}^*_{t-1}) \right) p_*(\boldsymbol{f}^*_1) \\
    & \propto \prod_{l,\tau} \exp \left[ \beta_* f^*_{l,t} f^*_{l, t-1}\right],
  \end{aligned}
\end{equation}
where we set $f^*_{l, 0} = 1$.

Then, they generate unreliable syndromes $s_{v, t}$ at vertex $v$ by randomly flipping the true syndrome $s^*_{v,t} = f^*_{l_\mathrm{L}, t} f^*_{l_\mathrm{R}(v) ,t}$ as 
\begin{align} \label{eq:repetition-code-measurement-model}
    p_*(s_{v, t} \vert f^*_{l_\mathrm{L}(v), t}, f^*_{l_\mathrm{R}(v), t}) \propto \exp\left[ \beta_* s_{v,t} f^*_{l_\mathrm{L}(v),t} f^*_{l_\mathrm{R}(v),t}\right],
\end{align}
where $l_{\rm{L}(v)}$ and $l_{\rm{R}(v)}$ are left and right links of vertex $v$, respectively. Then the observation model for the entire trajectory is
\begin{align}
    p_*(S \vert {F}^*) = \prod_{v, \tau} p_*(s_{v, t} \vert f^*_{l_\mathrm{L}(v), \tau} f^*_{l_\mathrm{R}(v), \tau}).  
\end{align}
The student could then infer a posterior distribution over the whole trajectory through Bayes' rule, 
\begin{align}
    p_s({F} \vert {S}) = \frac{p_s({S} \vert {F}) p_s({F})}{p_s({S})},
\end{align}
using assumed fidelity $\uppi \propto e^{-\beta}$. To infer the current errors in real time, the student may use the forward equation to infer $p_\mathrm{s}(\bm{f}_t \vert \bm{s}_{1:t})$, then take its $\mathrm{argmax}_{\bm{f}_t}$ to determine the error correction operation.

The natural order parameter in this case would be the mean-squared error of current bit-flips $\E[(f_{l, t} - f_{l, t}^*)^2] = 2 - 2 \mathbb{E}[f_{l, t} f^*_{l, t}]$, or equivalently the equal-bond correlation between that of the student and teacher, $\E \left[ f_{l, t} f_{l, t}^* \right]$.

To massage the above to the 2D RBIM, we picture spacetime as a 2D lattice. Now the original links on the 1D chain are horizontal links $l_\rightarrow = (l, t)$ in the 2D lattice. Vertical links $l_\uparrow = (v, t)$ connect the vertex at current timestep to that of the previous timestep. Note that the lattice is periodic in the horizontal direction, but in the vertical direction, starts with horizontal links and terminates with vertical links. We denote $e^*_{l_\rightarrow}=-1$ as a bit-flip error occurring at time $t$, and $e^*_{l_\uparrow} = -1$ to denote if a syndrome measurement is corrupted by a read-out error. In \cref{apdx:repetition-code}, we show that for $E^* = E^*_{\uparrow} + E^*_{\rightarrow}$, and similarly for $E$, $p(E \vert E^*)$ is again the 2D RBIM partition function \cref{eq:rbim-boltzmann-prob} but with periodic boundary conditions in space, fixed boundary conditions $\sigma=1$ at both $t=0$ and $t=t_\mathrm{f}$. Furthermore, the correlation order parameter becomes
\begin{align}
    \E [f_{l, t} f^*_{l, t}] = \E\left[ \prod_{\tau=1}^t \sigma_{v_\mathrm{L}(l), \tau} \sigma_{v_\mathrm{R}(l), \tau} \right],
\end{align}
a many-body correlation function equal to one if all spins are aligned. Ref. \cite{dennis2002topological} claims that the error-correctability transition in this case also coincides with that of ferromagnetism.

\subsection{Implication of the 2D random bond Ising model phase diagram}
\begin{figure}[t!]
    \centering
    \includegraphics[width=\linewidth]{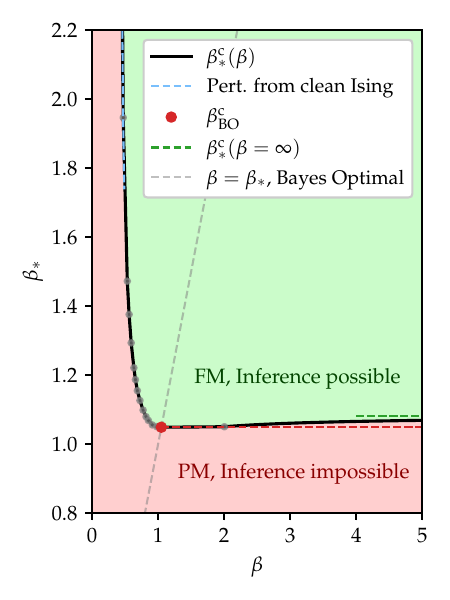}
    \caption{Phase diagram of the 2D random bond Ising model drawn in student-teacher (assumed-true) fidelity $\beta-\beta^*$ parameter space, which also applies to static QEC problems on the toric code and dynamic QEC the quantum repetition code. Light green region denotes ferromagnetic (FM) phase and the inference- or QEC-possible phase. Light red region denotes the paramagnetic (PM) phase and inference- or QEC-impossible phase. Black line: Fidelity thresholds, linearly interpolated from the datapoints in the commonly discussed $p-T$ parameters. Grey dashed line: Bayes optimal line $\beta = \beta^*$. Light blue dashed line: exact perturbation from clean 2D Ising model $\beta_*=\infty$ where $\beta^\rm{c}(\beta_*=\infty) \approx 0.4407$ \cite{onsager1944crystal}. Grey markers: fidelity thresholds obtained in Ref. \cite{merz2002two}. Red circle and dashed line: fidelity threshold on the Bayes optimal line $\beta^\mathrm{c}_\mathrm{BO} = 1.048 \pm 0.001$, obtained in Ref. \cite{ohzeki2015high}. Green dashed line: fidelity threshold at infinite student fidelity $\beta^\mathrm{c}_*(\beta =\infty) = 1.081 \pm 0.001$, obtained in Ref. \cite{fujii2012error}, which implies a slight re-entrance in the phase diagram. Note that the $y$-axis starts at $\beta_* = 0.8$ and that the aspect ratio is not equal between $\beta$ and $\beta_*$, to accentuate the re-entrance.} \label{fig:2d-rbim-phase-diagram}
\end{figure}

In the two cases, the information about the true errors $E^*$ is `planted' in the syndromes $S$. Therefore RBIMs can also be referred to as `planted Ising models' \cite{zdeborova2016statistical}.

Conventionally, the phase diagram of the 2D RBIM is drawn in the $p-T$ parameter space, where $p=\uppi_*$ is the probability of ferromagnetic bonds and $T=1/\beta$ is the temperature. To interpret it in the inference setting, however, it is instructive to plot it in the student-teacher fidelity space $\beta-\beta_*$. In this case, the Nishimori line, equal to the Bayes optimal line, is a straight line $\beta^* = \beta$. The phase diagram is plotted with these variables in Fig. \ref{fig:2d-rbim-phase-diagram} interpolated from numerically accurate datapoints obtained from previous studies \cite{merz2002two, ohzeki2015high, fujii2012error}.

Though previously debated, it is now established that there exists a slight re-entrance in the phase diagram, such that the critical line between $\beta_*^c(\beta = \beta_*)$ and $\beta_*^c(\beta = \infty)$ is not horizontal \cite{ohzeki2015high, fujii2012error}. This has implications on the thresholds of different decoders. The `decoder' discussed in quantum error correction literature is equivalent to the classifier $h(Y)$ onto the most likely error chain. As shown in \cref{sec:bayes-optimality}, the optimal classifier is the Bayes classifier - $\mathrm{argmax}$ of the true posterior $h_\mathrm{optimal}(S) = \mathrm{argmax}_E p_*(E \vert S)$. In a more realistic setting, however, the evaluation of the posterior is too time-consuming. Therefore, the minimum-weight perfect matching (MWPM) decoder is used \cite{edmonds1965paths}. In the toric code setting, This is when error chains with the shortest length between the syndromes is assumed to correct the error. It turns out that this is equivalent to the $\mathrm{argmax}$ of the posterior at $\beta \rightarrow \infty$ \cite{dennis2002topological}. Therefore, this suggests that the threshold for the MWPM decoder is slightly lower in $p$ (or higher in $\beta_*$) than the Bayes optimal setting, though not by much.

One order parameter not yet discussed for the 2D RBIM is the observable variance \cref{eq:dO2} of the error chain in the student's distribution, i.e. $\delta E_l^2$ and $\delta f_{l, t}^2$ for the toric and repetition code case, respectively. Although the error chain variance is equal to $\rm{MSEM}$ and $\rm{MSE}/2$ on the Bayes optimal line, it is not necesarily related off of it. In the FM/inference possible phase, it is clear that the student's variance is small. However, a priori, we cannot rule out that it is small in the inference impossible phase. Writing out the variance for both cases results in quantities that only depend on Edwards-Anderson-like (EA) order parameters, $1 - \E [\langle \sigma_{P_1(l)} \sigma_{P_2(l)} \rangle^2]$ and $1 - \E [\langle \prod_{\tau=1}^t \sigma_{v_\rm{L}(l), \tau} \sigma_{v_\rm{R}(l), \tau}\rangle^2]$ for the toric and repetition code, respectively (The EA order parameter is $\E[\langle \sigma_l \rangle^2]$). In the disordered systems literature, the spin glass phase (SG) is characterised by no FM order and presence of EA order \cite{nishimori2001statistical}. In the 2D RBIM, the SG phase does not exist for any $\beta < \infty$ \cite{fisher1988equilibrium}. Therefore, the error-chain fuzzy and sharp phases coincide with the inference impossible and possible phases, respectively.

\section{The planted SSEP and the Haar-random $\rm{U}(1)$-symmetric monitored quantum circuit} \label{sec:planted-ssep}
\begin{figure}[h!]
  \includegraphics[width=1.0\columnwidth]{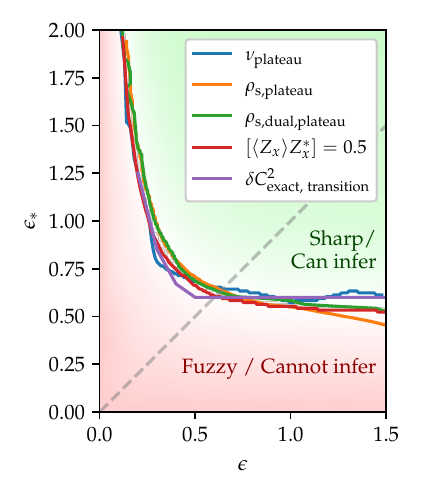}
  \caption{Phase diagram of the planted SSEP deduced from various methods. From TEBD, we obtain blue, orange, and green lines which delineate where extracted exponent $\nu$ and extracted superfluid stiffnesses $\rho_\mathrm{s}$, $\rho_\mathrm{s, dual}$ approache plateau values, respectively. Red line is contour for $\E[\langle s_x \rangle s^*_x] = 0.5$ from TEBD. Purple line is critical point determined by point where $\delta C^2(t)$ turns from a power law in time to exponential decay from exact calculation of the posterior. The $45^{\circ}$ grey dashed line in parameter space corresponds to Bayes optimality $\epsilon=\epsilon_*$. Shaded colors are given by the alignment $\E[\langle s_x \rangle s^*_x]$. The good agreement with phenomenology predicted by replica field theory suggests that the whole phase boundary is modified-KT.} \label{fig:planted-ssep-phase-diagram}
\end{figure}

In this section we discuss a weak-measurement variety of the recently studied Haar-random $\rm{U}(1)$-symmetric monitored quantum circuit \cite{agrawal2022entanglement}. This is a quantum inference problem on a qHMM, which using the arguments of \cref{sec:haar-case} can be mapped to a classical inference problem on a HMM that we call the \emph{planted SSEP}, where the state of a SSEP is inferred from its noisy `images'. In turn, the original quantum model can be referred to as the \emph{quantum planted SSEP}. If the student is disregarded, the model can be thought of as a standalone circuit operated by an experimentalist/teacher. We introduce the quantum model in \cref{sec:quantum-planted-ssep} and the classical counterpart in \cref{sec:classical-planted-ssep}. In \cref{sec:classical-planted-ssep-analysis}, we study the planted SSEP. After discussing the general considerations, we develop discuss its replica field theory away from Bayes optimality (\cref{sec:replica-field-theory}), whose predictions are verified numerically (\cref{sec:planted-ssep-numerics}) to obtain phase boundary in \cref{fig:planted-ssep-phase-diagram} and its universality class.

\subsection{Quantum circuit model} \label{sec:quantum-planted-ssep}
Consider a local Hilbert space of qubits and qudits, $\mathcal{H}_{\mathrm{loc}} = \mathbb{C}^{2} \otimes \mathbb{C}^d$, with qubit computational basis states $\lvert s \rangle \in \{ \lvert -1 \rangle, \lvert 1 \rangle \}$ and qudit basis states $\ket{g} \in \{\ket{0}, \dts, \ket{d-1}\}$. Now we consider a 1D chain of such local degrees of freedom living at sites $x$ with even system size $L$ and periodic boundary conditions. The teacher (or the experimentalist), prepares an initial state $\rho_0$, then runs the quantum circuit which we will describe below.

We study two scenarios for the initial states. The first scenario is where the teacher prepares a pure state $\rho_0 = \ketbra*{\psi_0}{\psi_0}$, applies the circuit, then measures the total charge of the system afterwards. The final measured charge is then the hidden register. For simplicity, we take the initial state to be an equal superposition of all qubit and qudit basis states, which can be prepared as a product state $\ket{\psi_0} = \ket{+}^{\otimes L} \otimes \ket{\chi}^{\otimes L} = (2d)^{-L/2} \sum_{\bm s, \bm g} \ket{\bm s} \otimes \ket{\bm g}$. Again, we used the bold symbol notation to denote configurations over all space, $\ket{+} := 2^{-1/2} (\ket{-1} + \ket{1})$ is the Pauli-$x$ eigenstate, and $\ket{\chi} := d^{-1/2} \sum_{g} \ket{g}$ is the equal superposition of qudit computational basis states.

The second scenario is where the teacher randomly picks from an ensemble of states $\{\ket{\psi_{C^*}}\}$, each with a definite total charge $C^*$, where charge in the system is defined as $\hat C = \sum_x Z_x$. The hidden register then becomes the randomly picked charge. Therefore the initial ensemble can be written as $\rho_0 = \sum_{C^*} p_*(C^*) \ketbra{\psi_{C^*}}{\psi_{C^*}} \otimes \ketbra{C^*}{C^*}$. In order that $\tr[\bb{P}_{\bm{s}} \rho_0]$ is the same as the previous scenario, we choose the charge to be binomially distributed $p_*(C^*) = \frac{1}{2^L}{L \choose (C+L)/2}$ and $\ket{\psi_{C^*}} \propto \left(\sum_{\bm s : C[\bm s] = C^*} \ket{\bm s}\right) \otimes \ket{\chi}^{\otimes L}$.

Although both of these scenarios can be used in the inference and standalone settings, the first scenario is the most natural when the student is disregarded, and the second is the most natural in the teacher-student inference setting.

After preparation of the state, the teacher runs the quantum circuit for timesteps in the order of system size $t_\mathrm{f} \sim L$. At each timestep $t$, the teacher applies unitary gates on neighbouring qubit-qudit pairs in a brickwork fashion, 
\begin{align}
    {\bm{u}}_t = \bigotimes_{x \in \mathcal{R}_t} u_{x:x+1, t},
\end{align}
where $\mathcal{R}_t$ are the even sites at even times and odd sites at odd times and each $u_{x:x+1, t}$ is a Haar-random two-qubit $u_{\rm{U}(1)}$ as \cref{eq:u-u-one}. We will denote all instances of unitaries as $U := \bm{u}_{1:t_\rm{f}}$. This layer of unitary gates is followed by weak measurements on each qubit at every site $Q(y_{x, t})$ as \cref{eq:weak-measurement}. We will use the notation $\bm{y}_t = (y_{x,t})_{x \in 1:L}$ and $Y=\bm{y}_{1:t_\rm{f}}$ to denote the collection of all measurements gained at timestep $t$ and at all timesteps, respectively.

In the inference setting, the teacher would then send the student the measurements $Y$ and possibly their sampled Haar unitaries $U$. The student then infers the total charge in the system $C^*$ with their model which has the same functional form as the teacher's but with the student's assumed parameters $\epsilon$, $\sigma$.

Thought of as a standalone circuit, the quantity of interest would be the charge variance $\delta C^2_*$, i.e. \cref{eq:quantum-dO2} for observable $O = \hat C$. 

At Bayes optimality, $\delta C^2_* = \delta C^2_\rm{s}$. This immediately implies the charge sharpening transition, which is the most naturally formulated in the standalone circuit coincides with the learnability transition in the inference setting here.

\subsection{Classical inference model} \label{sec:classical-planted-ssep}
Using the arguments of \cref{sec:haar-case}, in cases where the choice of Haar unitaries is hidden or qudits have large Hilbert space dimension $d \gg \rm{e}^{Lt_\rm{f}}$, the quantum model can be mapped to a classical inference problem on a HMM which we call the \emph{planted SSEP}. Here, the teacher generates trajectories of bitstrings $S^*$ via the SSEP \cite{binder2010theoretical}. Then, they generate noisy `images' of the configurations, $Y \sim p_*(\, \cdot \, \vert S^*)$ (which could be done at each timestep), which they hand to the student. The student then uses their own model to infer a posterior given the measurements $p_\mathrm{s}(S \vert Y)$. 

Concretely, we consider the teacher drawing a brickwork SSEP \cite{liggett1985interacting} on a 1D lattice of size $L$ with periodic boundary conditions. At $t=1$ the teacher draws initial Ising (equivalently Boolean $b = (s+1)/2$) variables $\bm{s}^*_1 = (s^*_{1, 1}, \dts, s^*_{L, 1})$ from a separable distribution $p_*(\bm{s}^*_1) = \prod_x p_*(s^*_{x, 1})$ (the initial state considered in the quantum model maps to an initial distribution of equally likely charge configurations, $p_*(\bm{s}_1) = 2^{-L}$). For subsequent times the teacher state evolves as a Markov process
\begin{align}
    p_*(\bm{s}^*_t \vert \bm{s}^*_{t-1}) = \prod_{x \in \mathcal{R}_t} p_*(s^*_{x:x+1, t} \vert s^*_{x:x+1, t-1}).
\end{align}
Denoting the incoming and outgoing states as $(s_{x, t-1}, s_{x+1, t-1})$, $(s_{x, t}, s_{x+1 ,t})$, respectively the two-site kernel is given by the SSEP kernel \cref{eq:ssep-kernel}.
The probability distribution for the whole trajectory is
\begin{align}
    p_*(S^*) = \left(\prod_{t=2}^{t_\mathrm{f}} p_*(\bm{s}^*_t \vert \bm{s}^*_{t-1}) \right) p_*(\bm{s}^*_1).
\end{align}

The `noisy image' generated at each timestep, $\bm{y}_t$, is given by the pixel values 
\begin{align}
    y_{x, t} = \epsilon_* s_{x, t} + \sigma_* \eta_{x, t},
\end{align}
where $\epsilon_*$ is the true/teacher's signal strength, and $\sigma_*$ is the true/teacher's noise strength, and $\eta_{x, t} \sim \mathcal{N}(0, 1)$. Therefore
\begin{align} \label{eq:teacher-measurement-model}
    p_*(y_{x, t} \vert s^*_{x, t}) = \frac{1}{\sqrt{2 \pi \sigma_*^2}} \exp\left( -\frac{(y_{x, t} - \epsilon_* s^*_{x, t})^2}{2 \sigma_*^2}\right),
\end{align}
and for set of all images, is $p_*(Y \vert S^*) = \prod_t p_*(\bm y_{t} \vert \bm{s}_t) = \prod_{x, t} p_*(y_{x, t} \vert s^*_{x, t})$. Again, we will consider the regime where the total number of timesteps is on the order of magnitude as the system size $t_\mathrm{f} \sim L$.

Given $Y$, the student infers a posterior distribution $p_\mathrm{s}(S \vert Y)$ using their own model with parameters $\epsilon$ and $\sigma$, with all other functional forms equal.

The joint distribution is then given by the planted ensemble \cref{eq:planted-ensemble} for $X, X^* \rightarrow S, S^*$.
In the Bayes optimal case, where $* = \rm{s}$, it is symmetric under $S \leftrightarrow S^*$. Since $S^*$ evolves as the SSEP, this means that $S$ must also evolve as the SSEP in this case. Away from this point, $p(S) \neq p_\mathrm{s}(S)$ generally.

As in the quantum problem, we define total charge as $C[\bm{s}_t] := \sum_x s_{x, t}$. Then, a natural quantity to study in the Bayesian inference setting is the $\mathrm{MSE}$ and $\mathrm{MSEM}$ of the total charge $C$. As we are interested in observables at the final time, we can marginalise over configurations at previous time and look at the posterior distribution for the filtering task, $p_\rm{s}(\bm{s}_t \vert \bm{y}_{1:t})$, which will evolve according to the forward equation \eqref{eq:forward-equation}.

This model is natural generalisation of the recently studied planted directed polymer \cite{pkim2025planted}. In that case the trajectory of one particle was inferred, equivalent to restricting the prior to the one-charge sector in the planted SSEP.

\subsection[Analysis of the planted SSEP]{Analysis of the planted SSEP} \label{sec:classical-planted-ssep-analysis}
For simplicity, we consider the case when all initial charge configuration is equally likely, $p_\mathrm{s}(\bm{s}_1) = p_*(\bm{s}^*_1) = 2^{-L}$. This is half-filling on average, with the total charge distributed binomially. Unless specified, we set $\sigma=\sigma_*=1$ and consider the behaviour of the planted SSEP across the $\epsilon-\epsilon_*$ plane.

Along the $\epsilon=0$ line, no information is assimilated by the student's posterior. Therefore the posterior is decoupled from the measurements and evolves as the SSEP $p_\mathrm{s}(S \vert Y) = p_\mathrm{SSEP}(S)$. Our uniform initial state is the stationary state of the SSEP, which has no second-order connected correlations, and therefore $\langle s_{x, t} s_{x', t} \rangle_\mathrm{c} = 0$. Here, the angled brackets denote expectation on the posterior. At average half filling we also have $\langle s_{x, t} s_{x', t} \rangle = \langle s_{x, t} \rangle = 0$.

On the Bayes optimal line $\epsilon_* = \epsilon$, due to the symmetry between the teacher and student in the planted ensemble, the marginalised distribution of $S$ follows that of the SSEP, $p(S) = p_\mathrm{SSEP}(S)$. Therefore observables linear with the student's posterior, such as the averaged disconnected correlator, is equal to that of the SSEP, i.e. zero due to the choice of average half-filling $\E_{Y,S^*} [\langle s_{x, t} s_{x', t} \rangle] = 0$. However, higher moments, such as the variance of the total charge, can still deviate from those of the SSEP. 

Remaining on the Bayes optimal line, existing numerics for the projective measurement case and analytic replica field theory for weak measurements suggest that there is a phase transition between a charge `fuzzy' and `sharp' phase, exhibited in the higher moment observables in the posterior (or unravelled density matrix of each trajectory) \cite{agrawal2022entanglement, barratt2022field}. The known phenomenology is as follows: the variance of the total charge of the student's posterior, averaged over $Y,S^*$, is finite for $t < t_\#$ and is exponentially decaying for $t > t_\#$ where $t_\#$ is the sharpening time. In the charge fuzzy phase, $t_\# \sim L$, whereas it is sublinear in the sharp phase. Our equation \eqref{eq:accuracy-precision} immediately implies that there is also a corresponding phase transition between a hard inference phase, where the time required for the mean-squared-error of the total charge to exponentially decay grows as $t_\mathrm{inference} \sim L$, and an easy phase, where it is sublinear. In the replica field theory, the phase transition is predicted to be a modified-KT transition \cite{barratt2022field}. Combined, existing works suggest a transition around projective measurement rate $\uppi^\#_\mathrm{BO} \in [0.2, 0.3]$ \cite{agrawal2022entanglement,barratt2022field}. From arguments of inferrability on a single site (\cref{apdx:inferrability-argument}), we expect the transition for weak measurements to be around $\epsilon^\#_\mathrm{BO} \in [0.49, 0.63]$.

We now go beyond Refs. \cite{agrawal2022entanglement,agrawal2023observing,barratt2022field} by moving off the Bayes optimal line. Here, $p(S)$ deviates from that of the SSEP. Treating $\epsilon$ as a small variable and $r_\epsilon := \epsilon_* / \epsilon$ as a factor of order 1 and using the identities of the SSEP, we can expand various correlators in orders of $\epsilon$, (\cref{apdx:perturbation-theory}). To smallest non-zero order, we find that the leading order terms scale with $\epsilon, \epsilon^*$ as
\begin{align}
  \mathbb{E}_{Y,S^*} [\langle s_{x, t} s_{0, t} \rangle_\mathrm{c}] & \propto - \epsilon^2, \label{eq:planted-ssep-connected-correlator} \\
  \mathbb{E}_{Y,S^*} [\langle s_{x, t} s_{0, t} \rangle] & \propto \epsilon^4 \left(r_\epsilon ^2 - 1 \right), \label{eq:planted-ssep-disconnected-correlator}  \\
  \mathbb{E}_{Y,S^*} [\langle s_{x, t} \rangle s^*_{0, t}] & \propto \epsilon \epsilon_*. \label{eq:planted-ssep-alignment} 
\end{align}
Along the Bayes optimal line, the averaged disconnected correlation function vanishes as required by the choice of initial condition and that $p(S) = p_\mathrm{SSEP}(S)$. We also note that its sign changes as the line is crossed. As expected, the correlation between the student's inferred and teacher's true states is only non-zero and positive only if both $\epsilon, \epsilon_* >0$.


\subsubsection[Replica field theory of the planted SSEP]{Replica field theory} \label{sec:replica-field-theory}
In this subsection we outline the derivation of the replica field theory to identify the universality class of the phase transition away from Bayes optimality. We will compute various correlators, and combine these with numerical results, in order to determine a phase diagram for the quantum inference problem. See Appendix~\ref{apdx:replica-field-theory} for details.

Observables, both linear and non-linear in the student's posterior (or student's density matrix in the quantum language), which could also be linear or independent on the teacher distribution (teacher's density matrix), can be written as
\begin{align}
  O & = \frac{1}{\mathcal{Z}} \sum_{S^{1:n},S^*} \int dY \\
  & \times O(S^{1:n},S^*) \frac{q_\rm{s}(S^1, Y) \cdts q_\rm{s}(S^n, Y)}{q_\rm{s}(Y)^n} q_*(S^*, Y) \nonumber,
\end{align}
where $q_\rm{s}(Y) = \sum_S q_\rm{s}(S, Y)$. Here, we replaced the probability distribution with unnormalised ones at the cost of introducing the partition function $\mathcal{Z} = \sum_{S^{1:n},S^*} \int dY \frac{q_\rm{s}(S^1, Y) \cdts q_\rm{s}(S^n, Y)}{q_\rm{s}(Y)^n} q_*(S^*, Y)$. As examples, $O(S^{1:n}, S^*)$ could be
\setlength{\leftmargini}{1em}
\begin{itemize}
  \item $\delta C^2_\rm{s}(\bm{s}^{1:2}_t) = C^2[\bm{s}^1_t] - C[\bm{s}^1_t] C[\bm{s}^2_t]$ \newline for the student's charge variance, 
  \item $\rm{MSE}(\bm{s}^1_t, \bm{s}^*_t) = (C[\bm{s}^1_t] - C[\bm{s}^*_t])^2$ \newline for the MSE of the charge, or
  \item $\rm{MSEM}(\bm{s}^{1:2}_t, \bm{s}^*_t) = C[\bm{s}^1_t] C[\bm{s}^2_t] - 2 C[\bm{s}^1_t] C[\bm{s}^*_t] + C^2[\bm{s}^*_t]$ \newline for the MSEM of the charge.
\end{itemize}

Note that we are free to write $q_\rm{s}(Y)^{-n} = \lim_{k \rightarrow -1} \left(\sum_S q_\rm{s}(S, Y)\right)^{nk}$. Defining $Q=n+nk+1$, we will adopt the replica trick and switch the limit and integrals, then assume that $nk$ is an integer, resulting in the following:
\begin{align}
  O & = \lim_{Q \rightarrow 1} \frac{1}{\mathcal{Z}_Q} \sum_{S^{1:Q}} O(S^{1:n}, S^Q) \\
  & \times \int dY q_\rm{s}(S^1, Y) \cdts q_\rm{s}(S^{Q-1}, Y) q_*(S^Q, Y).\nonumber
\end{align}
Throughout, we will refer to the first $Q-1$ replicas as the `student replicas', and the $Q$th replica as the `teacher replica'. Taking the rate of exchange between neighbouring particles in the SSEP kernel as a tunable parameter $J$, then going to continuous time $\tilde{t} = \mathsf{b} t$, where $\mathsf{b}$ is the temporal lattice constant, and finally removing the tildes, one can find that
\begin{align}
  \mathcal{Z}_Q & = \sum_{S^{1:Q}} \int dY q_\rm{s}(S^1, Y) \cdts q_\rm{s}(S^{Q-1}, Y) q_*(S^Q, Y) \\
  & \rightarrow \langle + \rvert^{\otimes QL} \exp\left(-\int_0^{t_\mathrm{f}} dt H_Q \right) \lvert + \rangle^{\otimes QL},
\end{align}
where $\lvert + \rangle$ is the Pauli-$x$ eigenstate, with replica Hamiltonian
\begin{align}
  {H}_Q = -J \sum_{A=X,Y,Z}\sum_{a, x} \hat{A}^a_x \hat{A}^a_{x+1} + \sum_{a,b,x} \hat{Z}^a_x M_{ab} \hat{Z}^b_x,
\end{align}
interreplica interaction matrix
\begin{align}
  M = \mathrm{diag}\left[ \vec{\epsilon^2} \right] - \frac{1}{Q} \vec{\epsilon} \, \vec{\epsilon}^{\, T},
\end{align}
and $\vec{\epsilon^m} = (\epsilon^m \; \cdts \; \epsilon^m \; \epsilon^m_*)$ is a $Q$-dimensional vector in replica space. Throughout this subsection we will adopt the notation that arrowed vectors refer to vectors in replica space.

We now adopt the spin-coherent path integral \cite{altland2010condensed} and take the continuum limit.  Introducing compact fields $(\theta^a_x(t), \phi^a_x(t))^{a=1:Q,}_{x=1:L}$ with correspondence $(\hat{X}, \hat{Y}, \hat{Z}) \rightarrow (\cos\phi\sin\theta, \cos\phi\sin\theta, \cos\theta)$ turns the partition function into a path integral $\mathcal{Z} = \int D[\theta^a_x, \phi^a_x]^a_x e^{-\mathcal{S}}$ with action $\mathcal{S} = \int_t (-\frac{i}{2} \sum_{a, x} \cos(\theta^a_x) (\partial_t \phi^a_x) + H_Q(\theta, \phi))$. As we are working at half-filling, we expand $\theta = \theta_0 + \delta \theta$ at $\theta_0=\pi/2$ and treat $\delta \theta \approx -\hat{Z}$ as a Gaussian variable, neglecting its compactness. Introducing continuum position $\tilde{x} = \mathsf{a} x$, where $\mathsf{a}$ is the spatial lattice constant, expanding the action to smallest relevant orders in $\mathsf{a}$, and removing the tildes, results in the partition function $\mathcal{Z} = \int D [\theta^a, \phi^a]^a e^{-\int_{t, x} \mathcal{L}}$ with Lagrangian density
\begin{align}
  \mathcal{L} & = \frac{i}{2 \mathsf{a}} \sum_{a=1}^Q \delta \theta^a \partial_t {\phi}^a 
  + \frac{1}{2\mathsf{a}} \sum_{a,b=1}^Q \delta \theta^a M_{ab} \delta \theta^b \nonumber \\
  & + \frac{J\mathsf{a}}{2} \sum_{a=1}^Q \left[ (\partial_x \delta \theta^a)^2 
  + (\partial_x {\phi}^a)^2 \right],
\end{align}
The interreplica interaction matrix can be diagonalised as $M = V \Lambda V^T$. There is one zero mode $\lambda_0 = 0$ with $\vec{v}_0 \propto \vec{\epsilon^{-1}}$ and one mode with eigenvalue $\lambda_K = \left({\epsilon}/{\sigma}\right)^2 \frac{1 + K r_\epsilon^2}{1 + K r_\sigma^2}$ and eigenvector $\vec{v}_K \propto (\epsilon \; \cdts \; \epsilon \;\; - K \epsilon_*)$ where $K := Q-1$ is the number of student replicas and $r_\sigma := \sigma_* / \sigma$. The rest of the modes are massive with eigenvalues $\lambda_{1 \leq \alpha \leq Q-2} = (\epsilon/\sigma)^2$. Integrating out the massive modes of $\delta \theta_\alpha$, doing a spacetime transformation to make the action isotropic $\tilde{x} = x / \sqrt{v}$ and $\tilde{t} = \sqrt{v} t$ with $v = 2 \mathsf{a} (\epsilon/\sigma) \sqrt{J}$, removing the tildes, applying a Hubbard-Stratonovich transformation, separating the azimuthal angle field into smooth and vortex parts $\phi^a = \phi^a_\mathrm{s} + \phi^a_\mathrm{v}$ then integrating out the smooth part, we retrieve the dual Lagrangian
\begin{align}
  \mathcal{L}_\mathrm{d} & = -i \sum_{a=1}^{Q} \chi^a \rho^a_{\mathrm{v}} + \mathcal{L}_\mathrm{G},
\end{align}
where the Gaussian part is
\begin{align}
  \mathcal{L}_\mathrm{G} & = \frac{1}{8 \pi^2 \rho_s} \left[ (\partial_t \chi_0)^2 + D^2 (\partial_x^2 \chi_0)^2 \right] \nonumber \\
  & + \frac{1}{8 \pi^2 \rho_s} \sum_{\mu=t, x} \sum_{a,b=1}^Q (\partial_\mu \chi^a) \Phi_{ab} (\partial_\mu \chi^b) \nonumber \\
  & + \frac{1}{8 \pi^2 \rho_s} \left[ (\partial_t \chi_{K})^2 + \frac{1+r_\epsilon^2 K}{1+r_\sigma^2 K} (\partial_x \chi_{K})^2 \right],
\end{align}
$\Phi = \mathbb{1}_Q - \vec{v}_0 \vec{v}_0^T - \vec{v}_{K} \vec{v}_{K}^T$ is a projector out of the zeroth and $K$th modes, the non-compact dual fields related to the original variables as $\hat{Z} \approx - (\mathsf{a}/\pi)\partial_x \chi$, the vortex densities are $\rho^a_\mathrm{v} = \sum_{i=1}^{N_a} m_{i}^a \delta(\mathbf{r}-\mathbf{r}_{i}^a)$ with vorticities $m_{i}^a$, space-time coordinate $\mathbf{r}=(t, x)$, ``superfluid stiffness'' $\rho_s \sim \sqrt{J\sigma^2/\epsilon^2}$ and diffusion coefficient $D^2 \sim \sqrt{J \epsilon^2/\sigma^2}$.

We would like to identify the most relevant term after integrating out $\rho^a_\mathrm{v}$. At Bayes optimality, it was shown in Ref. \cite{barratt2022field} that the minimal vortex configurations are interreplica vortex-antivortex pairs between all replicas, 
leading to the minimal Lagrangian $\mathcal{L}_\mathrm{BO} = \mathcal{L}_\mathrm{G} + \lambda \sum_{a > b} \cos(\chi_a - \chi_b)$. From renormalisation analysis, it can be shown that the minimal interaction becomes relevant at $\rho_\mathrm{s}^\# = \pi^{-1}$, corresponding to a modified Kosterlitz-Thouless (KT) transition \cite{tong2017statistical}.

Away from Bayes optimality, we use a `Landau approach' to determine the minimal interaction term. Writing down the most general term satifying the symmetries then identifying the most relevant minimal vortex term \cite{tong2017statistical}, we find that away from Bayes optimality, the teacher vorticity is forced to be zero, as adding terms that correspond to introducing teacher vorticity are irrelevant away from commensurate points, but also always less relevant than the term that follows. The minimal vortex configurations are interreplica vortex-antivortex pairs involving \emph{only} the student replicas.
This leads to the minimal Lagrangian $\mathrm{L}_\mathrm{BnO} = \mathcal{L}_\mathrm{G} + \lambda \sum_{a > b \; : \; a, b \leq K} \cos(\chi_{a} - \chi_{b})$. However, as vortex anti-vortex pairs are still the minimal configurations, we still predict a transition at the identical $\rho_\rm{s}^\# = \pi^{-1}$. 

Inside the charge-fuzzy phase, where the vortex terms are irrelevant, we can use the Gaussian theory to predict the scaling of correlators. Away from the Bayes optimal point, the connected correlator can be nonzero. Using the aforementioned operator relations we find
\begin{align}
  \E_{Y,S^*} [\langle s_{x} s_{0}\rangle_\c] & = - 4 \pi \rho_s \left(\frac{\mathsf{a}}{x} \right)^2, \\
  \E_{Y,S^*} [\langle s_{x} s_{0}\rangle] & = 4 \pi \rho_s (r_\epsilon^2 - 1) \left(\frac{\mathsf{a}}{x} \right)^2, \\
  \E_{Y,S^*} [\langle s_{x}\rangle s^*_{0}] & = 4 \pi \rho_s r_\epsilon \left(\frac{\mathsf{a}}{x} \right)^2.
\end{align}
Although the exact scaling does not agree with the lowest-order expansion \cref{eq:planted-ssep-connected-correlator,eq:planted-ssep-disconnected-correlator,eq:planted-ssep-alignment}, the sign of the quantities do. In particular, both methods agree that the sign of the disconnected correlator switches sign at the Bayes optimal point $r_\epsilon = 1$.

In the charge-sharp phase, connected correlators decay exponentially, $\E_{Y,S^*} [\langle s_{x} s_{0}\rangle_\c] \sim - e^{x/x_0}$. As for the disconnected correlator, in the crudest approximation, since we expect the student to closely follow the teacher, we expect $\E_{Y,S} [\langle s_x s_0 \rangle] \approx \E_{Y,S} [s^*_x s^*_0]$. Since the teacher configuration evolves as the SSEP and due to the choice of the prior at initial time, we expect $\E_{Y,S} [\langle s_x s_0 \rangle] \approx 0$ and have no sign on average.

The above field theory analysis is only valid for $\epsilon > 0$. At the exceptional line $\epsilon=0$, all three of the correlators should vanish for $x \neq 0$.

\subsubsection{Numerical results} \label{sec:planted-ssep-numerics}
\paragraph*{Time-evolving block decimation (TEBD) ---}
\begin{figure*}[t!]
  \centering
  \includegraphics[width=\linewidth]{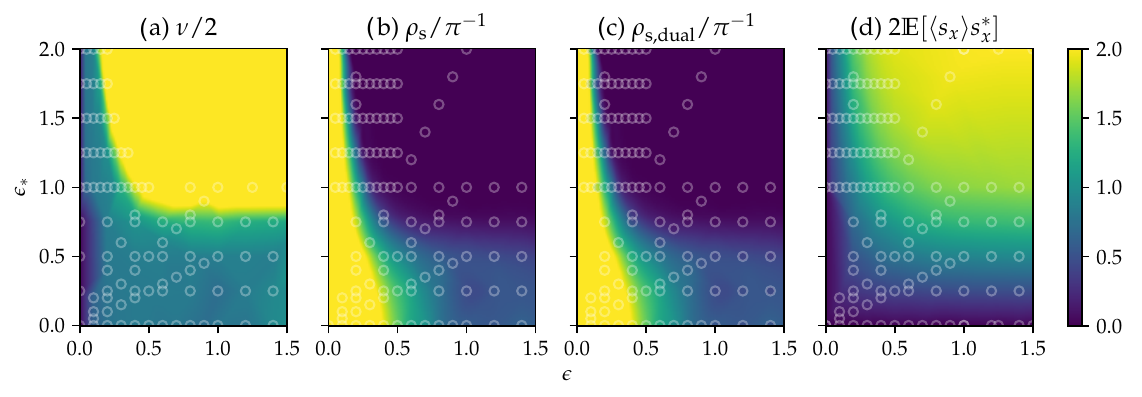}
  \caption{Heatmap of observables computed via TEBD, for $L=60$, $t_\mathrm{f}=4L$ and $\#_\text{samples} \geq 41$. a) Decay of connected correlator, $\E[\langle s_x s_0 \rangle_\mathrm{c}] \sim \lvert x\rvert^{-\nu}$, expected to be $\nu/2 \approx 1$ in the fuzzy phase. Note that vanishing $\nu$ as the $\epsilon=0$ axis is approached reflects vanishing correlations in the SSEP. b) Superfluid stiffness extracted from local charge variance $\delta C^2_{[0, x]} \sim (8 \rho_\mathrm{s}/\pi) \ln \lvert x \rvert$, c) from string operator $W^2_{[0, x]} \sim \lvert x \rvert^{- 2 \pi \rho_{\mathrm{s, dual}}}$, expected $\rho^\#_\mathrm{s}/\pi^{-1} = 1$ at the transition. d) Student-teacher alignment $\E[\langle s_x \rangle s^*_x]$, expected to be $\sim 1$ in the easy inference phase and $\sim 0$ in the hard phase. White circles indicate points in parameter space obtained, whilst the rest of the plot is interpolated. To make the features visible, we cap the dynamic range of the heatmap at $2$.} \label{fig:tebd-observables}
\end{figure*}
To test the long-distance predictions of the replica field theory, we must study larger system sizes. We therefore simulate the posterior using TEBD \cite{vidal2004efficient} for $L=60$ and $t_\mathrm{f}=4L$ using the \texttt{iTensor} library \cite{itensor}.

We fit a power law exponent $\nu$ to the decay of the averaged connected correlator, $\E_{Y, S^*}[\langle s_x s_0 \rangle_\mathrm{c}] \sim \lvert x \rvert^{-\nu}$. We expect $\nu$ to plateau to a value $\approx 2$ inside of the charge fuzzy phase with $\epsilon > 0$. 
In the charge sharp phase, we expect an exponential decay. Therefore, a fit to power law decay will give $\nu > 2$. In Fig. \ref{fig:tebd-observables}(a) we plot a heatmap of $\nu/2$ in the $\epsilon-\epsilon_*$ plane. We can clearly see two regions where $\nu$ plateaus to value $\nu_\mathrm{plateau} \approx 0.9 \times 2$, indicating a charge-fuzzy phase, and one where $\nu \gg 2$, indicating a charge-sharp phase.

We can extract superfluid stiffness $\rho_s$ from two different observables. The first is via the local charge variance, which in the charge-fuzzy phase, can be predicted to be
\begin{align}
  \delta C^2_{[0, x]} & = \E_{Y, S^*} \sum_{x', x'' \in [0, x]} [\langle s_{x'} s_{x''}\rangle_\mathrm{c}] \\
  & \approx \frac{8 \rho_\mathrm{s}}{\pi} \ln \lvert x \rvert + \dts
\end{align}
This is shown in Fig. \ref{fig:tebd-observables}(b). The second is via the string order parameter
\begin{align}
  W^2_{[0, x]} & = \E_{Y, S^*} [ \langle s_0 s_1 \cdts s_x \rangle^2] \\
  & \propto \lvert x \rvert^{-2 \pi \rho_\mathrm{s, dual}},
\end{align}
shown in Fig. \ref{fig:tebd-observables}(c). In the charge sharp phase, both quantities are exponentially decaying, so we expect both of the extracted stiffnesses to be small, $\rho_\mathrm{s} \ll 1$. We see good agreement between the two methods of extraction. At Bayes optimality, we see good evidence of phase transition at $\epsilon^\#_\mathrm{BO} \approx 0.6$, consistent with the expected location when mapping from projective to weak measurements as discussed in the beginning of the section. At fixed $\epsilon_* < \epsilon^\#_\mathrm{BO} \approx 0.6$, as $\epsilon$ is increased, we see that $\rho_\mathrm{s}$ plateaus instead of decaying to a value $\rho_\mathrm{s} \ll 1$ as is the case for $\epsilon_* > \epsilon^\#_\mathrm{BO}$. Presumably due to finite size effects, the plateau value is smaller than critical value predicted by the field theory $\rho_\mathrm{s,plateau}(\epsilon_* < \epsilon^\#_\mathrm{BO}) \approx 0.25 \rho_\mathrm{s}^\#$, but is much larger than $\rho_\mathrm{s}(\epsilon_* > \epsilon^\#_\mathrm{BO}) \approx 10^{-3} \rho_\mathrm{s}^\#$. 

In the Bayes optimal case, due to Eq. \eqref{eq:accuracy-precision}, the transition in the charge variance also corresponds to the transition in the mean-squared-error of the charge. We expect the phase diagram of the easy and hard regimes of inference to continue away from the Bayes optimal point. As a measure of the quality of the inference, we look at the alignment between the teacher and the student, $\E_{Y, S^*} [\langle s_x \rangle s^*_x]$ in Fig. \ref{fig:tebd-observables}(d). We see that it follows the general profile of the other three observables. This indicates that the charge-fuzzy and sharp regimes also corresponds to the inference impossible and possible regimes, respectively.

To determine the phase boundary from each of the computed observables, in \cref{fig:planted-ssep-phase-diagram}, we overlay the boundaries extracted from the four observables. For $\nu$, $\rho_\mathrm{s}$, $\rho_\mathrm{s, dual}$, we plot the points when the observables reach the plateau value. For $\E[\langle s_x \rangle s_x^*]$, we plot the contour when it is equal to $0.5$. We see that the phase boundary suggested by the four observables has good agreement. First, along the Bayes optimal line, we see a phase transition at around $\epsilon^\#(r_\epsilon=1) = \epsilon^\#_\mathrm{BO} \approx 0.6$. Looking at fixed $\epsilon_*$ and varying $\epsilon$, we see that there appears to be a threshold again at around $\epsilon_* = \epsilon^\#_\mathrm{BO} \approx 0.6$. 

\paragraph*{Exact calculation of the posterior ---}
\begin{figure*}[t!]
  \centering
  \includegraphics[width=\linewidth]{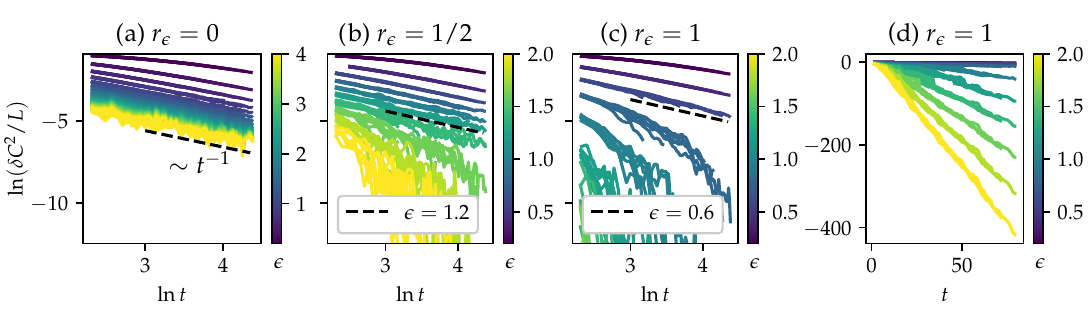}
  \caption{Time evolution of normalised charge variance $\delta C^2/L$ as function of time via exact calculation of the posterior, for $6 \leq L \leq 22$, $\#_\mathrm{samples}=100$. Data from all system sizes $L$ are overlaid on top of each other, showing good collapse. (a)-(c): log-log plot for (a) $r_\epsilon := \epsilon_*/\epsilon =0$, (b) $r_\epsilon=1/2$ and (c) Bayes optimal case $r_\epsilon=1$. For $r_\epsilon=0$, we see data consistent with a power law $\sim t^{-1}$ even for high $\epsilon = 4$. For $r_\epsilon=1/2, 1$, the scaling switches from $\sim t^{-1}$ to exponential decay $\mathrm{e}^{-t}$ at critical $\epsilon^\#(r_\epsilon)$, as shown by semilog plot in (d).} \label{fig:charge-variance}
\end{figure*}
The remaining question is whether the phase boundary hits the $\epsilon_* = 0$ line as $\epsilon$ is further increased. The replica field theory cannot give predictions on whether the phase transition occurs at finite $\epsilon$, since we don't have an explicit map to $\rho_\mathrm{s}$. From TEBD, we see no evidence it for $\epsilon < 1.5$. However, higher values of $\epsilon$ are unachievable at this system size using TEBD. This can be understood in the `forced' measurement limit ($r_\epsilon=0$) \cite{nahum2021measurement,feng2023measurement}, where the measurements are iid Gaussian noise. When $\epsilon$ is large, consider receiving the first measurement $\bm{y}_1$ is compatible with total charge $C_1 = \sum_r \mathrm{sgn}(y_{r,1})$. Consider when $\epsilon$ is large enough such that a TEBD algorithm with a finite cutoff for truncation (or even exact calculation of the posterior with finite precision) will discard other charge sectors not equal to $C_1$. In the next timestep however, we may receive $\bm{y}_{2}$ with $C_2 \neq C_1$ that has higher likelihood than the first image, $\exp(\epsilon \sum_{r} y_{r, 2}) \gg \exp(\epsilon \sum_{r} y_{r, 1})$. However, due to the truncation at the first timestep, the weight in charge sector $C_2$ is zero and stays zero even after applying the likelihood with large weight. The $C_1$ sector on the other hand may be killed as the likelihood has very small weight. Normalisation then leads to numerical instability due to division by zero.

We therefore resort to exact calculation of the prior for small system sizes $6 \leq L \leq 22$ using quadruple-precision floating numbers, capable of expressing floating points of the range $[10^{-4932},10^{4932}]$ compared to normal double-precision floats with range $[10^{-308},10^{308}]$. This allows us to reach higher values of $\epsilon$ without suffering from numerical instabilities.

We first study the charge variance. In Fig. \ref{fig:charge-variance}, we plot the dynamics of the normalised charge variance $\delta C^2 / L$ up to $t_\mathrm{f} = 4L$. We overlay data for all system sizes, which shows good collapse, indicating that $\delta C^2(t \sim L) \sim L$. We see that for $r_\epsilon = 1/2, 1$, in the charge fuzzy phase, the decay of the charge variance is consistent with $\delta C^2 \propto L t^{-1}$, while in the charge-sharp phase $\epsilon > \epsilon^\#(r_\epsilon)$, the dynamics switches to an exponential decay $\delta C^2 \propto L \mathrm{e}^{-t}$ as evidenced by Fig \ref{fig:charge-variance}(d). In the forced measurement case $r_\epsilon=0$, we see no deviation from the power law up to $\epsilon \leq 4$.

A simple tractable model to get an intuition on the power-law/exponential-decay transition is when the charges are either static or always swap, i.e. $J=0,\infty$, see Appendix \ref{apdx:static-saddle-point} for details. In this case the asymptotic behaviour for charge variance can be calculated exactly via saddle-point approximation. For $\epsilon > 0$ and $\epsilon_* = 0$, $\delta C^2 \propto L \epsilon^{-1} t^{-1/2}$, whilst for any $\epsilon_* > 0$, we have exponential decay with time, with $\delta C^2 \propto L \epsilon^{-1} t^{-1/2} e^{-r_\epsilon^2 t/2}$ when $r_\epsilon \leq 2 \epsilon$ and $\delta C^2 \propto L \exp(-2\epsilon(r_\epsilon - 2 \epsilon)t)$ when  $r_\epsilon > 2 \epsilon$. The reason for the change in scaling with $t$ is due to the fact that the saddle point is at zero for $\epsilon_*=0$ and therefore no leading order term is present. When $\epsilon_*>0$, the saddle point shifts away from zero resulting in leading order factor $e^{-r_\epsilon^2 t/2}$. Therefore the simple model has no phase transition in the forced measurement case and transitions into the exponentially decaying phase as soon as $\epsilon_* > 0$. In \cref{fig:planted-ssep-phase-diagram}, we also overlay the boundary where the charge variance switches from a power law decay in time to an exponential one. We see good agreement with the boundaries found from TEBD.

\begin{figure}[h!]
  \includegraphics[width=1.0\columnwidth]{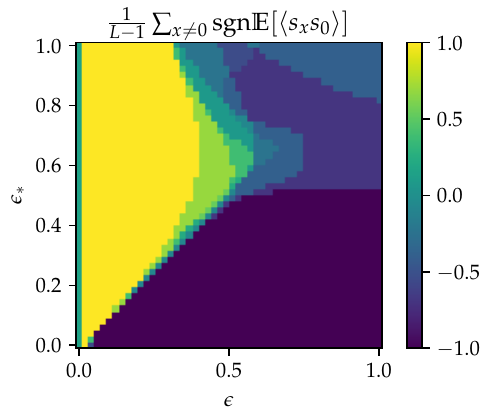}
  \caption{Average sign of the disconnected charge correlator for $L=14$ and $\#_\mathrm{samples}=100$, showing clear delination between $r_\epsilon \lessgtr 1$ charge-fuzzy regions.} \label{fig:signs}
\end{figure}

Next we look at the sign of the averaged disconnected charge correlator, $\frac{1}{L-1} \sum_{x\neq 0} \mathrm{sgn} \E_{Y, S^*} [\langle s_x s_{0} \rangle]$. At $\epsilon=0$, it is exactly zero due to it corresponding to SSEP at half filling. In the charge-fuzzy phase, both perturbation theory and replica field theory predicts that sign is positive for $r_\epsilon > 1$ and negative for $r_\epsilon < 1$. In the charge-sharp phase, we expect it to be zero. Therefore the sign of the disconnected correlator distinguishes between the three regions. The predictions are well supported by the numerics as shown in Fig. \ref{fig:signs}, showing a sharp delination between $r_\epsilon \lessgtr 1$ charge-fuzzy regions, even for smaller system sizes.

\section{The planted XOR and the Haar-random $\rm{U}(1)$-symmetric monitored quantum tree} \label{sec:planted-xor}
In the previous section, we started with a monitored quantum circuit and derived a classical inference problem. Here, we go the other way around, starting with a classical inference problem on a HMM we call \emph{the planted XOR} in \cref{sec:classical-planted-xor} and convert it to the Haar-random $\rm{U}(1)$ monitored quantum tree, which can also be referred to as \emph{the quantum planted XOR}, in \cref{sec:quantum-planted-xor}. For not only the original HMM (\cref{sec:classical-planted-xor-analysis}) but also for the fully quantum $d=1$ limit (\cref{sec:quantum-planted-xor-analysis}), we exploit the structure of the tree and analytically deduce the full phase diagram in the student-teacher parameter space as \cref{fig:planted-xor-phase-diagram}. Remarkably, we uncover a re-entrance in the phase diagram in the quantum case, which has been previously seen numerically in the 2D RBIM, cf. \cref{fig:2d-rbim-phase-diagram}.

\subsection{Classical inference model} \label{sec:classical-planted-xor}
Consider the teacher, who prepares $2^N$ hidden Boolean variables $\bm{b}^*_1 = b^*_{1:2^N, 1}$, each chosen randomly with equal probability. At each timestep, they apply the XOR gate on each neighbouring pair, yielding $\bm{b}^*_2 = b^*_{1:2^{N-1}, 2}$ at the next timestep. This is repeated until there's only one Boolean $b^*_{1, N+1}$ left at final timestep $N+1$. Throughout this section, we will interchange between Ising and Boolean variables as $s_{i, t} = 2b_{i, t} - 1$.

Although the evolution for the Booleans is deterministic, it can still be described as an HMM. The prior for the entire trajectory $B^* := \bm{b}^*_{1:t}$ would be $p_*(B^*) = \left[\prod_\tau p_*(\bm{b}^*_\tau \vert \bm{b}^*_{\tau-1})\right] p_*(\bm{b}^*_1)$. The initial choice of Booleans is still random, $p_*(\bm{b}^*_1) = 2^{-N}$, but the `Markov' kernel is deterministic and separable as $p_*(\bm{b}^*_t \vert \bm{b}^*_{t-1}) = \prod_{i} p_*(b^*_{i, t} \vert b^*_{i_\mathrm{L}, t-1}, b^*_{i_\mathrm{R}, t-1})$, where
\begin{align}
    p_*(b^*_{i, t} \vert b^*_{i_\mathrm{L}, t-1}, b^*_{i_\mathrm{R}, t-1}) = \delta\{b^*_{i, t}, \mathrm{XOR}(b^*_{i_\mathrm{L}, t-1}, b^*_{i_\mathrm{R}, t-1})\}.
\end{align}
Here, we recall that $\delta \{\cdot , \cdot\}$ is the Kronecker delta.

At each timestep, the teacher also generates observations on each Boolean $b^*_{i, t}$ $y_{i, t} = \epsilon_* s^*_{i, t} + \sigma_* \eta_{i, t}$, yielding the measurement model \cref{eq:teacher-measurement-model} as in the planted SSEP case.

The goal of the student is to determine the last bit $b^*_{1, N+1}$, or equivalently, the parity of the starting bits. They can  do this by inferring a posterior for the filtering task, $p_\rm{s}(b_{1, N+1} \vert Y)$. The natural order parameter then would be the $\rm{MSEM}$ or the $\rm{MSE}$ of the final bit.

\subsection{Quantum tree model} \label{sec:quantum-planted-xor}
To map the above classical inference problem into a quantum one, we first note that given input bits $b_\mathrm{L} b_\mathrm{R}$, the SSEP kernel \cref{eq:ssep-kernel} allows for the possible transitions $00 \rightarrow 00$, $01 \rightarrow 01$, $01 \rightarrow 10$, $10 \rightarrow 01$, $10 \rightarrow 10$, $11 \rightarrow 11$. Then, we observe that if we condition on the left output bit to be a $0$, the right output bit is $\mathrm{XOR}(b_\mathrm{L}, b_\mathrm{R})$. Conversely, if we condition the left output bit to be a $1$, then the right output bit is $\mathrm{NOT XOR}(b_\mathrm{L}, b_\mathrm{R})$. Then, applying the SSEP gate then giving the information about the left output bit is equivalent to applying a $\mathrm{XOR}$ or $\mathrm{NOT XOR}$ and revealing which was used. As this information is completely specified, the model can be exactly mapped to the above situation of applying the $\mathrm{XOR}$ each time (and the student knowing this fact). 

\begin{figure}[h]
  \def\u1gate[#1](#2,#3)
      { \draw[fill=white, #1] (#2-.4,#3-.4) rectangle (#2+.4,#3+.4) node[pos=.5] {$\rm{U}(1)$}; }

  \def\projmeas[#1](#2,#3)
      { \draw[fill=white] (#2-.2,#3-.2) rectangle (#2+.2,#3+.2) node[pos=.5] {$/$};
        \draw [domain=55:125] plot ({.25*cos(\x)+#2}, {.25*sin(\x)+#3-0.25}); }

  \def\weakmeas[#1](#2,#3)
      {
        \draw[fill=white] (#2-.2,#3-.2) rectangle (#2+.2,#3+.2) node[pos=.5] {};
        \draw plot [smooth] coordinates {(#2-.15,#3-.15) (#2-.07,#3-.1) (#2,#3+.15) (#2+.07,#3-.1) (#2+.15,#3-.15)};
      }
    \begin{circuitikz}[]
      \draw (-2, 1) node[] {\large {(a)}};
      \draw (2, 1) node[] {$\quad\quad$};
      \draw (0,0) node[xor port, rotate=90] (xor1) {\rotatebox{-90}{$\mathrm{XOR}$}}
      (-1, -1.8) node[xor port, rotate=90] (xor21) {\rotatebox{-90}{$\mathrm{XOR}$}}
      (1, -1.8) node[xor port, rotate=90] (xor22) {\rotatebox{-90}{$\mathrm{XOR}$}}
      (-1.5, -2*1.8) node[xor port, rotate=90] (xor31) {\rotatebox{-90}{$\mathrm{XOR}$}}
      (1.5, -2*1.8) node[xor port, rotate=90] (xor32) {\rotatebox{-90}{$\mathrm{XOR}$}}
      (xor21.out) -- (xor1.in 1)
      (xor22.out) -- (xor1.in 2);
      \draw[thick, dotted] (xor31.out) -- (xor21.in 1)
      (xor32.out) -- (xor22.in 2);
      \draw[] (xor1.out |- 0, 0.7) -- (xor1.out);
      \draw (xor1.out |- 0, 0.7) node[above] {$b_{1, N+1}$}
      (xor31.in 1) node[below] {$b_{1, 1}$}
      (xor31.in 2) node[below] {$b_{2, 1}$}
      (xor32.in 1) node[below] {$\cdots$}
      (xor32.in 2) node[below] {$\quad \; b_{2^N, 1}$};
      \draw (0, -2.5*1.8) node[] {\Huge $\; \cdots$};
      \draw[thick, dotted] (xor21.in 2 |- 0, -2.1*1.8) -- (xor21.in 2);
      \draw[thick, dotted] (xor22.in 1 |- 0, -2.1*1.8) -- (xor22.in 1);
      \weakmeas[](0, 0.3)
      \weakmeas[](-.65, -1.55)
      \weakmeas[](.65, -1.55)
    \end{circuitikz}

    \bigskip

    \begin{tikzpicture}[]
      \draw (-2, 1.5*.8) node[] {\large {(b)}};
      \draw (2, 1.5*.8) node[] {$\quad\quad\quad$};

      \draw (-0.25-1, -6*.8) -- (-0.25-1, 1.5*.8);

      \draw (0.25-1, -2.75*.8) -- (0.25-1, 1.5*.8);
      \draw[dotted, thick] (0.25-1, -3.5*.8) -- (0.25-1, -2.75*.8);
      \draw (0.25-1, -6*.8) -- (0.25-1, -3.5*.8);

      \draw (0.25, -2.75*.8) -- (0.25, 1.5*.8);
      \draw (-0.25, -2.75*.8) -- (-0.25, 1.5*.8);
      \draw[dotted, thick] (0.25, -3.5*.8) -- (0.25, -2.75*.8);
      \draw[dotted, thick] (-0.25, -3.5*.8) -- (-0.25, -2.75*.8);
      \draw (0, -5*.8) node[] {\huge $\; \cdots$};

      \draw (-0.25+1, -2.75*.8) -- (-0.25+1, 1.5*.8);
      \draw[dotted, thick] (-0.25+1, -3.5*.8) -- (-0.25+1, -2.75*.8);
      \draw (-0.25+1, -6*.8) -- (-0.25+1, -3.5*.8);

      \draw (0.25+1, -6*.8) -- (0.25+1, 1.5*.8);
      \projmeas[](-.25, 1*.8)
      \weakmeas[](.25, 1*.8)

      \u1gate[](0, 0)
      
      \projmeas[](-.75, -1*.8)
      \weakmeas[](-.25, -1*.8)
      \weakmeas[](.25, -1*.8)
      \projmeas[](.75, -1*.8)

      \u1gate[](-.5, -2*.8)
      \u1gate[](.5, -2*.8)

      \projmeas[](-.75-.5, -4*.8)
      \weakmeas[](-.25-.5, -4*.8)
      \weakmeas[](.25+.5, -4*.8)
      \projmeas[](.75+.5, -4*.8)

      \u1gate[](-1, -5*.8)
      \u1gate[](1, -5*.8)

    \end{tikzpicture}
  {\centering
  \caption{(a) The planted XOR model and (b) The Haar-random $\rm{U}(1)$-symmetric quantum tree. Box with a meter represents projective measurements. Box with a bell curve represents a noisy observations in the classical inference case, and continuous weak measurement in the quantum case. Box with $\rm{U}(1)$ denotes a Haar unitary acting on qubit-qudit pairs that conserves charge on the qubit sector.}
  \label{fig:xor} }
\end{figure}
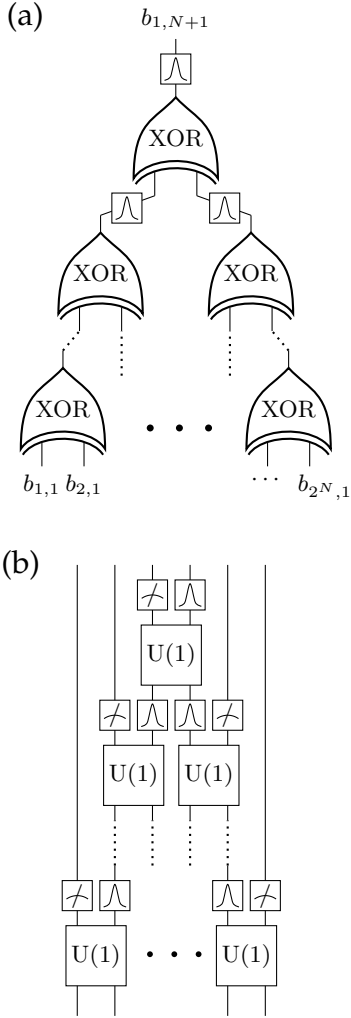

To construct the quantum tree model, we do the opposite procedure of \cref{sec:haar-case}. Since the SSEP kernel \cref{eq:ssep-kernel} is obtained from ${u}_{U(1)}$, we replace each $\mathrm{XOR}$ gate with a ${u}_{U(1)}$ gate with one output qubit projectively measured. The observation model is replaced by weak gaussian measurements \cref{eq:weak-measurement}. This results in a quantum tree as sketched in \cref{fig:xor}(b).

As for the initial state, we will select the scenario where teacher randomly chooses from an equally weighted ensemble of product states of qubit-strings and equal superposition of qudit strings in the computation basis,
\begin{align}
  \{ \ket{\psi_0} \} = \left\{\ket{b^*_{1:2^N}} \otimes \ket{\chi}^{\otimes d^{2^N}} : b^*_{1:2^N} \in \{0, 1 \}^N\right\}.
\end{align}

The teacher evolves the quantum tree, sends the choice of unitary gates sampled from the Haar distribution and the projective and weak measurement outcomes to the student. At the end, the teacher measures the final qubit in the computational basis to obtain $Z_*$. The student then tries to guess the final projective qubit measurement performed at the end by the teacher. 

In the case of general local qudit dimension $d < \infty$ and revealed choice of Haar unitaries, the inference problem on this quantum tree is not reducible to a classical HMM. A similar model with both measurements being projective with a finite rate was studied in Ref. \cite{feng2024charge} at Bayes optimality.

In the inference setting, the natural order parameter to study would be the mean-squared-errors or the alignment between the true and inferred final qubit,
\begin{equation}
\begin{aligned} \label{eq:mse-planted-xor}
  \mathrm{MSE} & = \E_{U, S, Y, S^*}[ (s_{1, N+1} - s^*_{1 + N+1})^2 ] \\
  & = 4 - 4 \E_{U, Y,S^*}[p_\mathrm{s}(s_{1, N+1}={s_{1, N+1}^*} \vert Y)].
\end{aligned}
\end{equation}
\begin{align} \label{eq:msem-planted-xor}
  \mathrm{MSEM} & = \E_{U, Y, S^*}[ (\expval{s_{1, N+1}} - s^*_{1 + N+1})^2 ] \\
  & = 4 - 4 \E_{U, Y, S^*}[p_\rm{s}(s_{1, N+1} = s^*_{1, N+1} \vert Y)] \\
  & + 4 \E_{U, Y, S^*}[p_\rm{s}(s_{1, N+1} = -1 \vert Y) p_\rm{s}(s_{1, N+1} = 1 \vert Y)], \nonumber
\end{align}
and the student's final qubit variance,
\begin{align} \label{eq:dZ2-planted-xor}
  \delta Z^2_\rm{s} & = \E_{U, Y}[ \langle Z^2 \rangle - \langle Z \rangle^2 ] \\
  & = 4 \E_{U, Y}[p_\rm{s}(s_{1, N+1} = -1 \vert Y) p_\rm{s}(s_{1, N+1} = 1 \vert Y)].
\end{align}
In the standalone quantum tree case, the natural order parameter would be the teacher's final qubit variance $\delta Z^2_*$, equal to \cref{eq:dZ2-planted-xor} at Bayes optimality.

\begin{figure}[h]
  \begin{flushleft}
    $\quad\quad\quad\quad$ \large (a) $\; d=\infty$ / planted XOR
  \end{flushleft}
  \center
  \vspace{-0.2cm}
  \includegraphics[width=.95\linewidth]{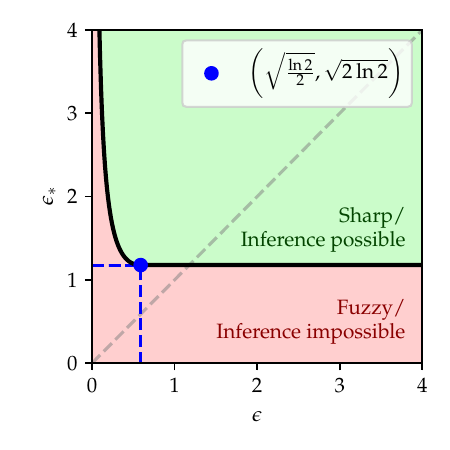}
  \vspace{-0.5cm}
  \begin{flushleft}
    $\quad\quad\quad\quad$ \large (b) $\; d=1$ quantum planted XOR
  \end{flushleft}
  \vspace{-0.2cm}
  \includegraphics[width=.95\linewidth]{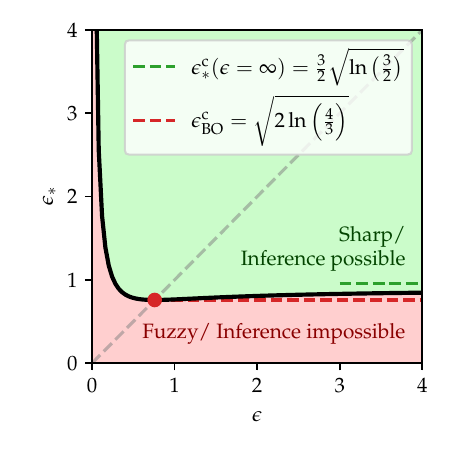}
  \caption{Exact expanded phase diagrams for the (a) classical planted XOR or the $d=\infty$ Haar-random $\rm{U}(1)$-symmetric monitored quantum tree and (b) quantum XOR or the $d=1$ Haar-random $\rm{U}(1)$-symmetric monitored quantum tree, the latter of which is given by \cref{eq:d=1-xor-boundary}. Note the re-entrance of the phase diagram, similar to that of 2D RBIM \cref{fig:2d-rbim-phase-diagram}.}
  \label{fig:planted-xor-phase-diagram}
\end{figure}

\subsection{Analysis of the planted XOR} \label{sec:classical-planted-xor-analysis}
Due to the tree structure and Booleanity of the model, the model can be studied without resorting to the replica trick. The key observation is that the student's posterior, or the density matrix, can be treated as a random variable dependent on the observations $Y$. Furthermore, due to the forward equation and the tree structure, the posterior for a particular bit at a particular time $p(b_{i, t} \vert \bm{y}_{1:t})$ only depends on the posterior for the left and right bits at previous times, $p(b_{i_\mathrm{L}, t-1} \vert \bm{y}_{1:t-1})$, $p(b_{i_\mathrm{R}, t-1} \vert \bm{y}_{1:t-1})$, which are independent, and likewise for the density matrix. Due to Booleanity, one only needs to keep track of the posterior for one value of the Boolean, since $p(b_{i, t} \vert \bm{y}_{1:t}) = 1 - p(\overline{b_{i, t}} \vert \bm{y}_{1:t})$.

\subsubsection{Finding the critical parameters}
Then, we can apply the method of travelling waves used to study the directed polymer on a tree in Ref. \cite{derrida1988polymers}. In that case, the partition function at time $t$, $Z_t$, evolved linearly with independent partition functions at previous times as $Z_t = e^{-\beta V} [Z^{(1)}_{t-1} + Z^{(2)}_{t-1}]$, where $V \sim \mathcal{N}(0, 1)$ in their case. Then, a generating function $G_t(x) := \E_V [\exp(-e^{\beta x} Z_t)]$ is defined and evolves as the FKPP (Fisher-Kolmogorov-Petrovsky-Piskunov) equation. In our case, the posterior at time $t$ evolves non-linearly as a function of the posterior at previous time $t-1$. However, if there exists a transition as parameters are varied, we expect the posterior of the student evaluated at the true Boolean to be close to one, $p_\mathrm{s}(b_{i, t} = b^*_{i, t} \vert \bm{y}_{1:t}) \approx 1$, or equivalently, the posterior evaluated at the opposite value $\overline{b^*_{i, t}}$ to be small $p_\mathrm{s}(b_{i, t} = \overline{b^*_{i, t}} \vert \bm{y}_{1:t}) \ll 1$. Therefore we can linearise the evolution equation and continue with the same analysis. This is the same strategy used in Ref. \cite{feng2023measurement}.

\newcommand\lft{\mathrm{left}}
\newcommand\rgt{\mathrm{right}}

Concretely, consider the posterior for a boolean $b = b_{i, t}$ on a particular branch at a given timestep, $p(b \vert \bm{y}_{1:t})$. Due to the structure of the tree, it is in fact only dependent on observations from branches connected to it from the past, i.e. $p(b \vert \bm{y}_{1:t}) = p(b \vert \bm{y}^\mathrm{c}_{1:t})$. Using the forward equation, this can be related to the posterior for booleans connected from left $b_{\mathrm{L}} = b_{i_\mathrm{L}, t-1}$ and the right $b_{\mathrm{R}} = b_{i_\mathrm{R}, t-1}$ from the previous timestep,
\begin{align}
  p(b \vert \bm{y}^\mathrm{c}_{1:t}) = \frac{p(y_t \vert b) \sum_{b_{\mathrm{L}:\mathrm{R}}} p(b \vert b_{\mathrm{L}:\mathrm{R}}) p(b_\mathrm{L} \vert \bm{y}_{1:t-1}^\mathrm{L}) p(b_\mathrm{R} \vert \bm{y}_{1:t-1}^\mathrm{R})}{p(\bm{y}^\mathrm{c}_{1:t})}.
\end{align}
Here, the marginalised likelihood is $p(\bm{y}^\mathrm{c}_{1:t}) = \sum_{b, b_{\mathrm{L}:\mathrm{R}}} p(y_t \vert b) p(b \vert b_{\mathrm{L}:\mathrm{R}}) p(b_\mathrm{L} \vert \bm{y}_{1:t-1}^\mathrm{L}) p(b_\mathrm{R} \vert \bm{y}_{1:t-1}^\mathrm{R})$.
Now, $y = y_t$ is not iid but is distributed as $y=\eta + \epsilon_* s^*$, with $s^* = -s^*_\mathrm{L} s^*_\mathrm{R}$. Now, near the phase transition, we assume that $p(b=b^* \vert \bm{y}^\mathrm{c}_{1:t})$ is large and $p(b=\overline{b^*} \vert \bm{y}^\mathrm{c}_{1:t})$ is small, and likewise for the posteriors for $b_\mathrm{L}$ and $b_{\mathrm{R}}$. If $b^* = 1$, we can either have $(b_\mathrm{L}, b_\mathrm{R}) = (0, 1)$ or $(1, 0)$. If $b^* = 0$, we can have $(1 ,1)$ or $(0, 0)$. For every case, denoting the smaller posterior value as $p_t = p(b=\overline{b^*} \vert \bm{y}^\mathrm{c}_{1:t})$ and likewise for $p_t^{\mathrm{L}, \mathrm{R}} = p(b_{\mathrm{L}, \mathrm{R}}=\overline{b^*_{\mathrm{L}, \mathrm{R}}} \vert \bm{y}^\mathrm{L,R}_{1:t-1})$, the posterior evolution equation is the same for every 4 possibility mentioned above (up to changing sign of $\eta$),
\begin{align}
    p_t = \frac{(p_{t-1}^\mathrm{L}+p_{t-1}^\mathrm{R}-2p_{t-1}^\mathrm{L} p_{t-1}^\mathrm{R}) e^{\epsilon  \left(\eta -\epsilon _*\right)}}{
    \left(\begin{aligned}
      (p_{t-1}^\mathrm{L}+p_{t-1}^\mathrm{R}-2p_{t-1}^\mathrm{L}p_{t-1}^\mathrm{R}) & e^{\epsilon  \left(\eta -\epsilon _*\right)} \\ 
      + (1-p_{t-1}^\mathrm{L}-p_{t-1}^\mathrm{R}+2p_{t-1}^\mathrm{L}p_{t-1}^\mathrm{R}) & e^{-\epsilon  \left(\eta -\epsilon _*\right)}
    \end{aligned}\right)},
\end{align}
which can be expanded to linear order as
\begin{align}
    p_t = A (p_{t-1}^\mathrm{L} + p_{t-1}^\mathrm{R}),
\end{align}
with $A = e^{2 \epsilon (\eta - \epsilon^*)}$, which is importantly iid.

To understand the evolution of typical $p_t$ with increasing $t$, we define the generating function $G_t(x) = \langle \exp(-e^{-x} p_t)\rangle$. The linearised recursion can be expressed as
\begin{align}
    G_{t+1}(x)=\E_\eta \left[ G_t\left(x-\ln A\right) G_t\left(x-\ln A\right) \right],
\end{align}
with traveling wave ansatz $G_t(x)=G^{(\lambda)}\left(x-v_{\epsilon, \epsilon^*}(\lambda) t\right)$. $\lambda \in [0, 1]$ parameterises a family of possible traveling wave solutions and will be chosen to be $\lambda_\rm{c}$ which minimises the velocity $v_{\epsilon, \epsilon^*}(\lambda)$, subject to the initial condition $G_0(x)= \exp(-e^{-x}/2) \stackrel{x \gg 1}{\rightarrow} 1 - (1/2)e^{-x}$ \cite{derrida1988polymers,nahum2021measurement}. Using the exponential decay of the wave ansatz at large arguments $G^{(\lambda)}(u) \simeq 1-e^{-\lambda u}$, and that $\E_\eta[A^\lambda]= e^{2 \lambda  \epsilon  \left(\lambda  \epsilon -\epsilon _*\right)}$, we can find the wavefront velocity as
\begin{align}
    v_{\epsilon, \epsilon^*}(\lambda)=\frac{1}{\lambda} \ln \left(2 \E_\eta[A^\lambda]\right) = \frac{\ln 2}{\lambda} + 2 \epsilon(\lambda \epsilon - \epsilon_*).
\end{align}
minimising it respect to $\lambda \in [0, 1]$, we find that
\begin{align}
  \lambda_\rm{c} = \min\left(1, \sqrt{\frac{\ln 2}{2}} \frac{1}{\epsilon}\right).
\end{align}
Therefore the velocity is given by $v_{\epsilon, \epsilon_*} = v_{\epsilon, \epsilon_*}(\lambda_\rm{c})$. If the velocity is positive, then it means that $p_t$ is exponentially decaying as we increase $t$. If velocity is negative, then $p_t$ is exponentially growing with time and the non-linear part of the evolution becomes important. Then the critical point is the velocity changes sign, i.e. when $v_{\epsilon, \epsilon_*} = 0$.
Combining the two conditions, we find
\begin{align}
  \epsilon_*^\rm{c}(\epsilon) = 
  \begin{cases}
    \frac{\ln 2}{2 \epsilon} + \epsilon \approx 0.346574 \epsilon^{-1} + \epsilon & \text{ for } \epsilon < \sqrt{\frac{\ln 2}{2}}, \\
    \sqrt{2 \ln 2} \approx 1.17741 & \text{ for }  \epsilon \geq \sqrt{\frac{\ln 2}{2}}.
  \end{cases}
\end{align}

Using this result we can deduce that in the regime where $\epsilon > 0$, $\epsilon_* > \epsilon_*^\mathrm{c}$, as the student correctly deduces the final Boolean, both $\mathrm{MSE}$, $\mathrm{MSEM}$ and $\delta Z^2_\rm{s}$ would decay exponentially with time and tend to zero as $t \rightarrow \infty$. On the other hand, for $\epsilon > 0$, $\epsilon_* < \epsilon_*^\mathrm{c}$, we would expect to converge to a non-zero value. The derived phase diagram is plotted in \cref{fig:planted-xor-phase-diagram}(a).

\subsubsection{Power-law scaling of bit-variance in the $\epsilon_* = 0$ limit}
The scaling of the long-time bit-variance can be analysed in the $\epsilon_* \rightarrow 0$ and large $\epsilon$ limit. Here, we cannot apriori expect either of $p_\mathrm{s}(1 \vert Y)$ or $p_\mathrm{s}(0 \vert Y)$ to be larger than the other. Instead we will exploit the iid nature of the disorder to extract the distribution of the posterior.

Let us arbitrarily choose one of the posterior $\uppi := p_\mathrm{s}(1 \vert \bm{y}_{1:t}^\mathrm{c})$ to study. We will denote $\uppi' := p_\mathrm{s}(1 \vert \bm{y}^\mathrm{L}_{1:t-1})$ and $\uppi'' := p_\mathrm{s}(1 \vert \bm{y}^\mathrm{R}_{1:t-1})$. Then the evolution is
\begin{align}
    \uppi & = \frac{e^{\epsilon y} A_+}{e^{\epsilon y} A_+ + e^{-\epsilon y} A_-},
\end{align}
where $A_+ = \uppi' (1 - \uppi'') + (1 - \uppi') \uppi''$ and $A_- = \uppi' \uppi'' + (1-\uppi')(1-\uppi'')$. For forced measurements $y$ is iid and uncorrelated with $\uppi'$ and $\uppi''$. 

The probability distribution for $\uppi$ then evolves as
\begin{align}
    p(\uppi) = \int dy \frac{e^{-y^2/2}}{\sqrt{2 \pi}} & \int_0^1 dp(\uppi') dp(\uppi'') \nonumber \\
    \times & \delta\left(\uppi - \frac{e^{\epsilon y} A_+}{e^{\epsilon y} A_+ + e^{-\epsilon y} A_-}\right).
\end{align}
Define $g(y) := \uppi - \frac{e^{\epsilon y} A_+}{e^{\epsilon y} A_+ + e^{-\epsilon y} A_-}$. The delta function of a function is given by $\delta(g(y)) = \sum_i \frac{1}{\lvert \partial_y g(y_i) \rvert} \delta(y - y_i)$, where $y_i$'s are zeros of $g(y)$. Firstly solving for $g(y)=0$ we get $y_0 = \frac{1}{2 \epsilon}\ln \left(\frac{\uppi _1 A_-}{\left(1-\uppi _1\right) A_+}\right)$. Differentiating $g(y)$ then substituting $y_0$ we get $\partial_y g(y_0) = 2 \epsilon \uppi(1-\uppi)$. Therefore we have
\begin{align}
    p(\uppi) 
    & = \frac{1}{2 \epsilon \uppi(1-\uppi) \sqrt{2 \pi}} \int_0^1 dp(\uppi') dp(\uppi'') e^{ -\frac{1}{8 \epsilon^2} \ln^2 \left(\frac{\uppi A_-}{\left(1-\uppi\right) A_+}\right)}. \label{eq:evolution}
\end{align}
We now look for the steady state, that is $p(\uppi) = p(\uppi'=\uppi) = p(\uppi''=\uppi)$. Then the above equation becomes a self-consistent equation/Neumann-type equation and we have a systematic way of expanding the series, if desired. Expanding with respect to large $\epsilon$ or small $1/\epsilon$ we find that
\begin{align}
    p(\uppi) & = \frac{1}{2 \epsilon \uppi(1-\uppi) \sqrt{2 \pi}} + \mathcal{O}(\epsilon^{-5}).
\end{align}
Although the series is not normalisable, we can calculate the leading order of the student's bit-variance, $\delta Z^2_\rm{s} = 4 p_0 p_1$. Using that $p_0 p_1 = \uppi(1-\uppi)$, we find that
\begin{align}
    \delta Z^2_\rm{s} = 4 \mathbb{E} [p_0 p_1]
    & = 4 \int_0^1 d\uppi \cdot \uppi (1-\uppi) p(\uppi) \nonumber \\
    & = \frac{2}{\epsilon \sqrt{2\pi}} + \mathcal{O}(\epsilon^{-5}),
\end{align}
a power-law in the student's assumed SNR, $\sim \epsilon^{-1}$. This indicates that the bit-variance does not vanish at long times on the $\epsilon_*=0$ line.

\subsection{Analysis of the $d=1$ quantum planted XOR} \label{sec:quantum-planted-xor-analysis}
Approaching the sharpening/inferrability transition from the qubit `sharp' phase, we expect the reduced density matrix of the student's density matrix of the final qubit to be collapsed to one of the two qubit values. Therefore we parametrise the student's reduced density matrix at a particular node as
\begin{align}
  \rho_{\rm{s}, t} = (1-z_t) \ketbra{b}{b} + z_t \ketbra{\bar b}{\bar b},
\end{align}
where we take $z_t$ to be a small parameter, and therefore $b$ to be the most likely qubit. The student's density matrix then evolves as
\begin{align}
  \rho_{\mathrm{s}, t+1} = \frac{(Q_{\mathrm{s}, y} \otimes P_{b'})U(\rho^\mathrm{L}_{\mathrm{s}, t} \otimes \rho^\mathrm{R}_{\mathrm{s}, t})U^\dag (Q_{\mathrm{s}, y} \otimes P_{b'})^\dag}{\tr[(Q_{\mathrm{s}, y} \otimes P_{b'}) U(\rho^\mathrm{L}_{\mathrm{s}, t} \otimes \rho^\mathrm{R}_{\mathrm{s}, t})U^\dag (Q_{\mathrm{s}, y} \otimes P_{b'})^\dag]}
\end{align}
where $Q_{\mathrm{s}, y}$ is the student's partial projector on (say) the left qubit given weak measurement outcome $y$, and $P_{b'}$ is the projector on the right qubit given strong measurement outcome $b'$, and $U$ is a Haar-random $\rm{U}(1)$-symmetric gate. The measurement outcomes have probability the teacher's system,
\begin{align}
  p_*(y, s' \vert U) = \tr[(Q_{*, y} \otimes P_{s'}) U(\rho^\mathrm{L}_{*, t} \otimes \rho^\mathrm{R}_{*, t})U^\dag (Q_{*, y} \otimes P_{s'})^\dag].
\end{align}
Here we recall that we will freely change notation between bits and Pauli-$z$ eigenvalues as $s = 2b -1$. We will assume that $\rho^\rm{L/R}_{*, t}$ has the same most likely qubit as that of the student, a sensible assumption in the inferrable phase. We would like to find the linearised evolution of $z_t$, and thereby construct the evolution for the generating function $G_t(x) = \E[\exp(-e^{-x} z_t)]$. Here we must keep track of both $(z_t, b)$ and consider all the cases $(b_\mathrm{L}, b_\mathrm{R}, b') = (0,0,0), (0,0,1), \dots, (1, 1, 1)$, and appropriately weight them by their probabilities. Here, $b_\rm{L/R}$ is the most likely qubit on the density matrices at timestep $t$ connected on the left/right site. 

There are three kinds of cases. The first is $b_\mathrm{L} = b_\mathrm{R} = b'$. In this case we have
\begin{align}
  z_{t+1} = e^{-{2 s' y \epsilon}} \left(\abs{u}^2 z^\rm{L}_t + \abs{u'}^2 z^\rm{R}_t\right),
\end{align}
where $u$ and $u'$ are elements from different columns of the Haar unitary on the $\ket{01}, \ket{10}$ sector. This occurs with probability up to zeroth order in $z_*$ as
\begin{align}
  p_*(y, s'=s_{\rm{L/R}} \vert U, s_\rm{L}=s_\rm{R}) = \frac{1}{\sqrt{2 \pi}} \exp \left( - \frac{(y-s'\epsilon_*)^2}{2} \right).
\end{align}
We neglect the case $b_\rm{L} = b_\rm{R} \neq b'$ as it occurs with probability $p_*(y, b' \vert U) \sim O(z_*)$. For the $b_\rm{L} \neq b_\rm{R}$ case, we find
\begin{align}
  z_{t+1} = \frac1{\abs{u}^2} e^{{2s'y\epsilon}} z_t^{\rm{L/R}},
\end{align}
occurring with probability up to zeroth order in $z_*$ as
\begin{align}
p_*(y, s' \vert U, s_\rm{L} \neq s_\rm{R}) = \frac{\abs{u}^2}{\sqrt{2 \pi}} \exp\left(-\frac{(y + s' \epsilon_*)^2}{2}\right).
\end{align}
Then the generating function is given by
\begin{align}
  G_{t+1}(x) & = \int_{U, y, z^{\rm{L}:\rm{R}}_t} \sum_{s_{\rm{L}:\rm{R}}, s'} \exp\left(-e^{-x} z_{t+1}(s_{\rm{L}:\rm{R}}, s')\right) \nonumber \\
  & \times p_*(y, s', U \vert s_{\rm{L}:\rm{R}}, z_{\rm{L}:\rm{R}}) p(s_\rm{L}, z^\rm{L}_t) p(s_\rm{R}, z^\rm{R}_t).
\end{align}
Substituting the linearised evolution and their probability for the two different cases and using the asymptotic form $G^{(\lambda)}(u) \approx 1 - e^{-\lambda u}$, we find the velocity
\begin{align}
  v_{\epsilon, \epsilon_*}(\lambda) = {2\epsilon(\lambda\epsilon- \epsilon_*)} + \frac{1}{\lambda}\ln \langle \abs{u}^{2 \lambda} + \abs{u}^{2-2\lambda} \rangle_u.
\end{align}
For a particular $(\epsilon, \epsilon_*)$, the velocity is given by $\lambda \in [0, 1]$ that minimises the velocity given $\lambda$. The phase transition occurs when $v_{\epsilon, \epsilon_*}(\lambda_\rm{c})= 0$. Using that $\langle \abs{u}^{2 n}\rangle = 1/(1+n)$, the exact phase boundary is then given by
\begin{align} \label{eq:d=1-xor-boundary}
  \epsilon_*^\mathrm{c}(\epsilon) = \min_{\lambda \in [0, 1]} \left[ \lambda \epsilon + \frac{1}{2 \lambda \epsilon} \ln \left(\frac{3}{2 + \lambda - \lambda^2}\right) \right],
\end{align}
while $\lambda_\rm{c}$ is given by the $\rm{argmin}$.
In general, this requires a solution to a trancendental equation. It is plotted in \cref{fig:planted-xor-phase-diagram}(b). However, in the Bayes optimal case $\epsilon_* = \epsilon$, \cref{eq:d=1-xor-boundary} can be found in closed form,
\begin{align}
  \epsilon^\mathrm{c}_*(\epsilon=\epsilon_*) = \sqrt{2 \ln \left(\frac{4}{3}\right)} \approx 0.758528,
\end{align}
with $\lambda_\rm{c}=1/2$, as well as in the limit of infinite student signal,
\begin{align}
  \epsilon^\mathrm{c}_*(\epsilon=\infty) = \frac{3}{2} \sqrt{\ln \left(\frac{3}{2}\right)} \approx 0.955142,
\end{align}
where $\lambda_\rm{c} \rightarrow 0$. Lastly, in the $\epsilon \rightarrow 0$ limit, $\lambda_\rm{c} \approx 0.916732$ \footnote{This is the solution to $(1-2 \lambda)\lambda + (2+\lambda-\lambda^2)\ln(3/[2 + \lambda - \lambda^2])$, which comes from minimising \cref{eq:d=1-xor-boundary} whilst ignoring the first term in the RHS.} and the critical teacher signal strength diverges as $\epsilon_*^\rm{c} \approx 0.200717 \epsilon^{-1}$. 

Similarly to the classical planted XOR case, we can deduce that in the regime where $\epsilon > 0$, $\epsilon_* > \epsilon_*^\mathrm{c}$, for a typical trajectory, the student's density matrix collapses onto one of the qubit and correctly deduces the final qubit. Therefore all of $\mathrm{MSE}$, $\mathrm{MSEM}$ and $\delta Z^2_{\rm{s}}$ would decay exponentially with time and tend to zero as $t \rightarrow \infty$. On the other hand, for $\epsilon > 0$, $\epsilon_* < \epsilon_*^\mathrm{c}$, we would expect to converge to a non-zero value.

The phase diagram for $d=1$ has two notable features. First, at fixed $\epsilon$, the SNR threshold is lower than that of $d=\infty$. Recall that the $d=\infty$ case also corresponds to the inference problem where $U$ is not provided (for any $d$), whereas in the $d=1$ case considered here, it is provided. Therefore, the lowering of the threshold is expected, as the student has more information. Secondly, the phase diagram features a re-entrance. This is qualitatively similar to the re-entrance seen in the QEC/2D RBIM, see \cref{fig:2d-rbim-phase-diagram}, but found exactly in our case.

\section{Discussion of the phase diagrams} \label{sec:discussion}
In this section we discuss the obtained phase diagrams. First, we discuss their general features followed by a rigorous argument as to why the phase boundary cannot curve downwards after hitting the Bayes optimal line. Then, focusing on quantum inference problems, we discuss the operational consequences of these phase diagrams.

In \cref{sec:bayes-optimality}, we introduced order parameters that pertain to inference (or `learnability'), such as $\rm{MSEM}$ and $\rm{MSE}/2$, and an order parameter related to the student's uncertainty (or `sharpness') in their posterior, $\delta O^2_\rm{s}$. Though these order parameters are equal at Bayes optimality, they need not be away from it. Therefore, there could be one phase boundary between observable sharp / fuzzy phases, and another between inference possible / impossible phases. However, in the three examples of (quantum) inference problems studied, they coincided, i.e. the sharp phase corresponded to the inference possible phase, and the fuzzy phase corresponded to the inference impossible phase. Recall from \cref{sec:2d-rbim} that in the RBIM case, quality of inference is quantified by FM order, whilst sharpness of the student's posterior is quantified by EA-like order parameters. The SG phase is characterised by the lack of FM order but the presence of EA order. Therefore, the SG phase can be intuitively thought of as a phase where the student is confident about their \emph{wrong} answer. In the language of spin glasses, the fact that the phases of inference coincide with those of sharpness means that there is no SG phase. Note that there are examples of inference models where SG-like phases exist, such as the planted RBIM on random graphs \cite{zdeborova2016statistical}, or the planted directed polymer on the tree \cite{pkim2025planted}.

The phase boundaries in these examples delineate when the student can infer the state despite the non-optimality/inaccuracy of their measurement model. Note the similarity of these phase diagrams (cf. \cref{fig:2d-rbim-phase-diagram,fig:planted-ssep-phase-diagram,fig:planted-xor-phase-diagram}). In all three cases, the studied parameters quantify the amount of true signal strength (quantified by $\epsilon_*$ or $\beta_*$) or assumed signal strength (quantified by $\epsilon$ or $\beta$) passed through the measurements. Below the Bayes optimal line, we find two scenarios: either the boundary plateaus (as for the classical planted XOR model or the planted directed polymer on a tree \cite{pkim2025planted}) or it curves slightly upwards to display a re-entrance (as shown numerically for the 2D RBIM or exactly for our $d=1$ quantum planted XOR model). 

That the phase boundary for the MSEM cannot curve down has the status of a theorem. It follows from the Bayes optimal estimator theorem \cref{eq:optimal-estimator} that for a given $\epsilon_*$ the MSEM is minimised at Bayes optimality ($\epsilon = \epsilon_*$). Thus if a given Bayes optimal point $(\epsilon=\epsilon_*,\epsilon_*)$ is in the inference impossible phase, then so are all points on the horizontal line passing through it. A corollary is that the phase boundary cannot pass through values of $\epsilon_*$ below the critical value $\epsilon_\rm{BO}^\#$ on the Bayes optimal line.

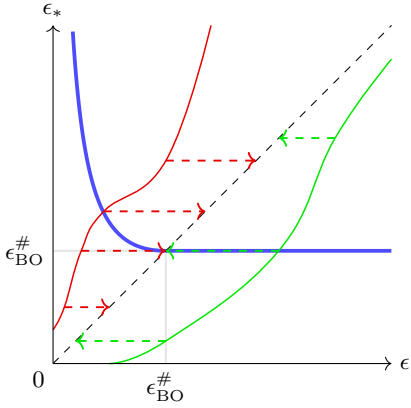
\begin{figure}[t]
  \center 
  \begin{tikzpicture}[scale=1.5, samples=100]
    \draw[smooth, domain = 0:1, color=black!10!white, thick] plot (1,\x);
    \draw[smooth, domain = 0:1, color=black!10!white, thick] plot (\x,1);

    \draw[smooth, domain = 0.175:1, color=white!30!blue, line width=0.5mm] plot (\x,{1/\x/2 + \x/2});
    \draw[smooth, domain = 1:3, color=white!30!blue, line width=0.5mm] plot (\x,{1});

    \draw[-] (1, 0) node[below] {$\epsilon^\#_\mathrm{BO}$};
    \draw[-] (0, 1) node[left] {$\epsilon^\#_\mathrm{BO}$};
    
    \draw[dashed] (0, 0) -- (3, 3);
    
    \draw[black!10!green, line width=0.2mm] plot [smooth, tension=0.75] coordinates {(0.5, 0) (1, 0.2) (2, 1) (2.5, 2.0) (3, 2.7)};
    \draw[->, dashed, black!10!green, line width=0.25mm] (1, 0.2) -- (0.2, 0.2);
    \draw[->, dashed, black!10!green, line width=0.25mm] (2, 1) -- (1, 1);
    \draw[->, dashed, black!10!green, line width=0.25mm] (2.5, 2.0) -- (2.0, 2.0);
    
    \draw[black!10!red, line width=0.2mm] plot [smooth, tension=0.75] coordinates {(0, 0.3) (0.1, 0.5) (0.25, 1) (0.45, 1.35) (1, 1.8) (1.4, 3)};
    \draw[->, dashed, black!10!red, line width=0.25mm] (0.1, 0.5) -- (0.5, 0.5);
    \draw[->, dashed, black!10!red, line width=0.25mm] (0.25, 1) -- (1, 1);
    \draw[->, dashed, black!10!red, line width=0.25mm] (0.45, 1.35) -- (1.35, 1.35);
    \draw[->, dashed, black!10!red, line width=0.25mm] (1, 1.8) -- (1.8, 1.8);

    \draw[->] (0, 0) -- (3, 0) node[right] {$\epsilon$};
    \draw[->] (0, 0) -- (0, 3) node[above] {$\epsilon_*$};
    \draw[-] (0, 0) node[below left] {$0$};
  \end{tikzpicture}
  \caption{Scenario where $\epsilon_*$ vary together with $\epsilon$ for a typical phase diagram of a quantum inference problem (red and green solid lines). Here we approximate that the phase boundary (blue line) flattens at the Bayes optimal transition point with no re-entrance. The phase that an experimentalist detects using their classical decoder is given by the phase on the red or green lines. The true state of the prepared state is given by the phase on the diagonal black dashed line horizontal to the point on the red or green lines, as connected by the red and green dashed lines. When measurement strength is overestimated (green line, $r_\epsilon < 1$), the detected phase on the decoder is the same as that of the prepared state. When the measurement strength is underestimated (red line, $r_\epsilon > 1$), as the measurement strength is increased, the phase transition is detected on the decoder \emph{after} the real one occurs.}
  \label{fig:quantum-non-optimal}
\end{figure}

Although not discussed earlier, other problems on quantum systems can also be viewed through the lens of the general quantum inference problem, such as quantum state preparation \cite{zhu2023nishimori} or quantum state tomography \cite{christandl2012reliable}. In the former case, an experimentalist wants to know whether it is possible to infer the correct operation to prepare the desired quantum state given measurements. The same teacher-student framework can be used to describe the underlying process generating the measurements and the experimentalists' measurement model.

What are the operational consequences of these phase diagrams? Remaining within the quantum state preparation setting, let us consider a situation where $\epsilon$ and $\epsilon^*$ vary together as the assumed signal strength is varied, meaning that we will move along a line $\epsilon^*(\epsilon)$ in a typical inference phase diagram like Fig. \ref{fig:quantum-non-optimal}. An example of such situation may be one where you (an experimentalist) prepare and evolve the quantum circuit with parameter $\epsilon$, and receive measurement outcomes $Y$. Then, you would assume that you prepared the state $\rho_t(\epsilon, Y)$. However, due to some systemic and/or random errors, the parameters are actually $\epsilon^*$ and in fact the state you prepared is $\rho_t(\epsilon^*, Y)$. Nevertheless, you may hope that $\epsilon$ and $\epsilon^*$ are in some way correlated, and in the simplest scenario $\epsilon^*$ is a simple function of $\epsilon$. Suppose then that you repeat the experiment for a range of $\epsilon$'s. You receive measurement outcomes from the system with parameter $\epsilon^*(\epsilon)$. Using your classical decoder, you detect a phase transition at $\epsilon^\#$.

Let us consider the case when $\epsilon^*(\epsilon)$ is monotonic, and $r_\epsilon < 1$, which corresponds to overestimating the measurement strength/rate in the quantum circuit. For simplicity, let us also approximate that the phase boundary flattens at the Bayes optimal transition point with no re-entrance. As shown in Fig. \ref{fig:quantum-non-optimal}, the relationship between the phase of the assumed and the true quantum states is given by drawing a horizontal line from the assumed point to the $45^\circ$ Bayes optimal line. In this case, we see that when the the classical decoder detects the phase transition at $\epsilon^\#$, the true quantum state is also prepared in the critical state. In fact, the phase detected by the classical decoder always coincides with that of the true quantum state.

Instead, assuming all of the above but for $r_\epsilon > 1$, which corresponds to underestimating the measurement strength, as $\epsilon$ is increased, the real quantum state undergoes the phase transition into the charge-sharp phase before the classical decoder detects it. When the classical decoder detects a phase transition, the real quantum state is already deep in the charge-sharp phase.

This would suggest that if the phase boundary is horizontal to the right of the Bayes optimal transition point with no re-entrance, then you should always overestimate the measurement strength if you want to prepare the true quantum state in the correct phase. Even if the phase boundary is not strictly horizontal, it is better to overestimate the measurement strength than underestimate it, because the curvature of the phase boundary means you have a better chance of preparing the quantum state in the intended phase. 

If the precise numerical value of the phase transition at Bayes optimality is already known (ex. through simulations), then you would see the phase transition detected by the classical decoder at the wrong value of $\epsilon$. This would provide the systematic error between the assumed parameters and true parameters. Lastly, in the special case of the planted SSEP, within the charge-fuzzy phase, the average sign of the disconnected correlator \cref{eq:planted-ssep-disconnected-correlator} could also be used to detect Bayes non-optimality (\cref{fig:signs}).

\section{Outlook}
In this work, we introduced \emph{the general quantum inference problem}, a generalisation of classical Bayesian inference to quantum systems. We also defined scenarios where the quantum inference model has the structure of quantum hidden Markov models (qHMMs), which generalised hidden Markov models (HMMs) in classical Bayesian inference. We discussed the requirements for qHMMs to reduce to classical HMMs and proved such cases for qHMMs with Haar-random unitary gates and measurements. Using the teacher-student scenario, we introduced the notion of Bayes non-optimality, where the model assumed by the student differs from the true model of the teacher. This allows us to go beyond existing work and expand the phase diagram in the student-teacher parameters. Under this lens we studied three explicit examples: 1) quantum error correction on the toric and repetition code under bit-flip errors, which map onto the 2D random-bond Ising model, 2) Haar-random $\rm{U}(1)$-symmetric monitored quantum circuit or the planted SSEP, and 3) Haar-random $\rm{U}(1)$-symmetric monitored quantum tree or the planted XOR. Then, we discussed the obtained phase diagrams, with a rigorous argument on their shape and their interpretation in quantum inference problems. 

Many avenues for future research exist. One could reverse the quantum to classical inference problem mapping discussed \cref{sec:haar-case} to study ``quantum'' versions of classical inference problems. One could explore the more general Bayes non-optimality where even the functional form of the student and teacher's models differ. This was done for the planted directed polymer on a tree, where the teacher could have any arbitrary noise model while the student assumes a Gaussian one \cite{pkim2025planted}. It is also a common case for many QEC problems where approximate decoders are used \cite{deMartiiOlius2024decodingalgorithms}. One could also explore non-Markovian quantum inference problems, such as non-Markovian errors in QEC problems \cite{kam2024detrimental}.

\section*{Acknowledgments}
S.W.P.K. acknowledge support from EPSRC DTP International Studentship Grant Ref. No. EP/W524475/1. C.vK. is supported by a UKRI FLF
MR/Z000297/1. A.L. acknowledge support from EPSRC Critical Mass Grant Ref. No. EP/V062654/1.

\bibliographystyle{apsrev4-2} 
\bibliography{main} 

\clearpage
\onecolumngrid

\appendix

\section{Derivation of the replica field theory for the planted SSEP} \label{apdx:replica-field-theory}
\subsection{Setup}
Consider a 1D lattice of Ising variables $\bm{s} = (s_1, \dots, s_L)$, picked from some separable initial distribution $p(\bm{s}) = \prod_x p(s_x)$, with $p(s_x) = \uppi \delta_{s_x, 1} + (1-\uppi) \delta_{s_x, -1}$. 

Let it evolve through charge-conserving Markovian dynamics, $p(\bm{s}_{t} \vert \bm{s}_{t-1}) = \prod_{x \in \mathcal{R}(t)} p(s_{x:x+1, t} \vert s_{x:x+1, t-1})$, where $\mathcal{R}(t)$ are even/odd sites for even/odd $t$, and the transition are given by
\begin{align}
  p(s_{x:x+1, t} \vert s_{x:x+1, t-1}) = \mel*{s_{x:x+1, t}}{u_{x:x+1}}{s_{x:x+1, t-1}},
\end{align}
where
\begin{align}
  u_{x:x+1} := \left( \frac{1-\sigma^z_x \sigma^z_{x+1}}{2} \right) \left(\frac{1}{2} \left(\sigma_x^+\sigma^-_{x+1} + \sigma^-_{x} \sigma^+_{x+1}\right) + \frac{1}{2} \right) + \frac{1+\sigma^z_x \sigma^z_{x+1}}{2}.
\end{align}

Let us denote the entire trajectory as $S := \bm{s}_{1:t}$. Then we have $p(S) = p(\bm{s}_t \vert \bm{s}_{t-1}) \cdts p(\bm{s}_2 \vert \bm{s}_1) p(\bm{s}_1)$.

Given the dynamics, we generate images over all times $Y := \bm{y}_{1:t}$, where each image consists of pixels $\bm{y}_t := (y_{1, t}, \dots, y_{L, t})$ from Ising configurations is given by $p(Y \vert S) = \prod_{x,t} p (y_{x,t} \vert s_{x,t})$, where
\begin{align}
  p (y_{x,t} \vert s_{x,t}) =  \frac{1}{\sqrt{2 \pi \sigma^2}} \exp\left( \frac{-(y_{x,t} - \epsilon s_{x,t})^2}{2 \sigma^2}\right) = \mel{s_{x, t}}{P_{x}(y_{x, t})}{s_{x, t}},
\end{align}
where $P_{x}(y_{x,t}) := \frac{1}{\sqrt{2 \pi \sigma^2}} \exp\left( \frac{-(y_{x,t} - \epsilon \sigma^z_{x})^2}{2 \sigma^2}\right)$.

Let the parameters of the evolution process and imaging process be unknown, and denote them via $(\uppi_*, \epsilon_*, \sigma_*)$. Assuming some other set of parameters $(\uppi, \epsilon, \sigma)$, the posterior distribution over $S$ given the images $Y$ is given by
\begin{align}
  p(S \vert Y) = \frac{p(Y \vert S) p(S)}{p(Y)},
\end{align}
where $p(Y) = \sum_{S} p(Y \vert S) p(S)$ is given by the assumed parameters. The distribution of $Y$ and true states $S^*$ is given by $p_*(Y, S) = p_*(Y \vert S^*) p_*(S^*)$ and $p_*(Y) = \sum_{S^*} p_*(Y \vert S^*) p_*(S^*)$, where the star denotes that the true parameters are used.

We have
\begin{align}
  p(Y,S) = p(Y|S)p(S) = \left( \prod_{t=2}^{t} ( \bm{s}_t \rvert P(\bm{y}_t) \bm{u}_t \lvert \bm{s}_{t-1} ) \right) \mel{\bm{s}_1}{P(\bm{y}_1)}{\bm{\uppi}},
\end{align}
where $P(\bm{y}_t) = \bigotimes_x P_x(y_{x, t})$ and $\bm{u}_t = \bigotimes_{x \in \mathcal{R}(t)} u_{x:x+1}$, and $\lvert \bm{\uppi} ) = \bigotimes_x \ket{\uppi}_x$ and $\ket{\uppi}_x = \uppi \ket{1}_x + (1-\uppi) \ket{-1}_x$.

Although we will sometimes keep track of normalisation, we will show now that the normalisation factor \emph{should} not matter in the replica calculation. For normalised probabilities we have
\begin{align}
  1 = \sum_{S^{1:n},S^*} \int dY p(S^{1:n},Y,S^*) = \sum_{S^{1:n},S^*} \int dY \frac{p(S^1, Y) \cdts p(S^n, Y)}{p(Y)^n} p_*(S^*, Y).
\end{align}
Let's consider unnormalised distributions $q(S, Y) := c \cdot p(S, Y)$ and $q_*(S^*, Y) := c^* \cdot p_*(S^*, Y)$, and define
\begin{align}
  \mathcal{Z} = \sum_{S^{1:n},S^*} \int dY \frac{q(S^1, Y) \cdts q(S^n, Y)}{q(Y)^n} q_*(S^*, Y) = c^*.
\end{align}
Then any mixed moment observable can be written as
\begin{align}
  O & = \frac{1}{\mathcal{Z}} \sum_{S^{1:n},S^*} \int dY O(S^{1:n}, S^*) \frac{q(S^1, Y) \cdts q(S^n, Y)}{q(Y)^n} q_*(S^*, Y) \nonumber \\
  & = \lim_{Q \rightarrow 1} \frac{1}{\mathcal{Z}_Q} \sum_{S^{1:Q}} O(S^{1:Q}) \int dY q(S^1, Y) \cdts q(S^{Q-1}, Y) q_*(S^Q, Y),
\end{align}
where $Q=nk+1$, with the $Q$th replica index being the teacher index. Throughout we will use $Q \leftrightarrow Q$ interchangably, and $\mathcal{Z}_Q = \sum_{S^{1:Q}} \int dY q(S^1, Y) \cdts q(S^{Q-1}, Y) q_*(S^Q, Y)$. 

For example student's variance or the `precision' on come observable $C_t$ can be written as
\begin{align}
  \delta C^2 & 
  = \E_{Y \sim p^*(Y)} \left[\V_{S \sim p(\cdot \vert Y)} [C(\bm{s}_t)] \right] 
  = \int dY p^*(Y) \sum_{S, S', S^*} \left( C_t C_t - C_t C_t'\right) \frac{p(S, Y)p(S', Y)}{p(Y)^2} \\
  & = \lim_{Q \rightarrow 1} \frac{1}{\mathcal{Z}_Q} \sum_{S^{1:Q}} \left( C^1_t C^1_t - C^1_t C^2_t\right) \int dY q(S^1, Y) \cdts q(S^{Q-1}, Y) q_*(S^Q, Y).
\end{align}

The mean-squared-error or the `accuracy', averaged over the joint distribution can also be written in above replica form as
\begin{align}
  \mathrm{MSE}_\mathrm{B} & = \E_{S,Y,S^* \sim p(S,Y,S^*)} \left[ (C_t - C^*_t)^2 \right] = \sum_S \int dY \sum_{S^*} p(S, Y, S^*) (C(\bm{s}_t) - C(\bm{s}^*_t))^2 \\
  & = \lim_{Q \rightarrow 1} \frac{1}{\mathcal{Z}_Q} \sum_{S^{1:Q}} \left( C^1_t C^1_t - 2 C^1_t C^Q_t + C^Q C^Q \right) \int dY q(S^1, Y) \cdts q(S^{Q-1}, Y) q_*(S^*, Y).
\end{align}

\subsection{Continuum limit in time}
We can write (unnormalised) expectation values of observables as
\begin{align}
  \sum_S C_t \cdot p(S, Y) = \langle \tilde{\Plus} \rvert C_t T_{1:t}(Y) \lvert \bm{\uppi} \rangle
\end{align}
and the `evidence' as
\begin{align}
  p(Y) = \langle \tilde{\Plus} \rvert T_{1:t}(Y) \lvert \bm{\uppi} \rangle,
\end{align}
where $T_{1:t}(Y) := T_t(\bm{y}_t) \cdts T_1(\bm{y}_1)$,
\begin{align}
  T_{t}(\bm{y}_t) = \begin{cases}
    P(\bm{y}_t) \bm{u}_t & t > 1, \\
    P(\bm{y}_t) & t = 1,
  \end{cases}
\end{align}
$\lvert \tilde{\Plus} \rangle := \bigotimes_x \lvert \tilde{+} \rangle_x$, $\lvert \tilde{+} \rangle_x = \lvert -1 \rangle_x + \lvert 1 \rangle_x$.

To go to the continuum limit, we slightly modify the 6-vertex transition kernel to the following:
\begin{align}
  u_{x:x+1} & \rightarrow \left( \frac{1-\sigma^z_x \sigma^z_{x+1}}{2} \right) \Big(4 J \mathsf{b} \left(\sigma_x^+\sigma^-_{x+1} + \sigma^-_{x} \sigma^+_{x+1}\right) + \left(1 - 4J \mathsf{b} \right) \Big) + \frac{1+\sigma^z_x \sigma^z_{x+1}}{2} \\
  & \approx \exp\left(2J [\vec{\sigma}_x \cdot \vec{\sigma}_{x+1} - 2] \mathsf{b} \right),
\end{align}
where $\mathsf{b}$ is the temporal lattice spacing. The above corresponds to setting a rate to the exchange.
We can perform integration over each measurement $y_{x, t}$. We consider
\begin{align} \label{eq:natural}
  \overline{P} & = \int dy P^1(y) \cdts P^Q(y) \\
  & = \int dy \frac{1}{(2 \pi \sigma^2)^{(Q-1)/2}} \frac{1}{(2 \pi \sigma^2_*)^{1/2}} \exp \left\{ - \sum_{a} \frac{1}{2 \sigma^2_a} (y - \epsilon^a \hat{Z}^a)^2 \right\} \\
 & = \int dy \frac{1}{(2 \pi \sigma^2)^{(Q-1)/2}} \frac{1}{(2 \pi \sigma^2_*)^{1/2}} \exp \left\{ - \sum_{a} \frac{\sum_c \sigma_c^{-2}}{2} y + \sum_{a} \frac{\epsilon_a}{\sigma_a^2} \hat{Z}^a - \frac{1}{2} \sum_{a} \frac{\epsilon^2_a}{\sigma_a^2} (\hat{Z}^a)^2 \right\} \\
 & = \int dy \frac{1}{(2 \pi \sigma^2)^{(Q-1)/2}} \left(\frac{\sigma_*^{-2}}{\sum_c \sigma_c^{-2}}\right)^{1/2} \exp \left\{ \frac{1}{2\sum_c \sigma_c^{-2}} \left( \sum_{a} \frac{\epsilon_a}{\sigma_a^2} \hat{Z}^a\right)^2 - \frac{1}{2} \sum_{a} \frac{\epsilon^2_a}{\sigma_a^2} (\hat{Z}^a)^2 \right\}.
\end{align}
Recall that we were allowed to replace the students' distributions with unnormalised ones. Since 
\begin{align}
  \exp\left\{ -(y - \epsilon \hat{Z})^2/2\sigma^2\right\} = \exp\left\{ -y^2/2\sigma^2 + y \epsilon \hat{Z}/\sigma^2 - \epsilon^2/2\sigma^2 \right\},
\end{align}
and only $\exp\left\{ y \epsilon \hat{Z}/\sigma^2\right\}$ depends on the state $\hat{Z}$, we could have ignored all factors independent of $\hat{Z}$ and just kept $\exp\left\{ y \epsilon \hat{Z} / \sigma^2 \right\}$, for example. In fact we could have had any factor of form $\exp \left\{ - \alpha y^2/2 + y \epsilon \hat{Z} / \sigma^2 - \beta/2 \right\}$ for each student replica. On the teacher replica we can have add any constant terms, so $\exp\left\{ -y^2/2\sigma_*^2 + y \epsilon_* \hat{Z}^Q /\sigma_* - \gamma/2 \right\}$. 

In general, we are free to substitute
\begin{align}
  \overline{P} & \rightarrow
   \int dy \left(\frac{\sigma_*^{-2}}{(Q-1)\alpha + \sigma^{-2}_*}\right)^{1/2} \exp \left\{ \frac{1}{2([Q-1]\alpha + \sigma_*^{-2})} \left( \sum_{a} \frac{\epsilon_a}{\sigma_a^2} \hat{Z}^a\right)^2 - \frac{1}{2} \left(\sum_{a=1}^{Q-1} \beta (\hat{Z}^a)^2 + \gamma (\hat{Z}^Q)^2 \right) \right\} \\
  & = \int dy \left(\frac{\sigma_*^{-2}}{(Q-1)\alpha + \sigma^{-2}_*}\right)^{1/2} \exp \left\{ - \frac{1}{2} \sum_{ab} M_{ab} \hat{Z}^a \hat{Z}^b \right\}, 
\end{align}
where
\begin{align}
  M = \begin{pmatrix}
    \beta & 0 & 0 & 0 \\
    0 & \ddots & 0 & 0 \\
    0 & 0 & \beta & 0 \\
    0 & 0 & 0 & \gamma
  \end{pmatrix}
  - \frac{1}{\sigma_*^{-2} + (Q-1)\alpha} \begin{pmatrix}
    \epsilon/\sigma^2 \\
    \vdots  \\
    \epsilon/\sigma^2 \\
    \epsilon_*/\sigma_*^2
  \end{pmatrix}
  \begin{pmatrix}
    \epsilon/\sigma^2 &
    \hdots &
    \epsilon/\sigma^2 &
    \epsilon_*/\sigma_*^2
  \end{pmatrix}.
\end{align}
The natural choice of $\alpha = \sigma^{-2}$ and $\beta = \epsilon^2 / \sigma^2$, $\gamma=\epsilon_*^2 / \sigma_*^2$ retrieves the result above. The choice of $\beta = \epsilon _*^2/\sigma _*^2$ and $\alpha = \sigma _*^2 \epsilon ^2 / \sigma ^4 \epsilon _*^2$, $\gamma = \epsilon_*^2/\sigma_*^2$ turns $M$ proportional to a projector.

However, we will choose the `natural' choice of $\alpha = \sigma^{-2}$ and $\beta = \epsilon^2 / \sigma^2$, $\gamma=\epsilon_*^2 / \sigma_*^2$ as \eqref{eq:natural}. Above we used that $\hat{Z} \hat{Z} = 1$ in order to argue that we can add any constant terms. However, below we will use the spin-coherent path integral formalism, which is valid for large spin, where $\hat{S}^z \hat{S}^z \not\propto 1$. Considering the model for general spin then interpolating to spin $1/2$ corresponds to fixing $\beta$ and $\gamma$ as above. Secondly, if we are studying the model near the Bayes optimal line, we must respect the symmetry between the student and teacher replica at the Bayes optimal point. Choosing a different $\alpha$ corresponds to breaking symmetry and altering original measurement operator on the student replicas, which also introduces additional $\hat{Z} \hat{Z}$ terms.

We renormalise $M_{ab} \rightarrow \mathsf{b} M_{ab}$, where $\mathsf{b}$ is the temporal lattice spacing. This can be also seen as renormalising the noise and signal $\sigma^2 \rightarrow \mathsf{b} \sigma^2$ and $\epsilon \rightarrow \mathsf{b} \epsilon$.

For small timesteps all operators commute so we can combine the even and odd gates into gates over all sites $\tilde{U}_t$ with half strength as 
\begin{align}
  \overline{P} U_t^{\otimes Q} \overline{P} U_{t-1}^{\otimes Q}
  & \propto \exp\left\{ -\mathsf{b} \left[ -\sum_{A=X,Y,Z} \sum_{x,a} 2J \hat{A}_x^a \cdot \hat{A}_{x+1}^a + \sum_{xab} M_{ab} \hat{Z}^a_x \hat{Z}^b_x \right]\right\} \\
  & \approx \prod_{t' = t, t-1} \exp\left\{ -\mathsf{b} \left[ - J \sum_{A=X,Y,Z}\sum_{x,a} \hat{A}^a_x \hat{A}^a_{x+1} + \frac{1}{2} \sum_{xab} M_{ab} \hat{Z}^a_x \hat{Z}^b_x \right]\right\} \\
  & \approx \overline{P} \tilde{U}_t \overline{P} \tilde{U}_{t-1}.
\end{align}
We define continuous time $\tilde{t} = \mathsf{b} t$. We send $t_\mathrm{f} \rightarrow \infty$ and $\mathsf{b} \rightarrow 0$ whilst keeping $\tilde{t}_\mathrm{f} = t_\mathrm{f} / \mathsf{b}$ constant. Doing so then removing the tildes,
\begin{align}
  \int dY \bigotimes_{a=1}^{Q}T^{(a)}_{1:t_\mathrm{f}} \propto \exp\left( -\int_0^{{t}_\mathrm{f}} d\mathsf{t} H_Q \right),
\end{align}
where
\begin{align}
  H_Q = -J \sum_{A=X,Y,Z}\sum_{ar} \hat{A}^a_x \hat{A}^a_{x+1} + \frac{1}{2} \sum_{abr} \hat{Z}^a_x M_{ab} \hat{Z}^b_x.
\end{align}
As we are interested in boundary observables at $t_\mathrm{f}$, we will shift time, such that starting time is at $-t_\mathrm{f}$ and end time at $0$. Therefore the partition function can be written as
\begin{align}
  \mathcal{Z}_Q = \langle \Hash \rvert \exp\left(-\int_{-{t}_\mathrm{f}}^0 d\mathsf{t} H_Q \right) \lvert \bm{\Uppi} \rangle,
\end{align}
and variance of an observable can be written as
\begin{align}
  \delta C^2
  & = \lim_{Q \rightarrow 1} \frac{1}{\mathcal{Z}_Q} \langle \Hash \rvert (C^1_0 C^1_0 - C^1_0 C^2_0) \exp\left(-\int_{-{t}_\mathrm{f}}^0 dt H_Q \right) \lvert \bm{\Uppi} \rangle,
\end{align}
where superscript above $C$ denotes the replica index, $\lvert {\Hash} \rangle := \lvert {\Plus} \rangle^{\otimes Q}$, and $\lvert \bm{\Uppi} \rangle := \lvert \bm{\uppi} \rangle^{\otimes Q}$.
Similarly, the mean-squared error can be written as
\begin{align}
  \mathrm{MSE}
  = \lim_{Q \rightarrow 1} \frac{1}{\mathcal{Z}_Q} \langle \Hash \rvert (C^1_0 C^1_0 - 2 C^1_0 C^*_0 + C^*_0 C^*_0) \exp\left(-\int_{-{t}_\mathrm{f}}^0 dt H_Q \right) \lvert \bm{\Uppi} \rangle.
\end{align}

\subsection{Coherent state path integral}
Doing the coherent state path integral involves replacing $(\hat X, \hat Y, \hat Z) \rightarrow (\sin \theta \cos \phi, \sin \theta \cos \phi, \cos \theta)$ in the Hamiltonian and adding the `kinetic' term $\frac{i}{2 \mathsf{a}} \sum_a \theta^a \partial_t \phi^a$. For simplicity, we will look at half filling, ($\cos(\theta_0) = 0$, $\sin(\theta_0) = 1$). We define continuous space $\tilde{x} := \mathsf{a} x$, expand $\theta = \theta_0 + \delta \theta$ around $\theta_0 = \pi/2$ and keeping the lowest order terms, then removing the tildes, we have
\begin{align}
  \mathcal{L} = & \frac{i}{2 \mathsf{a}} \sum_{a=1}^Q \delta \theta^a \partial_t {\phi}^a 
  + \frac{J\mathsf{a}}{2} \sum_{a=1}^Q \left( (\partial_x \delta \theta^a)^2 
  + (\partial_x {\phi}^a)^2 \right) 
  + \frac{1}{2\mathsf{a}} \sum_{a,b=1}^Q \delta \theta^a M_{ab} \delta \theta^b.
\end{align}
Then, we have the identification that $\hat{Z} \approx \cos \theta \approx - \delta \theta$. To keep track of the operators, we will add a source term $\sum_a h^a(x, t) \delta \theta^a(x, t)$ to the Lagrangian density and send $h^a \rightarrow 0$. We will not keep track of terms quadratic in $h$.

\subsection{Diagonalising $M$}
We can diagonalise $M$ as follows. For notational convenience, let's denote $K := Q-1$. We can write $M$ as
\begin{align}
  M = \begin{pmatrix}
    \frac{\epsilon^2}{\sigma^2} \mathbb{1}_{K} & \vec{0}_{K} \\
    \vec{0}_{K}^T & \frac{\epsilon_*^2}{\sigma_*^2}
  \end{pmatrix}
  -
  \frac{1}{2(K\sigma^{-2} + \sigma_*^{-2})}
  \begin{pmatrix}
    \frac{\epsilon^2}{\sigma^4} \vec{1}_{K} \vec{1}_{K}^T & \frac{\epsilon \epsilon_*}{\sigma^2 \sigma_*^2} \vec{1}_{K} \\
    \frac{\epsilon \epsilon_*}{\sigma^2 \sigma_*^2} \vec{1}_{K}^T & \frac{\epsilon_*^2}{\sigma_*^4}.
  \end{pmatrix}
\end{align}
On an ansatz of form $(v \; w)^T$, one can see that the solution only depends on $\vec{1}^T_K v$ and $w$. In the case $\vec{1}^T_K v = 0$, ex. non-zero wavevector Fourier modes on the $K$ student modes $\hat{v}_1, \dts \hat{v}_{Q-2}$, one finds that the eigenvalue is $\epsilon^2/\sigma^2$. In the other case, we have a $2 \times 2$ eigenvalue of the following form:
\begin{align}
    \begin{pmatrix}
        \frac{\epsilon ^2}{\sigma ^2}-\frac{K \epsilon ^2}{\sigma ^4 \left(\frac{K}{\sigma ^2}+\frac{1}{\sigma _*^2}\right)} & -\frac{K \epsilon  \epsilon _*}{\left(\sigma ^2 \sigma _*^2\right) \left(\frac{K}{\sigma ^2}+\frac{1}{\sigma _*^2}\right)} \\
        -\frac{\epsilon  \epsilon _*}{\left(\sigma ^2 \sigma _*^2\right) \left(\frac{K}{\sigma ^2}+\frac{1}{\sigma _*^2}\right)} & \frac{\epsilon _*^2}{\sigma _*^2}-\frac{\epsilon _*^2}{\sigma _*^4 \left(\frac{K}{\sigma ^2}+\frac{1}{\sigma _*^2}\right)}
    \end{pmatrix}
  \begin{pmatrix}
    \vec{1}^T_K v \\
    w
  \end{pmatrix}
  = 
  \lambda \begin{pmatrix}
    \vec{1}^T_K v \\
    w
  \end{pmatrix},
\end{align}
where eigenvalues and eigenvectors are given by
\begin{align}
  \lambda_0 = 0, \quad v_0^T = \frac{1}{\sqrt{K \epsilon^{-2} + \epsilon_*^{-2}}} (\epsilon^{-1} \; \cdts \; \epsilon^{-1} \; \epsilon_*^{-1})^T,
\end{align}
\begin{align}
    \lambda_{Q-1} = \frac{K \epsilon _*^2+\epsilon ^2}{K \sigma _*^2+\sigma ^2}, 
    \quad 
    v_{Q-1}^T = \frac{1}{\sqrt{K[\epsilon^2 + K \epsilon_*^2]}} (\epsilon, \; \cdts, \; \epsilon, \; - K \epsilon_*)^T,
\end{align}

So we have $V^T M V = \Lambda$ where $V = [\hat{v}_0 \; \hat{v}_1 \; \cdts \; \hat{v}_{Q-2} \; \hat{v}_{Q-1}]$ such that $\hat V_{a \alpha} = \hat{v}_\alpha(a)$ and 
$$\Lambda = \frac{\epsilon^2}{\sigma^2} \mathrm{diag}\left[0 \; 1 \cdts \; 1 \; \frac{1+(\epsilon_*/\epsilon)^2 K}{1+(\sigma_*/\sigma)^2 K} \right]^T.$$
Let $\delta \theta_\alpha = \sum_{a=1}^Q [\hat V^\dag]_{\alpha a} \delta \theta^a = \hat v^*_\alpha(a) \delta \theta^a$. We similarly define the transformed fields for $\phi$. As $\hat V$ is unitary, the already-diagonal terms stay diagonal. Then the Lagrangian density is (in diagonal basis)
\begin{align}
  \mathcal{L} 
  & =
  \frac{i}{2 \mathsf{a}} \sum_{\alpha=0}^{Q-1} \delta {\theta}_\alpha \partial_t {\phi}_\alpha 
  +
  \frac{J \mathsf{a}}{2} \sum_{\alpha=0}^{Q-1} \left[ (\partial_x \delta {\theta}_\alpha)^2 + (\partial_x {\phi}_\alpha)^2  \right] 
  +
  \frac{\epsilon^2}{2 \mathsf{a} \sigma^2} \left( \sum_{\alpha=1}^{Q-2} \delta \theta_\alpha^2 + \frac{1+(\epsilon_*/\epsilon)^2 K}{1+(\sigma_*/\sigma)^2 K} \delta \theta_{Q-1}^2\right) \\
  & + \sum_{\alpha=0}^{Q-1} h_\alpha \delta \theta_\alpha.
\end{align}
We use lower indices for the $\alpha$ basis and upper indices for $a$ basis throughout. 
Note that as long as we don't attempt to \emph{integrate} over $\phi$, we can write any linear combination of them such as above, as long as we transform back to the original basis when we consider its compactness.

\subsection{Integrating out massive modes of $\theta_\alpha$}
As we are treating $\delta \theta_\alpha$'s to be Gaussian fluctuations we can integrate over them as long as they are massive. Integrating over the $\alpha=1, \dts, Q-1$ modes, we get
\begin{align}
  \mathcal{L} 
  & = \frac{i}{2 \mathsf{a}} \delta {\theta}_0 (\partial_t {\phi}_0) 
  + \frac{\sigma^2}{8 \mathsf{a} \epsilon^2} \left( \sum_{\alpha=1}^{Q-2} (\partial_t {\phi}_\alpha)^2 
  + \frac{1+(\sigma_*/\sigma)^2 K}{1+(\epsilon_*/\epsilon)^2 K}(\partial_t {\phi}_{Q-1})^2 \right)
  + \frac{J\mathsf{a}}{2} \left[ (\partial_x \delta {\theta}_0)^2
  + \sum_{\alpha=0}^{Q-1} (\partial_x{\phi}_\alpha)^2 \right] \\
  & + h_0 \delta \theta_0 - \frac{i \mathsf{a} \sigma^2}{2 \epsilon^2} \left[ \sum_{\alpha=1}^{Q-2} h_\alpha (\partial_t \phi_\alpha) + \frac{1+(\sigma_*/\sigma)^2 K}{1+(\epsilon_*/\epsilon)^2 K} h_{Q-1} (\partial_t \phi_{Q-1}) \right].
\end{align}
We now rescale space and time to make the action as isotropic as possible. We transform $\partial_t \rightarrow \sqrt{v} \partial_t$ and $\partial_x \rightarrow (1/\sqrt{v}) \partial_x$ with $v = 2 \mathsf{a} (\epsilon/\sigma) \sqrt{J}$. Defining $\rho_s = \sqrt{J}/2 (\epsilon/\sigma)$, we have
\begin{align}
  \mathcal{L} 
  & = \frac{i \sqrt{v}}{2 \mathsf{a}} \delta {\theta}_0 (\partial_t {\phi}_0) + \frac{\rho_s}{2} (\partial_x {\phi}_0)^2
  + \frac{\rho_s}{2} (\partial_x \delta {\theta}_0)^2
  + \frac{\rho_s}{2} \sum_{\mu=t, x} \sum_{\alpha=1}^{Q-2} (\partial_\mu {\phi}_\alpha)^2 \nonumber\\
  & + \frac{\rho_s}{2} \frac{1+(\sigma_*/\sigma)^2 K}{1+(\epsilon_*/\epsilon)^2 K}  (\partial_t {\phi}_{Q-1})^2 + \frac{\rho_s}{2} (\partial_x {\phi}_{Q-1})^2 \\
  & + h_0 \delta \theta_0 - \frac{i \mathsf{a} \sigma^2 \sqrt{v}}{2 \epsilon^2} \left[ \sum_{\alpha=1}^{Q-2} h_\alpha (\partial_t \phi_\alpha) + \frac{1+(\sigma_*/\sigma)^2 K}{1+(\epsilon_*/\epsilon)^2 K} h_{Q-1} (\partial_t \phi_{Q-1}) \right].
\end{align}

\subsection{Duality transformation}
In preparation to the duality transformation, we will define $\j^t_0 := (\sqrt{v}/2\mathsf{a})\theta_0$. Defining $D^2 := v / 4 \mathsf{a}^2$, we have
\begin{align}
  \mathcal{L} 
  & = i \, \j^t_0 (\partial_t {\phi}_0) + \frac{\rho_s}{2 D^2} (\partial_x \j^t_0)^2 + \frac{\rho_s}{2} (\partial_x{\phi}_0)^2 \nonumber \\
  & + \frac{\rho_s}{2} \sum_{\mu=t, x} \sum_{\alpha=1}^{Q-1} (\partial_\mu {\phi}_\alpha)^2 
  + \frac{\rho_s}{2} \frac{1+(\sigma_*/\sigma)^2 K}{1+(\epsilon_*/\epsilon)^2 K} (\partial_t{\phi}_{Q-1})^2 + \frac{\rho_s}{2} (\partial_x {\phi}_{Q-1})^2 \\
  & + \frac{2 \mathsf{a}}{\sqrt{v}} h_0 \j^t_0 - \frac{i \mathsf{a} \sigma^2 \sqrt{v}}{2 \epsilon^2} \left[ \sum_{\alpha=1}^{Q-2} h_\alpha (\partial_t \phi_\alpha) + \frac{1+(\sigma_*/\sigma)^2 K}{1+(\epsilon_*/\epsilon)^2 K} h_{Q-1} (\partial_t \phi_{Q-1}) \right].
\end{align}
To disentangle the $(\partial_\mu \phi_\alpha)^2$ terms for $\alpha=1, \dts, Q-1$, we will introduce the Hubbard-Stratonovich currents $\j^\mu_\alpha$. To disentangle the $(\partial_x \phi_0)^2$ term, we introduce $\j^x_0$. Then we get
\begin{align}
  \mathcal{L} 
  & = i \sum_{\mu=t, x} \sum_{\alpha=0}^{Q-1} \j^\mu_\alpha (\partial_\mu \phi_\alpha)
  + \frac{1}{2 \rho_s} \sum_{\mu=t, x} \sum_{\alpha=1}^{Q-2} (\j^\mu_\alpha)^2 \nonumber \\
  & + \frac{1}{2 \rho_s} (\j^x_{Q-1})^2 + \frac{1}{2 \rho_s} \frac{1+(\epsilon_*/\epsilon)^2 K}{1+(\sigma_*/\sigma)^2 K} (\j^t_{Q-1})^2
  + \frac{1}{2 \rho_s} (\j^x_0)^2
  + \frac{D^2}{2 \rho_s} (\partial_x \j^t_0)^2 \\
  & + \frac{2 \mathsf{a}}{\sqrt{v}} h_0 \j^t_0 - \frac{i \mathsf{a} \sigma^2 \sqrt{v}}{2 \epsilon^2} \left[ \sum_{\alpha=1}^{Q-2} h_\alpha (\partial_t \phi_\alpha) + \frac{1+(\sigma_*/\sigma)^2 K}{1+(\epsilon_*/\epsilon)^2 K} h_{Q-1} (\partial_t \phi_{Q-1}) \right].
\end{align}

Now we can separate $\phi^a = \phi^a_\mathrm{s} + \phi^a_\mathrm{v}$ for smooth and vortex parts respectively and therefore $\phi_\alpha = \phi_\alpha^\mathrm{s} + \phi_\alpha^\mathrm{v}$, being careful that the vortex part is defined in the original $a$ basis. Integrating by parts to move the derivative to $\j^a_\mu$, then integrating over $\phi^a_\mathrm{s}$, we impose divergence-free condition on the current, $\sum_\mu \partial_\mu \j^a_\mu = 0 \forall a$ without the source term.

With the source term, we can use the ansatz
\begin{align}
  \j^t_\alpha = \frac{1}{2 \pi} \partial_x \chi_\alpha + \begin{cases}
    0 & \alpha = 0 \\
    \frac{\mathsf{a} \sigma^2 \sqrt{v}}{2 \epsilon^2} h_\alpha & \alpha = 1, \dts, Q-2 \\
    \frac{1 + (\sigma_*/\sigma)^2 K}{1 + (\epsilon_*/\epsilon)^2 K} \frac{\mathsf{a} \sigma^2 \sqrt{v}}{2 \epsilon^2} h_{Q-1} & \alpha = Q-1,
  \end{cases}
\end{align}
and $\j^x_\alpha = (-1/2\pi) \partial_t \chi_\alpha$. Substituting, we find
\begin{align}
  \mathcal{L} 
  & = -i \sum_{\mu=t, x} \sum_{a=1}^{Q} \chi^a \frac{(-1)^\mu}{2 \pi}(\partial_{\bar\mu}\partial_\mu \phi^a_\mathrm{v}) \\
  & + \frac{1}{8 \pi^2 \rho_s} \left[ 
  (\partial_t \chi_0)^2
  + D^2 (\partial_x^2 \chi_0)^2
  + \sum_{\mu=t,x}\sum_{\alpha=1}^{Q-2} (\partial_\mu \chi_\alpha)^2
  + \frac{1+(\epsilon_*/\epsilon)^2 K}{1+(\sigma_*/\sigma)^2 K} (\partial_x \chi_{Q-1})^2
  + (\partial_t \chi_{Q-1})^2 \right] \\
  & + \frac{\mathsf{a}}{\pi \sqrt{v}} \left[ h_0 (\partial_x \chi_0) + \sum_{\alpha=1}^{Q-2} h_\alpha (\partial_x \chi_\alpha) + \left(\frac{1+(\epsilon_*/\epsilon)^2 K}{1+(\sigma_*/\sigma)^2 K}\right)^2 h_{Q-1} \chi_{Q-1} \right].
\end{align}

Noticing that $\sum_\mu \frac{(-1)^\mu}{2 \pi} (\partial_\mu \partial_{\bar \mu} \phi^a_\mathrm{v}) = \rho^a_\mathrm{v}$ is the vortex density for the  $a$th replica, and partially transforming back to the original replica basis $a$,
\begin{align}
  \boxed{
  \begin{aligned}
    \mathcal{L}
    & = -i \sum_{a=1}^{Q} \chi^a \rho^a_{\mathrm{v}}
    + \frac{1}{8 \pi^2 \rho_s} \sum_{\mu=t, x} \sum_{a,b=1}^{Q} (\partial_\mu \chi^a) \Phi_{ab} (\partial_\mu \chi^b) \\
    & + \frac{1}{8 \pi^2 \rho_s} \left[ (\partial_t \chi_0)^2 + D^2 (\partial_x^2 \chi_0)^2\right]
    + \frac{1}{8 \pi^2 \rho_s} \left[(\partial_t \chi_{Q-1})^2 + \frac{1+(\epsilon_*/\epsilon)^2 K}{1+(\sigma_*/\sigma)^2 K} (\partial_x \chi_{Q-1})^2 \right]
  \end{aligned}
  },
\end{align}
where $\chi^a$ are non-compact variables, $\rho_s \sim \sqrt{J}/(\epsilon/\sigma)$, $\rho_0 \sim \epsilon^2$, $\Phi = \mathbb{1}_Q - \hat{v}_0 \hat{v}_0^T - \hat{v}_{Q-1} \hat{v}_{Q-1}^T$, and
\begin{align}
  \hat{v}_0 = \frac{1}{\sqrt{K \epsilon^{-2} + \epsilon_*^{-2}}} \begin{bmatrix} \epsilon^{-1} \\ \vdots \\ \epsilon^{-1} \\ \epsilon^{-1}_* \end{bmatrix} \implies \chi_0 = \frac{1}{\sqrt{K \epsilon^{-2} + \epsilon_*^{-2}}} \sum_{a=1}^{Q} \epsilon_a^{-1} \chi^a,
\end{align}
\begin{align}
  \hat{v}_{Q-1} = \frac{1}{\sqrt{K[\epsilon^2 + K \epsilon_*^2]}} \begin{bmatrix} \epsilon \\ \vdots \\ \epsilon \\ -K \epsilon_* \end{bmatrix} \implies \chi_{Q-1} = \frac{1}{\sqrt{K[\epsilon^2 + K \epsilon_*^2]}} \left( \epsilon \sum_{a=1}^{Q-1} \chi^a - \epsilon_* K \chi^Q \right).
\end{align}

\subsection{Renormalisation group analysis}
We have the partition function
\begin{align}
    \mathcal{Z}_Q = \int D[\chi^a]_{a=1}^{Q} D[\rho^a]_{a=1}^{Q} \exp \left\{- \int_{t, x} \mathcal{L}\right\},
  \end{align}
where $D[\rho^a] = \sum_{N_a=0}^\infty \sum_{i_a=1}^{N_a} \prod_{i_a} \left(\sum_{m_a=-\infty}^\infty \int_{x_{i_a}}\right)$ for each replica index $a$ and $\rho^a_\text{v} = \sum_{i_a} m_{i_a} \delta(r- r_{i_a})$.

We can see that the action is invariant under:
\begin{itemize}
  \item $\vartheta^a \rightarrow \vartheta^a + 2 \pi \forall a$,
  \item Replica permutation, $a \leftrightarrow b \forall a, b \neq Q$, i.e. for student replicas only.
  \item Symmetric under $\chi^a \rightarrow -\chi^a$.
\end{itemize}

We will try the `Landau approach' to determining relevant terms. We will try adding terms to the Lagrangian $\mathcal{L} = \mathcal{L}_\text{G} + \mathcal{V}$ that respects the symmetry of the problem. The minimal such terms are:
\begin{align}
  \mathcal{V} = -\lambda_0 \mathcal{S}_{Q-1} \cos \left( \sum_a n^a \chi^a \right) = -\lambda_0 \frac{1}{(Q-1)!} \sum_{\tau \in S_{Q-1}} \cos \left( \sum_a n^{\tau(a)} \chi^a \right),
\end{align}
with the understanding that $\tau(Q)=Q$, and $\{n^a\}_{a=1}^Q$ are integers.

Then, we will apply renormalisation and see if which terms are relevant under renormalisation group.

We will go through the following recipe.
\begin{enumerate}
  \item Integrate out the fast moving $+$ modes of wavevector $k \in (\Lambda/\zeta, \Lambda)$, where $\Lambda$ is the UV cutoff. Since $S_\text{G} = S^+_\text{G} + S^-_\text{G}$,
  \begin{align}
    Z = \int D[\chi^-] e^{-\int_{tx} \mathcal{L}^-_\text{G}} \left[ \int D[\chi^+] e^{-\int_{tx} (\mathcal{L}^+_\text{G} + V)}\right] = \int D[\chi^-] e^{-\int_{tx} \mathcal{L}^-_\text{G}} \left\langle e^{-\int_{tx} V}\right\rangle_\text{G+} =: \int D[\chi^-] e^{-\int_{tx} (\mathcal{L}^-_\text{G} + \Delta \mathcal{L}^-)},
  \end{align}
  where we equated the expectation value as additional terms in the new Lagrangian.
  \item Rescale $(t, x)$, such that $(t', x') = \frac{1}{\zeta}(t, x)$.
  \item If need be, rescale $\chi$ to keep the quadratic term equal.
\end{enumerate}

The interaction term integrated over the fast modes is
\begin{align}
  \left\langle \exp \left\{ - \lambda \mathcal{S}_{Q-1} \int_{tx} \cos\left(\sum_a n^a \vartheta^a(r)\right) \right\} \right\rangle_\text{G+} 
  & \approx 1 + \frac{\lambda}{2} \mathcal{S}_{Q-1} \int_{tx} \sum_{\sigma=\pm1} e^{i \sigma \sum_a n^a  \vartheta^a_-} \langle e^{i \sigma \sum_a n^a \vartheta^a_+} \rangle_\text{G+}.
\end{align}
Noting that $\vartheta^a_+ = \sum_{\alpha} U_{a\alpha} \vartheta_\alpha^+$ and using that $\langle e^{ib \vartheta_\alpha(r)} \rangle = e^{-\frac{b^2}{2} \langle \vartheta_\alpha(r) \vartheta_\alpha(r)\rangle}$, 
\begin{align}
  \langle e^{i \sigma \sum_a n^a \vartheta^a_+} \rangle_\text{G+} = \prod_\alpha \langle e^{i \sigma \sum_a n^a U_{a\alpha} \vartheta^+_\alpha} \rangle_\text{G+} 
  = \prod_\alpha \left\langle \exp\left\{-\frac{1}{2} \left(\sum_a n^a U_{a\alpha} \right)^2 \langle \vartheta^a_\alpha(r) \vartheta^a_+(r)\rangle \right\} \right\rangle_\text{G+}.
\end{align}

We have
\begin{align}
    \langle \vartheta^+_\alpha (r) \vartheta^+_\alpha \rangle_\text{G+} 
    = \int_{\Lambda/\zeta}^\Lambda d^2 k G_\alpha(k) 
    = \begin{cases}
      4 \pi^2 \rho_s \sqrt{4D} \bar{f} \zeta & \alpha = 0,\\
      2 \pi \rho_s \ln \zeta & \alpha = 1, \dots, Q-2, \\
      \sqrt{\frac{1+K(\sigma_*/\sigma)^2}{1+K(\epsilon_*/\epsilon)^2}} 2 \pi \rho_s \ln \zeta & \alpha = Q-1.
    \end{cases}
\end{align}

Therefore
\begin{align} \label{eq:scaling-factor}
    \langle e^{i \sigma \sum_a n^a \vartheta^a_+} \rangle_\text{G+}
    & = \exp\left\{ -\frac{1}{2} \left(\sum_a n^a V_{a0} \right)^2 4 \pi^2 \rho_s \sqrt{4D} \overline{f} \zeta \right\} \exp\left\{ -\frac{1}{2} \sum_{\alpha=1}^{Q-2} \left(\sum_{a} n^a V_{a\alpha}\right)^2 2 \pi \rho_s \ln \zeta \right\} \nonumber \\
    & \times \exp\left\{ -\frac{1}{2} \left(\sum_{a} n^a V_{a,Q-1}\right)^2 \sqrt{\frac{1+K(\sigma_*/\sigma)^2}{1+K(\epsilon_*/\epsilon)^2}} 2 \pi \rho_s \ln \zeta \right\} \\
    & = \exp\left\{ -\frac{1}{2[K\epsilon^{-2} + \epsilon_*^{-2}]} \left(\sum_a n^a (\epsilon^a)^{-1} \right)^2 4 \pi^2 \rho_s \sqrt{4D} \overline{f} \zeta \right\} \exp\left\{ -\frac{1}{2} \sum_{ab} n^a n^b \Phi_{ab} 2 \pi \rho_s \ln \zeta \right\} \\
    & \times \exp\left\{ -\frac{1}{2K[1+K(\epsilon_*/\epsilon)^2]} \left(\sum_{a=1}^{Q-1} n^a - K (\epsilon_*/\epsilon) n^Q \right)^2 \sqrt{\frac{1+K(\sigma_*/\sigma)^2}{1+K(\epsilon_*/\epsilon)^2}} 2 \pi \rho_s \ln \zeta \right\}.
\end{align}

Since the expectation value does not depend on $\sigma$, we see that
\begin{align}
  \left\langle \exp \left\{ - \int_{r} \mathcal{V} \right\} \right\rangle_\text{G+}
  & \approx 1 + \frac{\lambda \langle e^{i \sum_a n^a \vartheta^a_+} \rangle_\text{G+} }{2} \mathcal{S}_{Q-1} \int_r \sum_a \sum_{\sigma=\pm1} e^{i \sigma \vartheta^a_-} \\
  & \approx \exp\left\{ \lambda \langle e^{i \sum_a n^a \vartheta^a_+} \rangle_\text{G+} \mathcal{S}_{Q-1} \int_r \cos\left(\sum_a n^a \vartheta^a_-\right) \right\}.
\end{align}
Rescaling space-time, we get a factor of $\zeta^2$ and so the coupling constant flows as
\begin{align}
    \lambda \rightarrow \lambda \zeta^2 \langle e^{i \sum_a n^a \vartheta^a_+} \rangle_\text{G+}.
\end{align}
Looking at \eqref{eq:scaling-factor}, we see that in order to prevent the operator from being exponentially irrelevant, we have a hard constraint that we require $\sum_a n^a \epsilon_a^{-1} = 0$. Away from commensurate points, this forces the teacher vorticity to be zero, $n^Q = 0$. Then the minimal relevant configuration is the inter-replica vortex configuration $(1, -1, 0, \cdts, 0, 0)$, such that
\begin{align}
    \langle e^{i \sigma \sum_a n^a \vartheta^a_+} \rangle_\text{G+}
    & = \exp \left\{  -2 \pi \rho_s (\ln \zeta) \right\}.
\end{align}

Let us consider the best case scenario for commensurate points. For $r_\epsilon > 1$, we can consider $\epsilon^* = 2 \epsilon$. In this case, we can consider the configuration $(1, 0, \cdts, 0, -2)$. Here we have
\begin{align} 
  \langle e^{i \sigma \sum_a n^a \vartheta^a_+} \rangle_\text{G+}
  & = \exp\left\{ -\frac{1}{2} 2 \pi \rho_s (\ln \zeta) (-2) (\Phi_{1,Q} + \Phi_{Q, 1}) \right\} \\
  & \times \exp\left\{ -\frac{1}{2K[1+K(\epsilon_*/\epsilon)^2]} \left(1 + 2 K (\epsilon_*/\epsilon) \right)^2 \sqrt{\frac{1+K(\sigma_*/\sigma)^2}{1+K(\epsilon_*/\epsilon)^2}} 2 \pi \rho_s \ln \zeta \right\}.
\end{align}

For $r_\epsilon < 1$, we can consider $\epsilon^* = \frac{1}{2} \epsilon$. In this case, we can consider the configuration $(2, 0, \cdts, 0, -1)$. Then we have
\begin{align} 
  \langle e^{i \sigma \sum_a n^a \vartheta^a_+} \rangle_\text{G+}
  & = \exp\left\{ -\frac{1}{2} 2 \pi \rho_s (\ln \zeta) (-2) (\Phi_{1,Q} + \Phi_{Q, 1}) \right\} \\
  & \times \exp\left\{ -\frac{1}{2K[1+K(\epsilon_*/\epsilon)^2]} \left(2 - K (\epsilon_*/\epsilon) \right)^2 \sqrt{\frac{1+K(\sigma_*/\sigma)^2}{1+K(\epsilon_*/\epsilon)^2}} 2 \pi \rho_s \ln \zeta \right\}.
\end{align}

Therefore even at the commensurate points, the minimal relevant configuration is the inter-replica vortex configuration $(1, -1, 0, \cdts, 0, 0)$.



\subsection{Observables in the charge-fuzzy phase}
In the charge-fuzzy phase, where the vortex terms are irrelevant, we can calculate the expectation values using the Gaussian theory. 

As we are interested in the quantities at the end time $0$ (with starting time $-t_\mathrm{f}$), we need consider the end boundary conditions. This is given by $\theta^a(x, 0) = \pi/2 \forall a, x$, which corresponds to $\delta \theta^a(x, 0) = 0$, $\partial_x \phi^a(x, 0) = 0$ and $\partial_t \chi^a(x, 0)= 0$. Using the method of images and solving for the Green's functions one finds an overall factor of $2$ compared to bulk Green's function. The $(Q-1)$th mode has the same boundary Green's function due to the log multiply rule, so
\begin{align}
  \langle \chi_\alpha (x, 0) \chi_\alpha (x', 0) \rangle_\text{G} = \begin{cases}
    - 2 \times 2 \pi^2 \rho_s \lvert x - x' \rvert + \dts & \alpha = 0, \\
    - 2 \times 2 \pi \rho_s \ln \lvert x - x'\rvert + \dts & \alpha = 1, \dts Q-1.
  \end{cases}
\end{align}
Looking at the source term, we can deduce that $\hat{Z}_\alpha \approx - \frac{\mathsf{a}}{\sqrt{v} \pi}\partial_x \chi_\alpha$. Recall that we rescaled space, with $\sqrt{v} x_\text{iso.} = x$, and that the relations above are for $x_\text{iso.}$. In the non-isotropic coordinates, we have $\hat{Z}_\alpha \approx - \frac{\mathsf{a}}{\pi} \partial_x \chi_\alpha$. Using the diagonalisation transformation $Z^a = \sum_\alpha V_{a\alpha} Z_\alpha$, we have
\begin{align}
  \langle \hat{Z}^a_x \hat{Z}^b_{0} \rangle_\text{G} = - 4 \pi \rho_s \Psi_{ab} \left(\frac{\mathsf{a}}{x} \right)^2.
\end{align}
where $\Psi = \mathbb{1}_Q - \hat{v}_0 \hat{v}_0^T$ is the projector out of the zero mode. Then we have
\begin{align}
  \E_{Y,S^*} [\langle s_{x} s_{0}\rangle_\c] & = \lim_{Q \rightarrow 1} \left[ \langle \hat{Z}^1_x \hat{Z}^1_0 \rangle_\text{G} - \langle \hat{Z}^1_x \hat{Z}^2_0 \rangle_\text{G}\right] = - 4 \pi \rho_s \left(\frac{\mathsf{a}}{x} \right)^2, \\
  \E_{Y,S^*} [\langle s_{x} s_{0}\rangle] & = \lim_{Q \rightarrow 1} \left[ \langle \hat{Z}^1_x \hat{Z}^1_0 \rangle_\text{G} \right] = 4 \pi \rho_s (r^2 - 1) \left(\frac{\mathsf{a}}{x} \right)^2, \\
  \E_{Y,S^*} [\langle s_{x}\rangle s^*_{0}] & = \lim_{Q \rightarrow 1} \left[ \langle \hat{Z}^1_x \hat{Z}^*_0 \rangle_\text{G} \right] = 4 \pi \rho_s r \left(\frac{\mathsf{a}}{x} \right)^2.
\end{align}

\section{Perturbative analysis of the planted SSEP} \label{apdx:perturbation-theory}
We would like to calculate the various correlators at the limit of small $\epsilon$, whilst keeping $r_\epsilon := \epsilon_* / \epsilon$ in the order of 1. In tensor network/dirac notation, posterior distribution can be written as
\begin{align}
  p_\mathrm{s}(\bm{s}_{t_\mathrm{f}} \vert \bm{y}_{1:{t_\mathrm{f}}}) = \frac{\langle \bm{s}_{t_\mathrm{f}} \rvert M(\bm{y}_{t_\mathrm{f}}) \bm{u} \cdts M(\bm{y}_1) \bm{u} \lvert {\Plus}' \rangle}{\langle \tilde{\Plus} \rvert M(\bm{y}_{t_\mathrm{f}}) \bm{u} \cdts M(\bm{y}_1) \bm{u} \lvert {\Plus}' \rangle},
\end{align}
where $\lvert \tilde{\Plus} \rangle := \otimes_x \lvert \tilde{+} \rangle_x$, $\lvert \tilde{+} \rangle_x = \lvert -1 \rangle_x + \lvert 1 \rangle_x$, $\lvert {\Plus}' \rangle := \otimes_x \lvert {+}' \rangle_x$, $\lvert {+}' \rangle_x = \frac{1}{2}(\lvert -1 \rangle_x + \lvert 1 \rangle_x)$, and the measurement operator will be expanded as
\begin{align}
  M(\bm{y}_{t_\mathrm{f}}) = \,\, \otimes_x \exp(\epsilon y_{x,t} Z_x) \approx 1 + \epsilon y_{x,t} Z_x + \frac{1}{2} \epsilon^2 y_{x,t}^2 + \frac{1}{6} \epsilon^3 y_{x,t}^3 Z_x + \mathcal{O}(\epsilon^4).
\end{align}
Similarly, the joint distribution of measurement and true/teacher trajectories can also be written as
\begin{align}
  p_*(\bm{y}_{1:{t_\mathrm{f}}}, \bm{s}^*_{t_\mathrm{f}}) = e^{-L {t_\mathrm{f}} r_\epsilon^2 \epsilon^2 /2} \langle \bm{s}_{t_\mathrm{f}} \rvert M^*_{t_\mathrm{f}}(\bm{y}_{t_\mathrm{f}}) \bm{u} \cdts M^*_1(\bm{y}_1) \bm{u} \lvert {\Plus}' \rangle \pi(\bm{y}_{1:{t_\mathrm{f}}}),
\end{align}
where $\pi(\bm{y}_{1:{t_\mathrm{f}}})$ denotes the Gaussian measure, and $M^*$ denotes $M$ but with teacher parameters $\epsilon \rightarrow \epsilon_* = r_\epsilon \epsilon$.

We will use the fact that any state $\propto \lvert \Plus \rangle$ is an invariant state under the SSEP, that is $\bm{u} \lvert \Plus \rangle = \lvert \Plus \rangle$, and that the choice of half-filling initial condition has vanishing disconnected correlators, $\langle \Plus \rvert Z_x \lvert \Plus \rangle = \langle \Plus \rvert Z_x Z_0 \lvert \Plus \rangle = \langle \Plus \rvert Z_{x_1} \cdts Z_{x_n} \lvert \Plus \rangle = 0$.

We have
\begin{align}
  M(\bm{y}_{t_\mathrm{f}}) \bm{u} \cdts M(\bm{y}_1) \bm{u} \lvert \Plus' \rangle & = \left\{ 1 + \epsilon \underbrace{ \sum_{x,t} y_{x,t} Z_{x,t}}_{:=(1)} + \epsilon^2 \left( \underbrace{\frac{1}{2} \sum_{x,t} y^2_{x,t}}_{:=(2)} + \underbrace{\sum_{(x, t) \neq (x', t')} y_{x,t} y_{x't'} Z_{x,t} Z_{x't'}}_{:=(11)} \right) \right\} \bm{u} \cdts \bm{u} \lvert \Plus' \rangle,
\end{align}
where $Z_{x,t}$ denotes inserting the $Z_r$ operator in the middle of the string of $\bm{u}$'s at time $t$, and we introduced the bracket notation, where the each number $n$ in the bracket denotes including the $n$th order term of $M$. Continuing the series we get
\begin{align}
  M(\bm{y}_{t_\mathrm{f}}) \bm{u} \cdts M(\bm{y}_1) \bm{u} \lvert \Plus' \rangle & = \Bigg(1 + \epsilon (1) + \epsilon^2 \Big( (2) + (11) \Big) + \epsilon^3 \Big( (3) + (21) + (111)\Big) \nonumber \\
  & + \epsilon^4 \Big( (4) + (31) + (22) + (211) + (1111) \Big) \nonumber \\
  & + \epsilon^5 \Big( (5) + (41) + (32) + (311) + (22) + (211) + (11111)\Big) \nonumber \\
  & + \mathcal{O}(\epsilon^6) \Bigg) \bm{u} \cdts \bm{u} \lvert \Plus' \rangle.
\end{align}

\subsection{Disconnected correlator}
The disconnected correlator can be written as
\begin{align}
  \E [\langle s_{x, {t_\mathrm{f}}} s_{0, {t_\mathrm{f}}} \rangle] = e^{-L {t_\mathrm{f}} r_\epsilon^2 \epsilon^2 /2} \left[ \frac{A B}{C}\right],
\end{align}
where squared bracket will denote average over $Y$ under the Gaussian measure and
\begin{align}
  A & = \langle \tilde{\Plus} \rvert Z_x Z_0 M(\bm{y}_{t_\mathrm{f}}) \bm{u}_{t_\mathrm{f}} \cdts M(\bm{y}_1) \bm{u}_1 \lvert {\Plus}' \rangle, \\
  B & = \langle \tilde{\Plus} \rvert M^*(\bm{y}_{t_\mathrm{f}}) \bm{u}_{t_\mathrm{f}} \cdts M^*(\bm{y}_1) \bm{u}_1 \lvert {\Plus}' \rangle, \\
  C & = \langle \tilde{\Plus} \rvert M(\bm{y}_{t_\mathrm{f}}) \bm{u}_{t_\mathrm{f}} \cdts M(\bm{y}_1) \bm{u}_1 \lvert {\Plus}' \rangle.
\end{align}
Note that in the Dirac expectation, each $Z$ would flip $\lvert + \rangle \rightarrow \lvert - \rangle$, and $\bm{u}$ would generate a superposition of $\lvert - \rangle$ the diffuses. Since at the end, we bra with a $\langle \Plus \rvert$, each $Z$ must be able to be paired up with another $Z$ from previous timesteps. Expanding each factor as $A = A_0 + \epsilon A_1 + \epsilon^2 A_2$ and similarly for $B,C$, This implies that $A_\mathrm{odd} = B_\mathrm{odd} = C_\mathrm{odd} = 0$. Due to the choice of the initial condition and properties of the SSEP, we also have that $A_0 = 0$, and $B_0 = C_0 = 1$. Also, since we put dependence on $\epsilon$ and $\epsilon_*$ outside the definition of $A_n, B_n, C_n$, we have that $B_n = C_n$. Denote $(n \cdts m)_{A} := \langle \tilde{\Plus} \rvert Z_x Z_0 (n \cdts m) \bm{u} \cdts \bm{u} \lvert {\Plus}' \rangle$ and similarly for $B,C$. Then any lone $(n_\mathrm{odd})_{A,B,C} = 0$ there are odd numbers of $Z$'s, $(n_\mathrm{even})_{A} = \sum_{x,t} y_{x,t}^{n_\mathrm{even}} \langle Z_x Z_0 \rangle_\mathrm{SSEP} = 0$ due to the choice of the initial condition, and $(n_\mathrm{even})_{B, C} = \sum_{x,t} y_{x,t}^{n_{\mathrm{even}}}$.

Therefore
\begin{align}
  \left[ \frac{A B}{C}\right] & = \left[ \frac{(\epsilon^2 A_2 + \epsilon^4 A_4 + \cdts) (1 + r_\epsilon^2 \epsilon^2 B_2 + r_\epsilon^4 \epsilon^4 B_4 + \cdts)}{1 + \epsilon^2 B_2 + \epsilon^4 B_4 + \cdts}\right] \nonumber \\
  & = \Big[ (\epsilon^2 A_2 + \epsilon^4 A_4 + \cdts) (1 + r_\epsilon^2 \epsilon^2 B_2 + r_\epsilon^4 \epsilon^4 B_4 + \cdts) \nonumber \\
  & \times \Big(1 - (\epsilon^2 B_2 + \epsilon^4 B_4 + \cdts) + (\epsilon^2 B_2 + \epsilon^4 B_4 + \cdts)^2 + \cdts \Big) \Big] \\
  & = \Bigg[ \Big( \epsilon^2 (11)_A + \epsilon^4 \Big\{ (31) + (211) + (1111)\Big\}_A + \mathcal{O}(\epsilon^6) \Big) \\
  & \times \Big( 1 + r_\epsilon^2 \epsilon^2 \Big\{ (2) + (11) \Big\}_B + \mathcal{O}(\epsilon^4) \Big) \\
  & \times \Big( 1 - \epsilon^2 \Big\{ (2) + (11) \Big\}_B + \mathcal{O}(\epsilon^4) \Big)\Bigg].
\end{align}
When averaging over each $y_{x,t}$, $[y_{x,t}^{n_{\mathrm{odd}}}] = 0$, which means that each $y_{x,t}$ must pair up. Therefore
\begin{align}
  \left[ \frac{A B}{C}\right] & = \epsilon^4 (r_\epsilon^2 - 1) \left[ (11)_A (11)_B \right].
\end{align}
The exponential factor cannot pair up with any term to create factors of order smaller or equal to $\epsilon^4$, so we have
\begin{align}
  \ \E [\langle s_{x, {t_\mathrm{f}}} s_{0, {t_\mathrm{f}}} \rangle] & = \epsilon^4 (r_\epsilon^2 - 1) \left[ (11)_A (11)_B \right].
\end{align}
Now
\begin{align}
  \left[ (11)_A (11)_B \right] & = \sum_{(x, t) \neq (x', t')} \langle Z_{0, t_\mathrm{f}} Z_{l, t_\mathrm{f}} Z_{x, t} Z_{x', t'} \rangle_\mathrm{SSEP} \langle Z_{x, t} Z_{x', t'} \rangle_\mathrm{SSEP} \\
  & = \left(\sum_{t' \neq t, x, x'} + \sum_{t' = t, x' \neq x} \right) \langle Z_{0, t_\mathrm{f}} Z_{l, t_\mathrm{f}} Z_{x, t} Z_{x', t'} \rangle_\mathrm{SSEP} \langle Z_{x, t} Z_{x', t'} \rangle_\mathrm{SSEP}.
\end{align}
The second sum is zero again due to the half-filling initial condition. So
\begin{align}
  \left[ (11)_A (11)_B \right]
  & = 2 \sum_{t < t', x, x'} \langle Z_{0, t_\mathrm{f}} Z_{l, t_\mathrm{f}} Z_{x, t} Z_{x', t'} \rangle_\mathrm{SSEP} \langle Z_{x, t} Z_{x', t'} \rangle_\mathrm{SSEP}. \\
  & = 2 \sum_{t < t', x, x'} p( \mathbf{0}, \mathbf{l} \leftarrow \mathbf{r}, \mathbf{r}' \text{ or } \mathbf{0}, \mathbf{l} \leftarrow \mathbf{r}', \mathbf{r}; \text{$\mathbf{r}$, $\mathbf{r}'$ do not meet}) p(\mathbf{r}' \leftarrow \mathbf{r}),
\end{align}
where $\mathbf{r} = (x, t)$, $\mathbf{0} = (0, t_\mathrm{f})$, $\mathbf{l} = (l, t_\mathrm{f})$, and the probabilities refers to inserting a particle at $\mathbf{r}$ then at $\mathbf{r}'$ and evolving them according to the SSEP. This is a sum of probabilities, and therefore is greater than zero. In fact, all terms $[(n \cdts n')_X \cdts (m \cdts m')_Y]$ are non-negative quantities similarly to above. Therefore
\begin{align}
  \E [\langle s_{x, {t_\mathrm{f}}} s_{0, {t_\mathrm{f}}} \rangle] = 2 \epsilon^4 (r_\epsilon^2 - 1) \sum_{t < t', x, x'} p( \mathbf{0}, \mathbf{l} \leftarrow \mathbf{r}, \mathbf{r}' \text{ or } \mathbf{0}, \mathbf{l} \leftarrow \mathbf{r}', \mathbf{r}; \text{$\mathbf{r}$, $\mathbf{r}'$ do not meet}) p(\mathbf{r}' \leftarrow \mathbf{r}) + \mathcal{O}(\epsilon^6).
\end{align}

\subsection{Connected correlator}
To calculate the connected correlator, we look at the disconnected part,
\begin{align}
  \E[\langle s_{0, {t_\mathrm{f}}} \rangle\langle s_{l, {t_\mathrm{f}}} \rangle] = e^{-Lt_\mathrm{f} r^2 \epsilon^2 / 2}\left[ \frac{D D' C}{B^2}\right],
\end{align}
where
\begin{align}
  D & = \langle \tilde{\Plus} \rvert Z_l M(\bm{y}_{t_\mathrm{f}}) \bm{u}_{t_\mathrm{f}} \cdts M(\bm{y}_1) \bm{u}_1 \lvert {\Plus}' \rangle, \\
  D' & = \langle \tilde{\Plus} \rvert Z_0 M(\bm{y}_{t_\mathrm{f}}) \bm{u}_{t_\mathrm{f}} \cdts M(\bm{y}_1) \bm{u}_1 \lvert {\Plus}' \rangle.
\end{align}
In this case, for $D = D_0 + \epsilon D_1 + \epsilon^2 D_2 + \cdts$, only odd terms survive, since there odd numbers of $Z$'s must be inserted to have even numbers in the Dirac expectation. So we have
\begin{align}
  \left[ \frac{D D' C}{B^2}\right] & = \left[ \frac{(\epsilon D_1 + \epsilon^3 D_3 + \cdts)(\epsilon D'_1 + \epsilon^3 D'_3 + \cdts)(1+r^2 \epsilon^2 B_2 + r^4 \epsilon^4 B_4 + \cdots)}{(1+\epsilon^2 B_2 + r^4 \epsilon^4 B_4 + \cdots)(1 + \epsilon^2 B_2 + \epsilon^4 B_4 + \cdots)} \right] \\
  & = \left[  \epsilon^2 D_1 D'_1 + \epsilon^4 r^2 D_1 D'_1 B_2 - 2 \epsilon^4 D_1 D'_1 B_2 + \cdts \right] \\
  & = \epsilon^2 [(1)_D (1)_{D'}] + \cdts \\
  & = \epsilon^2 \sum_{(x, t)} \langle Z_{0, t_\mathrm{f}} Z_{x, t} \rangle_\mathrm{SSEP} \langle Z_{l, t_\mathrm{f}} Z_{x, t} \rangle_\mathrm{SSEP} + \cdts  \\
  & = \epsilon^2 \sum_{\mathbf{r}} p(\mathbf{0} \leftarrow \mathbf{r}) p(\mathbf{l} \leftarrow \mathbf{r}) + \cdts.
\end{align}
Again, any term from the expontial factor $e^{Lt_\mathrm{f} r_\epsilon^2 \epsilon^2 / 2}$ only creates higher order terms, so
\begin{align}
  \E[\langle s_{0, {t_\mathrm{f}}} \rangle\langle s_{l, {t_\mathrm{f}}} \rangle] = \epsilon^2 \sum_{\mathbf{r}} p(\mathbf{0} \leftarrow \mathbf{r}) p(\mathbf{l} \leftarrow \mathbf{r}).
\end{align}
Since the disconnected correlator is zero for order $\epsilon^2$, we have
\begin{align}
  \E[\langle s_{0, t_\mathrm{f}} s_{l, t_\mathrm{f}} \rangle_\mathrm{c}] = - \epsilon^2 \sum_{\mathbf{r}} p(\mathbf{0} \leftarrow \mathbf{r}) p(\mathbf{l} \leftarrow \mathbf{r}) + \mathcal{O}(\epsilon^6).
\end{align}
\subsection{Teacher-student alignment}
We have
\begin{align}
  \E[\langle s_{l, {t_\mathrm{f}}} \rangle s_{0, {t_\mathrm{f}}}^* ] = e^{-Lt_\mathrm{f} r^2 \epsilon^2 / 2}\left[ \frac{D E}{B}\right],
\end{align}
where
\begin{align}
  E & = \langle \tilde{\Plus} \rvert Z^*_0 M^*(\bm{y}_{t_\mathrm{f}}) \bm{u}_{t_\mathrm{f}} \cdts M^*(\bm{y}_1) \bm{u}_1 \lvert {\Plus}' \rangle.
\end{align}
Again only $E_{n_\mathrm{odd}}$ survive. We also have that $E_n = D'_n$ as we put dependence on $r_\epsilon, \epsilon$ outside, so
\begin{align}
  \left[ \frac{D E}{B}\right] & = \left[ \frac{(\epsilon D_1 + \epsilon^3 D_3 + \cdts)(r_\epsilon \epsilon D'_1 + r_\epsilon^3 \epsilon^3 D'_3 + \cdts)}{1+\epsilon^2 B_2 + r^4 \epsilon^4 B_4 + \cdots} \right] \\
  & = \left[  \epsilon^2 D_1 D'_1 + \cdts \right] = r_\epsilon \epsilon^2 [(1)_D (1)_{D'}] + \cdts \\
  & = r_\epsilon \epsilon^2 \sum_{(x, t)} \langle Z_{0, t_\mathrm{f}} Z_{x, t} \rangle_\mathrm{SSEP} \langle Z_{l, t_\mathrm{f}} Z_{x, t} \rangle_\mathrm{SSEP} + \cdts = r_\epsilon \epsilon^2 \sum_{\mathbf{r}} p(\mathbf{0} \leftarrow \mathbf{r}) p(\mathbf{l} \leftarrow \mathbf{r}) + \cdts,
\end{align}
so we have
\begin{align}
  \E[\langle s_{l, {t_\mathrm{f}}} \rangle s_{0, {t_\mathrm{f}}}^* ] = r_\epsilon \epsilon^2 \sum_{\mathbf{r}} p(\mathbf{0} \leftarrow \mathbf{r}) p(\mathbf{l} \leftarrow \mathbf{r}) + \mathcal{O}(\epsilon^4).
\end{align}

\section{Saddle point analysis of the static charge inference problem} \label{apdx:static-saddle-point}
We consider a simple model of inference, where the charges are static. In this case, the posterior is given by
\begin{align}
  p_\mathrm{s} (\bm{s} \vert \bm{y}_{1:t_\mathrm{f}}) = \frac{\exp\left\{\epsilon \sum_{x, t} y_{x, t} s_{x} \right\}}{Z(\bm{y}_{1:t_\mathrm{f}})},
\end{align}
where
\begin{align}
  Z(\bm{y}_{1:t_\mathrm{f}}) = \sum_{\bm{s}_{t_\mathrm{f}}}\exp\left\{\epsilon \sum_{x, t} y_{x, t} s_{r} \right\}.
\end{align}
As the partition function is separable we can sum over each site separately. Hence we have
\begin{align}
  \langle s_x \rangle = \tanh\left(\epsilon \sum_{t} y_{x, t}\right), \quad \langle s_x s_{x'} \rangle = \tanh\left(\epsilon \sum_{t} y_{x, t}\right) \tanh\left(\epsilon \sum_{t} y_{x', t}\right) \text{ for } x \neq x'.
\end{align}
Therefore we have that $\langle s_x s_{x'} \rangle_\mathrm{c} = 0$.
Using translation invarinace, the charge variance is given by
\begin{align}
  \delta C^2 & = \E\left[ \sum_{x, y} \langle s_x s_y \rangle_\mathrm{c} \right] = \E\left[ L (1 - \langle s_0\rangle^2 ) + 2 L \sum_{d = 1}^{L/2} \langle s_0 s_d \rangle_\mathrm{c} \right]
  = \E\left[ L (1 - \langle s_0\rangle^2 )\right] = L \times \E\left[ \mathrm{sech}^2\left(\epsilon \sum_{t} y_{0, t}\right)\right].
\end{align}

The teacher's distribution is given by
\begin{align}
  p_*(\bm{s^*}, \bm{y}_{1:t_\mathrm{f}}) = \frac{1}{2^L} \prod_{x, t} \left(\frac{\exp\left\{ -\frac{1}{2} (y_{x, t} - \epsilon_* s^*_x)^2 \right\}}{\sqrt{2 \pi}} \right).
\end{align}

To calculate the expectation value, first note that we can integrate over all $y_{x\neq 0,t}$. Then, we shift $y_{0, t} \rightarrow y_{0, t} + \epsilon_* s^*_0$. We are left with

\begin{align}
  \delta C^2
  & = \frac{L}{2} \sum_{s_0^*} \int \prod_{t} \left( dy_{0, t} \frac{e^{-y_{0, t}^2/2}}{\sqrt{2 \pi}} \right) \mathrm{sech}^2\left(\epsilon \sum_{t} y_{0, t} + \epsilon_* t_\mathrm{f} s_0^* \right).
\end{align}
Next we shift $y_{0, t} \rightarrow s^*_0 y_{0, t}$. Using the evenness of $\mathrm{sech}^2$, we get
\begin{align}
  \delta C^2
  & = L \int \prod_{t} \left( dy_{0, t} \frac{e^{-y_{0, t}^2/2}}{\sqrt{2 \pi}} \right) \mathrm{sech}^2\left(\epsilon \sum_{t} y_{0, t} + \epsilon_* t_\mathrm{f} \right).
\end{align}
We can treat $\sum_t y_{0, t} =: y$ as a single random variable with variance $t_\mathrm{f}$. Making this transformation then transforming $y \rightarrow t_\mathrm{f} y$ we get
\begin{align}
  \delta C^2
  & = L \sqrt{\frac{t_\mathrm{f}}{2 \pi}} \int dy \exp \left[ -t_\mathrm{f} f(y) \right],
\end{align}
where $f(y) = \frac{y^2}{2} - \frac{2}{t_\mathrm{f}} \ln \mathrm{sech} t_\mathrm{f} \left(\epsilon y + \epsilon_* \right)$. Then the saddle point approximation is
\begin{align}
  \delta C^2 = L \frac{1}{\sqrt{f''(y_0)}} e^{-t_\mathrm{f} f(y_0)} (1 + \dts).
\end{align}

Now $- (1/t_\mathrm{f}) \ln \mathrm{sech} (t_\mathrm{f}x) \rightarrow \lvert x \rvert$ as $t_\mathrm{f} \rightarrow \infty$, which we will use to identify the saddle point $y_0 + \mathrm{sgn}(\epsilon y_0 + \epsilon_*) = 0$, though we will keep the exact function for the curvature $f''(y) = 1 + 2 t_\mathrm{f} \epsilon ^2 \text{sech}^2\left(t_\mathrm{f} \left(y \epsilon +\epsilon _*\right)\right)$. 

At $\epsilon_*=0$, the saddle point is at $y_0=0$, giving $f(y_0 = 0) = 0$ with curvature $f''(0) = 1 + 2 \epsilon^2 t_\mathrm{f}$. 

For $0 < r_\epsilon < 2 \epsilon$, the saddle point is at $y_0 = -\epsilon_*/\epsilon$ so $f(y_0) = r_\epsilon^2/2 - (2 / t_\mathrm{f}) \ln \mathrm{sech} t_\mathrm{f} (0) = r^2_\epsilon/2$ and $f''(y_0) = 1 + 2 t_\mathrm{f} \epsilon^2$.

For $r_\epsilon > 2 \epsilon$, $y_0 = -2\epsilon$ and $f''(y_0) = 1 + 2 t_\mathrm{f} \epsilon^2 \mathrm{sech}^2(t_\mathrm{f}(\epsilon_* - 2 \epsilon^2)) \rightarrow 1$. 

Therefore for large $t_\mathrm{f}$ we have (for $\epsilon > 0$)
\begin{align}
  \delta C^2 = L \times \begin{cases}
    \epsilon^{-1} t_\mathrm{f}^{-1/2} & r_\epsilon = 0, \\
    \epsilon^{-1} t_\mathrm{f}^{-1/2} e^{-r_\epsilon^2 t_\mathrm{f}/2} & r_\epsilon < 2 \epsilon, \\
    e^{-2\epsilon(r_\epsilon - 2\epsilon) t_\mathrm{f}}& r_\epsilon > 2 \epsilon.
  \end{cases}
\end{align}

\section{Inferrability argument for mapping from projective to weak measurements} \label{apdx:inferrability-argument}
In projective measurement case, when a measurement does occur, then we know the true spin with probability $1$. When it doesn't occur, then the probability of guessing the true spin is $1/2$. Therefore with rate $\uppi$ the probability of guessing the correct spin is $p(\uppi)=\frac{1}{2} \times (1-\uppi) + 1 \times\uppi$. 

For the weak measurement case, the probability of guessing the correct spin should be given by $p(s = s' \vert s^* = s')$. Now
\begin{align}
  p(s=s' \vert s^*=s') & = \int_\phi p(s=s' \vert \phi) p(\phi \vert s^*=s')  = p(s \vert s^*)
  = \int_\phi \frac{p(\phi \vert s=s') p(\phi \vert s^*=s')}{p(\phi)} \\
  & = \int_\phi \frac{e^{\epsilon \phi}}{2 \cosh \epsilon \phi} \frac{e^{-(\phi-\epsilon)^2/2}}{\sqrt{2\pi}} =: p(\epsilon).
\end{align}

If we assume that only the probability of guessing the correct spin is important in the transition, then the measurement rate and signal strength variables should relate as $p(\epsilon)=p(\uppi)$.

In Ref. \cite{barratt2022field}, they approximate the critical rate to be around $\uppi^\#_\mathrm{BO} \in [0.2, 0.3]$ for the $\rm{U}(1)$ quantum circuit. Using above relations, this corresponds to $p^\#_\mathrm{BO} \in [0.60,0.65]$, which corresponds to $\epsilon^\#_\mathrm{BO} \in [0.49, 0.63]$.

\section{Exact relations between `sharpening' and `inference' moment order parameters} \label{apdx:moment-relatons}
Let us use $\langle \cdot \rangle$ to denote the average over the student's posterior. The first natural quantity is the mean-squared-error of the mean of the student's inferred charge distribution,
\begin{align}
  \mathrm{MSEM} & = \E_{Y, X^*} \left[ (\langle C \rangle - C^*)^2 \right] = \E_{Y, X^*} \left[ \langle C \rangle^2 - 2 \langle C \rangle C^* + (C^*)^2 \right].
\end{align}
The second is the mean-squared error on the joint distribution,
\begin{align}
  \mathrm{MSE} & = \E_{Y, X^*} \left[ \langle (C - C^*)^2 \rangle  \right] = \E_{Y, X^*} \left[ \langle C^2 \rangle - 2 \langle C \rangle C^* + (C^*)^2 \right].
\end{align}
The third is the charge variance of the student's inferred charge distribution,
\begin{align}
  \delta C^2 = \E_{Y, X^*} \left[ \langle C^2 \rangle - \langle C \rangle^2  \right].
\end{align}
In the Bayes optimal case, distribution of $X$ and $X^*$ is identical in the joint distribution. Therefore $[\langle C \rangle^2] = [\langle C \rangle C^*]$. This means that, at Bayes optimality,
\begin{align}
  \mathrm{MSEM}(\mathrm{Bayes \; optimal}) = \frac{1}{2} \mathrm{MSE}(\mathrm{Bayes \; optimal}) = \delta C^2(\mathrm{Bayes \; optimal}).
\end{align}

\section{Derivation of the repetition code case} \label{apdx:repetition-code}

We denote $e^*_{l_\rightarrow}=-1$ as a bit-flip error occurring at time $t$, and $e^*_{l_\uparrow} = -1$ to denote if a syndrome measurement is corrupted by a read-out error. Then, we have $e^*_{l_\rightarrow} = f^*_{l, t} f^*_{l, t-1}$ and $e^*_{l_\uparrow} = s_{v,t} s^*_{v, t} = s_{l_\uparrow} s^*_{l_\uparrow}$, where $s^*_{v, t} = f^*_{l_\mathrm{L}(v), t} f^*_{l_\mathrm{R}(v), t}$  is the 'true' syndrome. Then we can decouple the prior as

$$
p_*({F}^*) \rightarrow p_*({E}^*_\rightarrow) \propto \exp\left[ \beta_* \sum_{l_\rightarrow} e^*_{l_\rightarrow}\right].
$$
The likelihood for each syndrome can also be written as
$$
\begin{aligned}
p_*(s_{v, t} \vert f^*_{l_\mathrm{L}(v), t}, f^*_{l_\mathrm{R}(v), t})  & \rightarrow p_*(s_{l_\uparrow}, e^*_{l_\uparrow} \vert s^*_{l_\uparrow}) \propto \delta \{s_{l_\uparrow}, e^*_{l_\uparrow} s^*_{l_\uparrow}\} \exp\left[ \beta_* e^*_{l_\uparrow}\right] = p_*(s_{l_\uparrow} \vert e^*_{l_\uparrow} , s^*_{l_\uparrow}) p_*(e^*_{l_\uparrow}).
\end{aligned}
$$
Denoting all occurrences of error as ${E}^* = {E}^*_\rightarrow \odot {E}^*_\uparrow$, we then have
$$
p_*({S} \vert {F}^*) \rightarrow p_*\left({S} \vert {E}^*_\uparrow, {S}^*({E}^*_\rightarrow)\right) p_*({E}^*_\uparrow) = p_*({S} \vert {E}^*) p_*({E}^*).
$$
Let ${W}$ be the true 'worldlines', which includes the 'true' syndromes ${S}^*$ on the vertical links and the true bit-flips ${E}_\rightarrow^*$ on the horizontal links. Then ${W} = {S}^* \odot {E}^*_\rightarrow$. For the whole set of syndromes we have ${E}^*_\uparrow = {S} \odot {S}^*$. Then we have that
$$
{S} = {E}^* \odot {W}.
$$
Now, ${W}$ does not have a boundary, if we assert that vertical links at final time has no boundary above. Then this means that ${S}$ and ${E}^*$ must share a boundary, Therefore, likelihood for syndrome trajectory given all errors is
$$
p_*({S} \vert {E}^*) = \delta_{\partial {S}, \partial {E}^* }.
$$
In this case, the boundary of ${S}$ or ${E}^*$ now lives on \emph{vertices} of the 2D spacetime lattice.

Marginalising over the syndromes, the conditional probability of inferring ${E}$ given ${E}^*$ is
$$
\begin{aligned}
p({E} \vert {E}^*) & = \sum_{ S} p_\mathrm{s}({E} \vert {S}) p_*({S} \vert {E}^*) = \sum_{{S}} \frac{\delta_{\partial {S}, \partial {E}} p_\mathrm{s}({E}) \delta_{\partial {S}, \partial {E}^*}}{p_\mathrm{s}({S})} = \sum_{ S} \frac{\delta_{\partial {S}, \partial {E}} p_\mathrm{s}({E}) \delta_{\partial S, \partial E^*}}{\sum_{E'} \delta_{\partial S, \partial E'} p_s(E')}.
\end{aligned}
$$
Since in the sum over $E'$, only ${S}$ such that $\partial S = \partial E^*$ survive, we can replace $\delta_{\partial S, \partial E'} \rightarrow \delta_{\partial E^*, \partial E'}$. Then, this term does not depend on $S$ and can be pulled out of the sum. Therefore
$$
\begin{aligned}
p(E \vert E^*) 
& = \sum_S \frac{\delta_{\partial S, \partial E} p_\mathrm{s}(E) \delta_{\partial S, \partial E^*}}{\sum_{E'} \delta_{\partial E^*, \partial E'} p_s(E')} = \frac{p_\mathrm{s}(E)}{\sum_{E'} \delta_{\partial E^*, \partial E'} p_s(E')} \sum_S {\delta_{\partial S, \partial E} \delta_{\partial S, \partial E^*}} \\
& = \frac{p_\mathrm{s}(E)}{\sum_{E'} \delta_{\partial E^*, \partial E'} p_s(E')} \sum_S \delta_{\partial E, \partial E^*} \delta_{\partial S, \partial E^*} = p_\mathrm{s}(E)\delta_{\partial E, \partial E^*} \times \frac{\sum_S \delta_{\partial S, \partial E^*}}{\sum_{E'} \delta_{\partial E^*, \partial E'} p_s(E')}.
\end{aligned}
$$
Then, one notes that the second factor is independent of $E$, therefore 
$$
\begin{aligned}
p(E \vert E^*) 
& \propto p_\mathrm{s}(E)\delta_{\partial E, \partial E^*}.
\end{aligned}
$$
Simiarly to the toric code case, to respect the delta function, we only consider $E = E^* \odot C$, where $C$ is a cycle up to the boundary. Therefore we can parameterise $c_l = \sigma_{v_1(l)} \sigma_{v_2(l)}$ \emph{except} at the final time where $c_{l_\uparrow} = \sigma_{v_\mathrm{below}(l_\rightarrow)}$, and the vertex lives \emph{below} $l_\uparrow$. We could also set the boundary condition that the vertex that lives above has $\sigma_{v_\mathrm{above}} = 1$. Therefore we again have the 2D RBIM partition function \cref{eq:rbim-boltzmann-prob} with periodic boundary conditions in space, free open boundary at $t=1$, and fixed boundary condition $\sigma_{v_\mathrm{above}} = 1$ at $t=t_\mathrm{f}$. To calculate expectation values, we do so with respect to the \emph{true/teacher's} prior distribution $p_*(E^*) \propto \exp\left[\beta_* \sum_l e^*_l\right]$. 

Using that $f_{l, t} = e_{l, t} \cdts e_{l, 1}$, $f^*_{l, t} = e^*_{l, t} \cdts e^*_{l, 1}$, $e_{l, t} = c_{l, t} e^*_{l, t} = \sigma_{v_\mathrm{L}(l), t} \sigma_{v_\mathrm{R}(l), t} e^*_{l, t}$, the correlation order parameter becomes
\begin{align}
    \E [f_{l, t} f^*_{l, t}] = \E\left[ \prod_{\tau=1}^t \sigma_{v_\mathrm{L}(l), \tau} \sigma_{v_\mathrm{r}(l), \tau} \right].
\end{align}

\section{Proof of concentration of measure for Haar unitaries} \label{apdx:concentration-of-measure}
Consider a quantum circuit given by the evolved state $\rho_t =  \cl U \cdts \cl U \cl M \cdts \cl M \cdts \rho_0$, where each $\cl U$ is an instance Haar-random distributed local unitary gate, acting on (say) sites $\bm i = (i_1, i_2, \dots, i_k)$, i.e. on $k$ qubit-qudit pairs. It could have further block-diagonal structure, and its average is given by $\E_u \cl U = \bar{\cl U}$. $\cl M(y)$ is a local measurement map
\begin{align}
    \cl M(y) : \rho \rightarrow \frac{Q(y) \rho Q(y)}{\tr[Q(y) \rho Q(y)]},
\end{align}
where $Q(y)$ are partial projectors acting on qubits on sites $\bm i'$. We will suppress the argument $y$ in what follows, unless required.

We will denote state evolved by averaged Haar-unitary elements as $\bar{\rho}_t = \bar{\cl U} \cdts \bar{\cl U} \cl M \cdts \cl M \cdts \rho_0$.

We would like to show that the Bayes-optimal posterior distribution is concentrated around that given by the Haar-averaged elements, i.e.
\begin{align}
    \E_{U, Y} \abs{p(\bm s \vert Y) - \bar p(\bm s \vert Y)} \mathop{\longrightarrow}^{d \rightarrow \infty} 0,
\end{align}
where the posterior is given by $p(\bm s \vert Y) = \tr[\bb P_{\bm{s}} \rho_t]$ and the posterior from averaged elements, and $\bar p(\bm s \vert Y) := \tr[\bb P_{\bm{s}} \bar \rho_t]$. $Y = \{y\}_y$ is the set of all measurements. Precisely, we will prove the following.

\begin{theorem} \label{thm:concentration-of-measure}
    Consider a quantum circuit composed of Haar-randomly distributed local unitaries and local measurements. The unitaries acts on $k$ qudit-qubit pairs, and there is a computational qubit basis such that it is block diagonal, $u = u^{(1)} \oplus \cdts \oplus u^{(B)}$. Each sector has dimensions $\propto d^k$. The local measurements $\cl M$ measure $k'$ qubits and are also diagonal in the same computational qubit basis. Assume that there is an extensive number of such elements $\sim VT$, where $VT$ is the spacetime volume, and that measurements are performed after each layer of unitaries. Then there exists some constant $r$, such that
    \begin{align}
        \E_{U, Y} \abs{p(\bm s \vert Y) - \bar p(\bm s \vert Y)} \leq C \frac{VT r_1^V r_2^{VT}}{d^{k/2}}.
    \end{align}
\end{theorem}

\begin{corollary} \label{cor:concentration-of-measure-ssep}
    For a circuit with Haar-random $u_{U(1)}$ gates in brickwork fashion with single-site monitoring, we have
    \begin{align}
        \E_{U, Y} \abs{p(\bm s \vert Y) - \bar p(\bm s \vert Y)}
        \leq \frac{LT\phi}{2d} r_\rm{unitary}^{L/2} (r^{1/2}_\rm{unitary} r_\rm{meas})^{LT},
    \end{align}
    with $\phi = \sqrt{3}$, $r_\rm{unitary} = 10$, $r_\rm{meas} = 2(\sqrt{2} + 1)$.
\end{corollary}

\bigskip

\hrule

\bigskip

\begin{proof}[Proof of \Cref{thm:concentration-of-measure} and \Cref{cor:concentration-of-measure-ssep}]
  Our strategy is as follows. We would like to bound
  \begin{align}
      c := \E_{U, Y} \abs{p(\bm s \vert Y) - \bar p(\bm s \vert Y)} = \E_{U, Y} \abs{\tr[\bb P_{\bm s} (\rho - \bar \rho)]}.
  \end{align}
  We define an inner product between operators as $\BraKet{A}{B} := \tr[A^\dag B]$. By Cauchy-Schwartz inequality, we have that $\abs{\langle x, y\rangle} \leq \norm{x} \norm{y}$. Since we have $\BraKet{A}{A} = \tr[A^\dag A] = \norm{A}_\rm{F}^2$, the appropriate norm is the Frobenius norm. Therefore we have
  \begin{align}
      c \leq \E_{U, Y} \left[\norm{\bb P_{\bm s}}_\rm{F} \norm{\rho - \bar \rho}_\rm{F}\right] = \norm{\bb P_{\bm s}}_\rm{F} \E_{U, Y} \left[\norm{\rho - \bar \rho}_\rm{F}\right].
  \end{align}
  Looking at $\norm{\bb P_s}_\rm{F}$, it will grow as $d^{L}$. However, we also have that
  \begin{align}
      p(\bm s \vert Y) = \tr[\bb P_{\bm s}^{\sf S} \tr_{\backslash \sf S}[\rho]],
  \end{align}
  where $\bb P_{\bm s}^{\sf S}$ only acts on a reduced Hilbert space of qubits $\sf S$ and $\tr_{\backslash \sf S}$ denotes the trace over all degrees of freedom other than qubits. Denoting the space of qudits as $\sf G$, we have $\backslash \sf S = \sf G$. Following through the line of reasoning as before, we have
  \begin{align}
      c = \E_{U, Y} \abs{\tr[\bb P_{\bm s}^\sf{S} \tr_{\sf G}(\rho - \bar \rho)]} = \E_{U, Y} \abs{\norm{\bb P_{\bm s}^\sf{S}}_\rm{F} \norm{\tr_{\sf G}(\rho - \bar \rho)}_\rm{F}} = \E_{U, Y} \norm{\tr_{\sf G}(\rho - \bar \rho)}_\rm{F}.
  \end{align}
  Suppose that the last operation was an application of $\cl U$, supported on $\sf{SG}(\bm{i})$, the Hilbert space of qubits and qudits on sites $\bm i$. Then
  \begin{align}
      c & = \E_{U,Y} \norm{\tr_{\sf G}(\rho_t-\bar \rho_t)}_\rm{F} = \E_{U, Y} \norm{\tr_{\sf G}(\rho_t-\bar \rho_t)}_\rm{F} = \E_{U, Y} \norm{\tr_{\sf G}(\cl U \rho_{t-1} - \bar{\cl U} \bar \rho_{t-1})}_\rm{F} \\
      & = \E_{U, Y} \left[\norm{\tr_{\sf G} \bar{\cl U} (\rho_{t-1} - \bar \rho_{t-1}) + \tr_{\sf G} (\cl U - \bar{\cl U}) \rho_{t-1}}_\rm{F} \right]
      \leq \E_{U, Y} \left[\norm{\tr_{\sf G} \bar{\cl U} (\rho_{t-1} - \bar \rho_{t-1})}_\rm{F} \right] + \E_{U, Y} \left[\norm{\tr_{\sf G} (\cl U - \bar{\cl U}) \rho_{t-1}}_\rm{F} \right].
  \end{align}
  In Lemmas \ref{lemma:bound-op-diff-norm} and \ref{lemma:depolarising-channel-norm}, we will prove that the above is bounded by
  \begin{align}
      c & \leq r_\rm{unitary} \E_{U, Y} \left[\norm{\tr_{\left(\sf G \backslash \sf S(\bm i)\right) \cup \sf G(\bm i)} (\rho_{t-1} - \bar \rho_{t-1})}_\rm{F} \right] + \frac{\phi}{d^{k/2}}
      = r_\rm{unitary} \E_{U, Y} \left[\norm{\tr_{\sf G} (\rho_{t-1} - \bar \rho_{t-1})}_\rm{F} \right] + \frac{\phi}{d^{k/2}},
  \end{align}
  for some constant $r_\rm{unitary}$ and $\phi$. If the last operation was instead an application of $\cl M$, with support on qubits on sites $\bm{i}$, we would have
  \begin{align}
      c = \E_{U, Y} \norm{\tr_{\sf G}[\cl M[\rho_{t-1}] - \cl M[\bar \rho_{t-1}]]}_\rm{F}.
  \end{align}
  We will also prove in Lemma \ref{lemma:measurement-map-norm} that
  \begin{align}
      \E_{U, Y} \norm{\tr_{\sf G}[\cl M[\rho_{t-1}] - \cl M[\bar \rho_{t-1}]]}_\rm{F} 
      & \leq r_\rm{meas}^{(1)} \E_{U, Y} \norm{\tr_{\sf G \backslash \sf S(\bm i)} (\rho_{t-1} - \bar \rho_{t-1})}_\rm{F} + r_\rm{meas}^{(2)}  \E_{U, Y} \norm{\tr_{\backslash \sf S(\bm i)} (\rho_{t-1} - \bar \rho_{t-1})}_\rm{F} \\
      & = r_\rm{meas}^{(1)} \E_{U, Y} \norm{\tr_{\sf G} (\rho_{t-1} - \bar \rho_{t-1})}_\rm{F} + r_\rm{meas}^{(2)}  \E_{U, Y} \norm{\tr_{\sf G \cup \sf S(\backslash \bm i)} (\rho_{t-1} - \bar \rho_{t-1})}_\rm{F}.
  \end{align}
  Then, we can recursively apply the same procedure as above for $\E_{U, Y}[\norm{\tr_{\sf K} (\rho_\tau - \bar \rho_\tau)}_\rm{F}]$ for subspaces of the form $\sf K$. Lemmas \ref{lemma:bound-op-diff-norm} and \ref{lemma:depolarising-channel-norm} also prove that, for subspaces of the form $\sf K = \sf G \cup \sf S (\cdot)$, i.e. contains \emph{all} qudits,
  \begin{align}
      \E_{U, Y} \left[\norm{\tr_{\sf K}(\rho_\tau - \bar \rho_\tau)}_\rm{F} \right] 
      \leq 
      r_\rm{unitary} \E_{U, Y} \left[\norm{\tr_{(\sf K \backslash \sf{S}(\bm i)) \cup \sf{G}(\bm i)}(\rho_{\tau-1} - \bar \rho_{\tau-1})}_\rm{F} \right] + \frac{\phi}{d^{k/2}},
  \end{align}
  if the previous operation was a unitary. On the other hand, Lemma \ref{lemma:measurement-map-norm} proves that
  \begin{align}
      \E_{U, Y} \left[\norm{\tr_{\sf K}(\rho_\tau - \bar \rho_\tau)}_\rm{F} \right] \leq r^{(1)}_\rm{meas} \E_{U, Y} \norm{\tr_{\sf K \backslash \sf S(\bm i)} (\rho_{\tau-1} - \bar \rho_{\tau-1})}_\rm{F} + r^{(2)}_\rm{meas} \E_{U, Y} \norm{\tr_{\sf G \cup \sf S(\backslash \bm i)} (\rho_{\tau-1} - \bar \rho_{\tau-1})}_\rm{F},
  \end{align}
  if the previous operation was a measurement. Note that starting with $\sf K = \sf G$, we only keep generating subspaces of form $\sf G \cup \sf S(\cdot)$. We can continue these bounds recursively until $\tau=0$, where we have that $\rho_0 = \bar \rho_0$, and therefore $\E_{U, Y}[\tr_{\sf K}(\rho_0 - \bar \rho_0)] = 0$. 
  
  Then, if we assume that a layer of measurement follows each layer of unitaries, and that each layer of unitary contains $\alpha V$ unitaries and each layer of measurement contains measurements, then we have
  \begin{align}
    c \leq \frac{\phi}{d^{k/2}} \left( \sum_{x=1}^{\alpha V} (r_\rm{unitary})^x \right) \left( \sum_{t=0}^{T-1} \left((r_\rm{unitary})^{\alpha} (r_\rm{meas})^\gamma \right)^{Vt} \right) 
    \leq \frac{\phi}{d^{k/2}} \times \alpha V \times r_\rm{unitary}^{L/2} \times T \times \left((r_\rm{unitary})^{\alpha} (r_\rm{meas})^\gamma \right)^{VT}.
  \end{align}
  
  For example, for single site measurements interleaved between Haar-random brickwork circuit,
  \begin{align}
      c \leq \frac{\phi}{d} \left( \sum_{x=1}^{L/2} r_\rm{unitary}^x \right) \left( \sum_{t=0}^{T-1} (r_\rm{unitary}^{1/2} r_\rm{meas})^{Lt} \right) 
      \leq \frac{\phi}{d} \times \frac{L}{2} \times r_\rm{unitary}^{L/2} \times T \times (r^{1/2}_\rm{unitary} r_\rm{meas})^{LT},
  \end{align}
  with $\phi = \sqrt{3}$, $r_\rm{unitary} = 10$, $r_\rm{meas} = r_\rm{meas}^{(1)} + r_\rm{meas}^{(2)} =  2(\sqrt{2} + 1)$.
\end{proof}

As a first example, we consider a full block Haar unitary gate.

\begin{lemma}
    Consider a full block Haar unitary gate supported on Hilbert space $\sf N$ with dimension $N$, acting on full Hilbert space that is potentially larger. Then 
    \begin{equation*}
        \E_{U}\left[ \norm{\tr_{\sf K} (\cl U - \cl D) \rho}_\rm{F} \right] \leq \frac{1}{\sqrt{\abs{\sf N \cap \sf K}}}.
    \end{equation*}
    for any state $\rho$, and the trace is on any subspace $\sf K$, which includes the trivial case where there is no partial trace.
\end{lemma}

\begin{proof}
    Using Jensen's inequality, we have that
    \begin{align}
        \E_{U}\left[ \norm{\tr_{\sf K} (\cl U - \cl D) \rho}_\rm{F} \right] \leq \sqrt{\E_{U, Y} \left[\norm{\tr_{\sf K} (\cl U - \cl D) \rho}_\rm{F}^2 \right]}.
    \end{align}
    Now consider
    \begin{align}
        \E_{U, Y} \left[\norm{\tr_{\sf K} (\cl U - \cl D) \rho}_\rm{F}^2 \right] 
        & = \E_{U, Y} \left[\norm{\tr_{\sf K} (u \rho u^\dag) - \tr_{\sf K} \cl D[\rho]}_\rm{F}^2 \right] 
        = \E_{U, Y} \left[\tr[\left(\tr_{\sf K} (u \rho u^\dag) - \tr_{\sf K} \cl D[\rho]\right) \left(\tr_{\sf K} (u \rho u^\dag) - \tr_{\sf K}  \cl D[\rho]\right)]\right] \\
        & = \E_{U, Y}\left[ \tr[\tr_{\sf K}(u \rho u^\dag) \tr_{\sf K}(u \rho u^\dag)] - 2 \tr[\tr_{\sf K} \cl D (\rho) \tr_{\sf K}(u \rho u^\dag)] + \tr[\tr_{\sf K} \cl D (\rho) \tr_{\sf K} \cl D (\rho)]\right].
    \end{align}
    Now taking the expectation value for $u$ only, we have that $\E_{U, Y}[u \rho u^\dag] = \cl D[\rho]$, such that
    \begin{align}
        \E_{U, Y} \left[\norm{\tr_{\sf K} (\cl U - \cl D) \rho}_\rm{F}^2 \right]
        & = \E_{U, Y}\left[ \tr[\tr_{\sf K}(u \rho u^\dag) \tr_{\sf K}(u \rho u^\dag)] - \tr[\tr_{\sf K} \cl D (\rho) \tr_{\sf K} \cl D (\rho)]\right].
    \end{align}
    Let us consider the first term.
    
    The following identities will prove to be useful, for an $N \times N$ Haar random unitary \cite{collins2022weingarten}:
    \begin{align}
        \E_{u} [u_{ij} u^*_{kl}] & = \frac{1}{N} \delta_{(ij),(kl)}, \\
        \E_{u} u_{ij} u^*_{mn} u_{k\ell} u^*_{pq}
        & = \frac{1}{N^2} \left(1 + \frac{1}{N^2 - 1} \right)(\delta_{(ij),(mn)} \delta_{(kl),(pq)} + \delta_{(ij),(pq)} \delta_{(kl),(mn)}) \\ 
        & - \frac{1}{N(N^2-1)} (\delta_{im} \delta_{jq} \delta_{kp}\delta_{\ell n} + \delta_{ip} \delta_{jn} \delta_{km}\delta_{\ell q}) \\
        & = \frac{1}{N} \delta_{(ij),(mn)} \frac{1}{N} \delta_{(kl),(pq)} + \frac{1}{N^2} \delta_{(ij),(pq)} \delta_{(kl),(mn)} \\
        & + \frac{1}{N^2(N^2-1)} \Big( \delta_{(ij),(mn)} \delta_{(kl),(pq)} + \delta_{(ij),(pq)} \delta_{(kl),(mn)} - \delta_{im} \delta_{jq} \delta_{kp}\delta_{\ell n} - \delta_{ip} \delta_{jn} \delta_{km}\delta_{\ell q} \Big).
    \end{align}
    These can be written diagrammatically as the following:
    \begin{align}
        \E_{u} \left[

    \right).
\end{align}
    Here $\chi_N$ is the maximally mixed state on Hilbert space acted on by the Haar unitary, $\sf N$.  Now consider the partial trace over $\sf K$. Part of it will be within the $\sf N$ space, denoted by $\sf K \cup \sf N$ while the rest will be over $\backslash \sf N$ space, denoted by $\sf K \backslash \sf N$. To denote $\sf K \cup \sf N$ and $\sf N \backslash (\sf K \cup \sf N)$ separately, we can double the lines as following
    \begin{equation}
        \Ket{\rho} \leftrightarrow
        %
,
    \end{equation}
    where gray lines are in $\sf K$, black lines in $\backslash \sf K$, and solid lines are in $N$ and dotted lines are in $\backslash \sf N$. All the lines in the unitary relations can also be doubled to accomodate this, i.e.
    \begin{align}
        %
.
    \end{align}
    For example, we have
    \begin{align}
        \tr_{\sf K} \cl U [\rho] \leftrightarrow \tr_{\sf K} \cl U \Ket{\rho} \leftrightarrow 
        %
,
    \end{align}
    such that we have
    \begin{align}
        \tr[\left(\tr_{\sf K} \cl U [\rho]\right)^2] \leftrightarrow
        %
.
    \end{align}
    Then, we can apply the identities. For example, the first two terms in is equal to
    \begin{align}
        %
        \leftrightarrow
        \tr[\tr_{\sf K} \cl D[\rho] \tr_{\sf K}\cl D[\rho]] 
        +
        \tr_{\sf N \cap \sf K}[(\tr_{\sf N \backslash \sf K}[\chi_N])^2] \cdot \tr[\rho^2].
    \end{align}
    Additionally using that $\rho^\dag = \rho$ on the other terms, we find
    \begin{align}
        \E_{U, Y}\left[ \tr[\tr_{\sf K}(u \rho u^\dag) \tr_{\sf K}(u \rho u^\dag)] \right] 
        & = \tr[\tr_{\sf K} \cl D[\rho] \tr_{\sf K}\cl D[\rho]] 
        +
        \tr_{\sf N \cap \sf K}[(\tr_{\sf N \backslash \sf K}[\chi_N])^2] \cdot \tr[\rho^2]
        \\
        & + \frac{1}{N^2 - 1} \left(\tr[\tr_{\sf K} \cl D[\rho] \tr_{\sf K}\cl D[\rho]] + \tr_{\sf N \cap \sf K}[(\tr_{\sf N \backslash \sf K}[\chi_N])^2] \cdot \tr[\rho^2]\right) \\
        & - \frac{1}{N^2 - 1} \left(\tr[\tr_{\sf K} \cl D[\rho] \tr_{\sf K}\cl D[\rho]] + \tr_{\sf N \cap \sf K}[(\tr_{\sf N \backslash \sf K}[\chi_N])^2] \cdot \tr[\rho^2] \right).
    \end{align}
    The last two terms cancel out, and $\tr_{\sf N \cap \sf K}[(\tr_{\sf N \backslash \sf K}[\chi_N])^2] = \tr_{\sf N \cap \sf K} [\chi_{\sf N \cap \sf K}] = 1/\abs{\sf N \cap \sf K}$. Furthermore $\tr_{\sf K \backslash \sf N}[\rho] = \rho'$ is another state, and the trace of square of a density matrix is upper bounded by 1. The first term exactly cancels out the second term in $\E_{U, Y} \left[\norm{\tr_{\sf K} (\cl U - \cl D) \rho}_\rm{F}^2 \right]$, and therefore we may upper bound
    \begin{align}
        \E_{U, Y} \left[\norm{\tr_{\sf K} (\cl U - \cl D) \rho}_\rm{F}^2 \right] \leq \frac{1}{\abs{\sf N \cap \sf K}}
    \end{align}
    with no further terms. Using Jensen's inequality, we have
    \begin{align}
        \E_{U, Y} \left[\norm{\tr_{\sf K} (\cl U - \cl D) \rho}_\rm{F} \right] \leq \frac{1}{\abs{\sf N \cap \sf K}}.
    \end{align}
\end{proof}

\begin{corollary}
    For two-site full block Haar unitary acting on a qubit-qudit pair with $\sf K = \sf G \cup \sf S (\cdot)$, we have $\sf N = \sf G(i_1, i_2) \cup \sf S(i_1, i_2)$, and therefore the bound
    \begin{align}
        \E_{U, Y} \left[\norm{\tr_{\sf K} (\cl U - \cl D) \rho}_\rm{F} \right] \leq \frac{1}{d}.
    \end{align}
\end{corollary}

\begin{lemma}
    For the depolarising channel $\cl D$ acting on $\sf N$ and any matrix $A$ living in a potentially larger Hilbert space,
    \begin{equation}
        \norm{\tr_{\sf K} \cl D A}_\rm{F} \leq \frac{1}{\sqrt{\abs{\sf N \backslash \sf K}}} \norm{\tr_{\sf N}\tr_{\sf K \backslash \sf N} A}_\rm{F},
    \end{equation}
    for any subspace $\sf K$.
\end{lemma}

\begin{proof}
    \begin{align}
        \norm{\tr_{\sf K} \cl D A}^2_\rm{F} 
        & = \tr[\left(\tr_{\sf K \cap \sf N} \cl D \tr_{\sf K \backslash \sf N} A^\dag \right) \left(\tr_{\sf K \cap \sf N} \tr_{\sf K \backslash \sf N} A \right)] \\
        & = \tr[\left(\tr_{\sf K \cap \sf N} \chi_{\sf N} \otimes \tr_{\sf N}\tr_{\sf K \backslash \sf N}[A^\dag] \right) \left(\tr_{\sf K \cap \sf N} \chi_{\sf N} \otimes \tr_{\sf N}\tr_{\sf K \backslash \sf N}[A] \right)] \\
        & = 
        \tr_{\sf N \backslash \sf K}[\tr_{\sf K \cap \sf N} [\chi_{\sf N}] \tr_{\sf K \cap \sf N} [\chi_{\sf N}]] 
        \times
        \tr[\tr_{\sf N}\tr_{\sf K \backslash \sf N}[A^\dag]\tr_{\sf N}\tr_{\sf K \backslash \sf N}[A]] \\
        & = \frac{1}{\abs{\sf N \backslash \sf K}} \norm{\tr_{\sf N}\tr_{\sf K \backslash \sf N} A}^2_\rm{F}.
    \end{align}
\end{proof}

\begin{remark}
    For any block-diagonal Haar unitary $u = u^{(0)} \oplus \cdts \oplus u^{(B)}$ acting on $\sf N$, with $\sum_{\alpha=1}^B N_\alpha= N$, we have
    \begin{align}
        \E_u \left[\cl U\right] = \cl D_{\sf N_1} \oplus \cdts \oplus \cl D_{\sf N_B} = \sum_\alpha \frac{\bb P_\alpha^{\sf N}}{N_\alpha} \otimes \tr_{\sf N} \cl P_\alpha.
    \end{align}
\end{remark}

\begin{lemma} \label{lemma:bound-op-diff-norm}
    For a block-diagonal Haar unitary $u = u^{(0)} \oplus \cdts \oplus u^{(B)}$ with $\sf N = \sf m \otimes \sf n$, $\sf n = \bigoplus_\alpha \sf n_\alpha$ $\sum_\alpha N_\alpha = N$ where $N_\alpha = m n_\alpha$ and $\sf m \subset \sf K$,
    \begin{equation*}
        \E_{U}\left[ \norm{\tr_{\sf K} (\cl U - \cl D_{\sf N_1} \oplus \cdts \oplus \cl D_{\sf N_B}) \rho}_\rm{F} \right] \leq \sum_\alpha \tr_{\sf N \cap \sf K}\left[\tr_{\sf N \backslash \sf K}\left[\frac{\bb P^{\sf N}_\alpha}{N_{\alpha}} \right]^2\right] \leq \sqrt{\frac{B}{m}}.
    \end{equation*}
\end{lemma}

\begin{proof}
The block diagonal Haar unitary can be written as
\begin{align}
    u = \sum_\alpha \bb P_\alpha u_\alpha \bb P_\alpha \leftrightarrow
.
\end{align}
Here $u_\alpha = u^{(\alpha)} \oplus \bb 1_{N \backslash N_\alpha}$. Then we have
\begin{align}
    \left[ \norm{\tr_{\sf K} (\cl U - \cl D_{\sf N_1} \oplus \cdts \oplus \cl D_{\sf N_B}) \rho}_\rm{F} \right] = 
    %
.
\end{align}
Since there are projectors in $v_\alpha$, we have
\begin{align}
    = 
    \sum_{\alpha \beta \gamma \delta}
    %
.
\end{align}
From Haar averaging, one of (i) $\alpha = \beta$, $\gamma =\delta$, $\alpha \neq \gamma$, (ii) $\alpha = \gamma$, $\beta = \delta$, $\alpha \neq \beta$, or (iii) $\alpha = \beta = \gamma = \delta$ must be satisfied. (i) and (ii) are disallowed by projectors. Therefore only (iii) is allowed, and we have
\begin{align}
    \mathop{\longrightarrow}^{\E[\cdot]}
    \sum_\alpha \E \left[
    %
\right].
\end{align}
Therefore we have
\begin{align}
    \E_{U} \left[ \norm{\tr_{\sf K} (\cl U - \cl D_{m_1} \oplus \cdts \oplus \cl D_{m_B}) \rho}_\rm{F} \right]^2 
    & \leq
    \sum_\alpha \tr_{\sf N \cap \sf K}[\tr_{\sf N \backslash \sf K}[\chi_{N_\alpha}]^2] \cdot \tr[\cl P_{\alpha}[\rho]^2] 
    = \sum_\alpha \tr_{\sf N \cap \sf K}\left[\tr_{\sf N \backslash \sf K}\left[\frac{\bb P^{\sf N}_\alpha}{N_{\alpha}} \right]^2\right] \cdot \tr[\cl P_{\alpha}[\rho]^2] \\
    & \leq \sum_\alpha \tr_{\sf N \cap \sf K}\left[\tr_{\sf N \backslash \sf K}\left[\frac{\bb P^{\sf N}_\alpha}{n_{\alpha} m} \right]^2\right]
    \leq \sum_\alpha m \frac{1}{m^2} \leq \frac{B}{m}.
\end{align}
\end{proof}

\begin{corollary}
    For the $U(1)$-symmetric gate acting on two qubit-qudit pairs and $\sf K = \sf G \cup \sf S (\cdot)$, we have
    \begin{align}
        \E_{u} \left[ \norm{\tr_{\sf K} (\cl U_{U(1)} - \cl E_\rm{SSEP} \otimes \cl D_\rm{qudits}) \rho}_\rm{F} \right] \leq \frac{\sqrt{3}}{d}.
    \end{align}
\end{corollary}

\begin{lemma} \label{lemma:depolarising-channel-norm}
    Consider $(\cl D_{N_1} \oplus \cdts \oplus \cl D_{N_B})$ that decomposes as $\cl D_{m} \otimes (\cl D_{n_1} \oplus \cdts \oplus \cl D_{n_B})$, such that $\sum_\alpha n_\alpha = n$ and $nm = N$ $\sf n \otimes \sf m = \sf N$, and $\cl D_m$ acts on $\sf m$. Then we have
\begin{align}
    \norm{\tr_{\sf K} (\cl D_{\sf m} \otimes \bigoplus_\alpha \cl D_{\sf n_\alpha}) A}_\rm{F} \leq \frac{\sqrt{n \abs{\sf K \cap \sf n}}}{\abs{\sf m \backslash K}}\left(\sum_\alpha \frac{1}{n_\alpha} \right)\norm{\tr_{\sf m} \tr_{\sf K \backslash \sf n}[A]}_\rm{F}.
\end{align}
\end{lemma}
\begin{proof}
    Then we have that
    \begin{align}
        \norm{\tr_\sf{K} \left(\cl D_{\sf m} \otimes \bigoplus_\alpha \cl D_{\sf n_\alpha}\right) A}_\rm{F}^2 
        & = \norm{\tr_\sf{K} \left(\cl D_{\sf m} \otimes \bigoplus_\alpha \cl D_{\sf n_\alpha} \right) [A]}_\rm{F}^2 \\
        & = \norm{\tr_{\sf K \cap \sf m} \chi_m \otimes \tr_{\sf K \cap \sf n} \left(\bigoplus_\alpha \cl D_{\sf n_\alpha}\right) \tr_{\sf m} \tr_{\sf K \backslash \sf N}[A]}_\rm{F}^2 \\
        & = \frac{1}{\abs{\sf m \backslash \sf K}} \norm{\tr_{\sf K \cap \sf n} \sum_\alpha \frac{\bb P_\alpha^{\sf n}}{n_\alpha} \otimes \tr_{\sf n} \cl P_\alpha \tr_{\sf m} \tr_{\sf K \backslash \sf n}[A]}_\rm{F}^2 \\
        & \leq \frac{1}{\abs{\sf m \backslash \sf K}} \left( \sum_\alpha \norm{\tr_{\sf K \cap \sf n} \frac{\bb P_\alpha^{\sf n}}{n_\alpha} \otimes \tr_{\sf n}[ \cl P_\alpha \tr_{\sf m} \tr_{\sf K \backslash \sf n}[A]]}_\rm{F}\right)^2 \\
        & \leq \frac{1}{\abs{\sf m \backslash \sf K}} \left(\sum_\alpha \frac{\sqrt{\abs{\sf K \cap \sf n}}}{n_\alpha} \norm{\tr_{\sf n}[\bb P_\alpha \tr_{\sf m} \tr_{\sf K \backslash \sf n}[A]]}_\rm{F}\right)^2 \\
        & \leq \frac{1}{\abs{\sf m \backslash \sf K}} \left(\sum_\alpha \frac{\sqrt{\abs{\sf K \cap \sf n}}}{n_\alpha} \sqrt{n} \norm{\bb P_\alpha \tr_{\sf m} \tr_{\sf K \backslash \sf n}[A]}_\rm{F}\right)^2 \\
    \end{align}
    
    Now observe that $\norm{\bb P_\alpha A}^2_\rm{F} = \tr[A^\dag \bb P_\alpha A] \geq 0$. Also we have that $\norm{A^\dag A}^2_\rm{F} = \tr[A^\dag A] = \sum_\alpha \tr[A^\dag \bb P_\alpha A] = \sum_\alpha \norm{\bb P_\alpha A}^2_\rm{F}$. Therefore, $\norm{\bb P_\alpha A}_\rm{F} \leq \norm{A^\dag A}_\rm{F}$. Therefore we have
    \begin{align}
        \norm{\tr_\sf{K} \left(\cl D_{\sf m} \otimes \bigoplus_\alpha \cl D_{\sf n_\alpha}\right) A}_\rm{F}
        & \leq \frac{1}{\abs{\sf m \backslash \sf K}} \sum_\alpha \frac{\sqrt{\abs{\sf K \cap \sf n}}}{n_\alpha} \sqrt{n} \norm{\tr_{\sf m} \tr_{\sf K \backslash \sf n}[A]}_\rm{F} \\
        & = \frac{\sqrt{n \abs{\sf K \cap \sf n}}}{\abs{\sf m \backslash K}}\left(\sum_\alpha \frac{1}{n_\alpha} \right)\norm{\tr_{\sf m} \tr_{\sf K \backslash \sf n}[A]}_\rm{F}.
    \end{align}
\end{proof}

\begin{corollary}
    For the SSEP gate, we have
    \begin{align}
        \norm{\tr_{\sf K} \E_u[\cl U] A}_\rm{F} 
        \leq
        10 \norm{\tr_{\sf m} \tr_{\sf K \backslash \sf n}[A]}_\rm{F},
    \end{align}
    and therefore $r_\rm{unitary} = 10$.
\end{corollary}

\begin{lemma} \label{lemma:measurement-map-norm}
    Consider a measurement map $\cl M$ that acts on $\sf S (\bm i)$. Then we have
    \begin{align}
        \E_{U, Y} \norm{\tr_{\sf K}[\cl M[\rho_\tau] - \cl M[\bar \rho_\tau]]}_\rm{F} \leq \abs{\sf S (\bm i)} \left(\sqrt{\abs{\sf K \cap \sf S(\bm i)}} \E_{U, Y} \norm{\tr_{\sf K \backslash \sf S(\bm i)} (\rho_\tau - \bar \rho_\tau)}_\rm{F} + \E_{U, Y} \norm{\tr_{\backslash \sf S(\bm i)} (\rho_\tau - \bar \rho_\tau)}_\rm{F} \right).
    \end{align}
    For $\sf K = \emptyset$, $\abs{\sf K \cap \sf S(\bm i)} \rightarrow 1$. 
\end{lemma}
\begin{proof}
    For $\cl Q[\rho] := Q \rho Q$,
\begin{align}
    & \E_{U, Y} \norm{\tr_{\sf K}[\cl M[\rho_\tau] - \cl M[\bar \rho_\tau]]}_\rm{F} \\
    & = \E_{U, Y} \norm{\frac{\tr_{\sf K}[\cl Q \rho_\tau]}{\tr[\cl Q \rho_\tau]} - \frac{\tr_{\sf K}[\cl Q \bar \rho_\tau]}{\tr[\cl Q \bar \rho_\tau]}}_\rm{F}
    = \E_{U, Y} \left[\frac{1}{\tr[\cl Q \rho_\tau] \tr[\cl Q \bar \rho_\tau]} \norm{\tr[\cl Q \bar \rho_\tau] \tr_{\sf K}[\cl Q \rho_\tau] - \tr[\cl Q \rho_\tau] \tr_{\sf K}[\cl Q \bar \rho_\tau]}_\rm{F} \right] \\
    & = \E_{U, Y} \left[\frac{1}{\tr[\cl Q \rho_\tau] \tr[\cl Q \bar \rho_\tau]} \norm{\tr[\cl Q \bar \rho_\tau] \tr_{\sf K}[\cl Q (\rho_\tau - \bar \rho_\tau)] - (\tr[\cl Q \rho_\tau] - \tr[\cl Q \bar \rho_\tau]) \tr_{\sf K}[\cl Q \bar \rho_\tau]}_\rm{F} \right] \\
    & \leq \E_{U, Y} \left[\frac{1}{\tr[\cl Q \rho_\tau] \tr[\cl Q \bar \rho_\tau]} \left(\tr[\cl Q \bar \rho_\tau] \norm{\tr_{\sf K}[\cl Q (\rho_\tau - \bar \rho_\tau)]}_\rm{F} + \abs{\tr[\cl Q (\rho_\tau - \bar \rho_\tau)]} \norm{\tr_{\sf K}[\cl Q \bar \rho_\tau]}_\rm{F} \right)\right] \\
    & \leq \E_{U, Y} \left[\frac{1}{\tr[\cl Q \rho_\tau] \tr[\cl Q \bar \rho_\tau]} \left(\tr[\cl Q \bar \rho_\tau] \norm{\tr_{\sf K}[\cl Q (\rho_\tau - \bar \rho_\tau)]}_\rm{F} + \norm{\cal Q_\rm{loc}}_\rm{F} \norm{\tr_{\backslash \sf S(\bm i)} (\rho_\tau - \bar \rho_\tau)}_\rm{F} \norm{\tr_{\sf K}[\cl Q \bar \rho_\tau]}_\rm{F} \right)\right]
\end{align}
We first look at $\norm{\tr_{\sf K}[\cl Q \bar \rho_\tau]}^2_\rm{F}$. Note that $\bar \rho_\tau$ is a completely diagonal density matrix. Therefore
\begin{align}
    \norm{\tr_{\sf K}[\cl Q \bar \rho_\tau]}^2_\rm{F} & = \tr[(\tr_{\sf K}\cl Q \bar \rho_\tau)(\tr_{\sf K} \cl Q \bar \rho_\tau)] = \sum_{s, s', s'', s'''} \sqrt{p(y \vert s) p(y \vert s') p(y \vert s'') p(y \vert s''')} \tr[\tr_{\sf K}[\bb P_s \bar \rho_\tau \bb P_{s''}] \tr_{\sf K}[\bb P_{s''} \bar \rho_\tau \bb P_{s'''}]],
\end{align}
where $s,s',s'',s'''$ are the degrees of freedom involved in the measurement. 
Now $\bar \rho_\tau$ is completely diagonal. Therefore this gives $\delta_{s, s'}$, $\delta_{s'', s'''}$:
\begin{align}
    \norm{\tr_{\sf K}[\cl Q \bar \rho_\tau]}^2_\rm{F} & = \sum_{s, s''} p(y \vert s) p(y \vert s'') \tr[\tr_{\sf K}[\bb P_s \bar \rho_\tau \bb P_s] \tr_{\sf K}[\bb P_{s''} \bar \rho_\tau \bb P_{s''}]].
\end{align}
Now $\bb P_{s} = \bb P_{s \backslash \sf K} \bb P_{s \cap \sf K}$, and $\bb P_{s \backslash \sf K}$ commutes through the trace. Therefore
\begin{align}
    \norm{\tr_{\sf K}[\cl Q \bar \rho_\tau]}^2_\rm{F}
    & = \sum_{s, s''} p(y \vert s) p(y \vert s'') \tr[ \bb P_{s \backslash \sf K}\tr_{\sf K}[\bb P_{s \cap \sf K} \bar \rho_\tau \bb P_{s \cap \sf K}] \bb P_{s \backslash \sf K} \bb P_{s'' \backslash K} \tr_{\sf K}[\bb P_{s'' \cap \sf K} \bar \rho_\tau \bb P_{s'' \cap \sf K}] \bb P_{s'' \backslash \sf K}].
\end{align}
Using that $\bar \rho = \sum_{\bm s, \bm{g}} \bar{p}(\bm s, \bm{g}) \ketbra{\bm s, \bm g}{\bm s, \bm g}$, where $\ket{\bm s} \in \sf S$, $\ket{\bm g} \in \sf G$,
\begin{align}
    \norm{\tr_{\sf K}[\cl Q \bar \rho_\tau]}^2_\rm{F}
    & = \sum_{s, s''} p(y \vert s) p(y \vert s'') \delta_{s \backslash \sf K, s'' \backslash \sf K} \bar p(s) \bar p(s'')
    \leq \sum_{s, s''} p(y \vert s) p(y \vert s'') \bar p(s) \bar p(s'') = \bar p(y)^2 = \tr[\cl Q \bar \rho_\tau]^2.
\end{align}
Therefore
\begin{align}
    & \E_{U, Y} \norm{\tr_{\sf K}[\cl M[\rho_\tau] - \cl M[\bar \rho_\tau]]}_\rm{F} \\
    & \leq \E_{U, Y} \left[\frac{1}{\tr[\cl Q \rho_\tau] \tr[\cl Q \bar \rho_\tau]} \left(\tr[\cl Q \bar \rho_\tau] \norm{\tr_{\sf K}[\cl Q (\rho_\tau - \bar \rho_\tau)]}_\rm{F} + \norm{\cl{Q}^{\sf S (\bm i)}}_\rm{F} \norm{\tr_{\backslash \sf S(\bm i)} (\rho_\tau - \bar \rho_\tau)}_\rm{F} \tr[\cl Q \bar \rho_\tau] \right)\right] \\
    & = \E_{U, Y} \left[\frac{1}{\tr[\cl Q \rho_\tau]} \left(\norm{\tr_{\sf K}[\cl Q (\rho_\tau - \bar \rho_\tau)]}_\rm{F} + \norm{\cl{Q}^{\sf S(\bm i)}}_\rm{F} \norm{\tr_{\backslash \sf S(\bm i)} (\rho_\tau - \bar \rho_\tau)}_\rm{F} \right)\right].
\end{align}
Now consider
\begin{align}
    \norm{\tr_{\sf K}[\cl Q (\rho_\tau - \bar \rho_\tau)]}_\rm{F} & = \norm{\tr_{\sf K \cap \sf S(\bm i)} [\cl Q^{\sf S(\bm i)} \tr_{\sf K \backslash \sf S(\bm i)} (\rho_\tau - \bar \rho_\tau)]}_\rm{F}  \leq \norm{\tr_{K \cap s} [\cl Q^{\sf S(\bm i)} \tr_{K \backslash \sf S(\bm i)} (\rho_\tau - \bar \rho_\tau)]}_\rm{F} \\
    & \leq \sqrt{\abs{\sf K \cap \sf S(\bm i)}} \norm{\cl Q^{\sf S(\bm i)} \tr_{K \backslash \sf S(\bm i)} (\rho_\tau - \bar \rho_\tau)}_\rm{F} \leq \sqrt{\abs{\sf K \cap \sf S(\bm i)}} p_\rm{max} \norm{\tr_{K \backslash \sf S(\bm i)} (\rho_\tau - \bar \rho_\tau)}_\rm{F},
\end{align}
where $p_\rm{max} = \max_s p(y \vert s)$, and the third line is due to Rastegin '12. Similarly we have $\norm{\cl Q^{\sf S(\bm i)}}_\rm{F} \leq p_\rm{max}$.

Notice that $p(y_\tau) = \tr[\cl Q \rho_\tau]$, such that $\E_{y_\tau}[\frac{1}{\tr[\cl Q \rho_\tau]} \cdot] = \int d y_\tau [\cdot]$. Therefore
\begin{align}
    \E_{U, Y} \norm{\tr_{\sf K}[\cl M[\rho_\tau] - \cl M[\bar \rho_\tau]]}_\rm{F} \leq \left(\int_y p_\rm{max}(y)\right) \E_{U, y_{1:\tau-1}} \left[\sqrt{\abs{\sf K \cap \sf S(\bm i)}} \norm{\tr_{K \backslash \sf S(\bm i)} (\rho_\tau - \bar \rho_\tau)}_\rm{F} + \norm{\tr_{\backslash \sf S(\bm i)} (\rho_\tau - \bar \rho_\tau)}_\rm{F} \right].
\end{align}
Now $p_\rm{max}(y) = \max_s p(y \vert s) \leq \sum_s p(y \vert s)$. Therefore $\int_y p_\rm{max}(y) \leq \sum_{s} \int_y p(y \vert s) = \sum_s 1 = \abs{\sf S(\bm i)}$. Therefore
\begin{align}
    \E_{U, Y} \norm{\tr_{\sf K}[\cl M[\rho_\tau] - \cl M[\bar \rho_\tau]]}_\rm{F} 
    \leq
    \abs{\sf S(\bm i)} \left(\sqrt{\abs{\sf K \cap \sf S(\bm i)}} \E_{U, Y} \norm{\tr_{\sf K \backslash \sf S(\bm i)} (\rho_\tau - \bar \rho_\tau)}_\rm{F} + \E_{U, Y} \norm{\tr_{\backslash \sf S(\bm i)} (\rho_\tau - \bar \rho_\tau)}_\rm{F} \right).
\end{align}
\end{proof}

\begin{corollary}
    For on-site measurements, we have
    \begin{align}
        \E_{U, Y} \norm{\tr_{\sf K}[\cl M[\rho_\tau] - \cl M[\bar \rho_\tau]]}_\rm{F} 
        \leq
        2 \left(\sqrt{2} \E_{U, Y} \norm{\tr_{\sf K \backslash \sf S(\bm i)} (\rho_\tau - \bar \rho_\tau)}_\rm{F} + \E_{U, Y} \norm{\tr_{\backslash \sf S(\bm i)} (\rho_\tau - \bar \rho_\tau)}_\rm{F} \right),
    \end{align}
    and therefore $r_\rm{meas} = 2(\sqrt{2} + 1)$.
\end{corollary}

\end{document}